\documentclass{ws-ijmpe}

\usepackage{graphicx}
\usepackage{amsmath}
\newlength{\epigraphwidth}
\begin{document}

\markboth{T. Rauscher}
{The Path to Improved Reaction Rates for Astrophysics}

%
\catchline{}{}{}{}{}
%

\title{The Path to Improved Reaction Rates for Astrophysics}

\author{Thomas Rauscher}

\address{Department of Physics, University of Basel, Klingelbergstr.\ 82\\
4056 Basel, Switzerland\\
Thomas.Rauscher@unibas.ch}

\maketitle

\begin{history}
\received{5 Oct 2010}
\end{history}

\begin{abstract}
This review focuses on nuclear reactions in astrophysics and, more specifically, on reactions with light ions (nucleons and $\alpha$ particles) proceeding via the strong interaction.
It is intended to present the basic definitions essential for studies in nuclear astrophysics, to point out the differences between nuclear reactions taking place in stars and in
a terrestrial laboratory, and to illustrate some of the challenges to be faced in theoretical and experimental studies of those reactions. The discussion revolves around the relevant quantities for astrophysics, which are the astrophysical reaction rates.
The sensitivity of the reaction rates to the uncertainties in the prediction of various nuclear properties is explored and some guidelines for experimentalists
are also provided.

\end{abstract}

\ccode{PACS numbers: 26.50.+x, 26.20.-f, 26.30.-k, 26.35.+c, 95.30.Cq, 98.80.Ft, 24.60.Dr, 24.50.+g, 24.10.Eq}

\setcounter{secnumdepth}{4}
\setcounter{tocdepth}{4}

\newpage
\tableofcontents
\newpage

\markboth{T. Rauscher}
{The Path to Improved Reaction Rates for Astrophysics}

\hfill \begin{minipage}{1.2\epigraphwidth}
{\vspace{1cm}
\flushright \footnotesize \it
Caminante, son tus huellas el camino, y nada m\'as;\\
Caminante, no hay camino, se hace camino al andar.\\
Al andar se hace camino y al volver la vista atr\'as\\
se ve la senda que nunca se ha de volver de pisar.\\
Caminante, no hay camino sino estelas en la mar.\\
\hfill \textsc{Antonio Machado}
}
\end{minipage}

\section{Introduction}

Natural processes proceed simultaneously on many different scales interacting with each other. This is especially important in astrophysics, where an astronomical observation or a hypothesized system can only be understood by modeling processes on scales spanning many orders of magnitude. Thus, also nuclear physics and astrophysics are closely entwined. Nuclear reactions power quiescent burning of stars and cause the most powerful explosions known.
They not only release or transform energy but also change the composition of the matter in which they occur and thus are responsible for the range of chemical elements found on our planet and throughout the Universe. Finally, to understand the properties of matter at extreme density and/or density, the nuclear equation of state has to be known. It determines the properties of neutron stars and the latest stages of the life of stars with more than $8 M_\odot$ which end their lives in a supernova explosion.

It is evident that nuclear physics input is essential for many astrophysical models and this fact is represented in the field of Nuclear Astrophysics. There are different perceptions on how to define this field. Some limit it to the application of nuclear physics to reactions of astrophysical interest. A more comprehensive, and perhaps more adequate, definition includes the more astrophysical aspects in the investigation of nuclear processes and nucleosynthesis in astrophysical sites through reaction networks. In any case, the interests of astrophysics emphasize different aspects than those of basic nuclear physics and this makes Nuclear Astrophysics a distinct research area. From the nuclear point of view, astrophysics involves low-energy reactions with light projectiles on light, intermediate, and heavy nuclei. Although nuclear physics has moved to higher energies in the last decades, low-energy reactions are not well enough explored and still offer considerable challenges to both experiment and theory, even for stable target nuclei. Explosive conditions in astrophysics favor the production of nuclei far off stability and reaction rates for these have to be predicted across the nuclear chart. This proves very difficult, especially for first-principle methods, due to the complexity of the nuclear many-body problem. The calculation of astrophysical reaction rates also includes special requirements and processes not studied in nuclear physics so far. Among these are effects appearing in low energy, subCoulomb reactions and reactions on excited states of target nuclei. Depending on the conditions, plasma effects also have to be considered because these alter the reaction rates. This includes, e.g., the shielding of Coulomb barriers through electrons, pycnonuclear burning in the lattice of a high-density plasma, and the modification of nuclear partition functions at very high plasma temperatures.

This paper focuses on nuclear reactions in astrophysics and, more specifically, on reactions with light ions (nucleons and $\alpha$ particles) proceeding via the strong interaction.
It is intended to present the basic definitions essential for studies in nuclear astrophysics, to point out the differences between nuclear reactions taking place in stars and in
a terrestrial laboratory, and to illustrate some of the challenges to be faced in theoretical and experimental studies of those reactions. The sensitivity of the reaction rates to the uncertainties in the prediction of various nuclear properties is explored and some guidelines for experimentalists
are also provided. The discussion revolves around the relevant quantities for astrophysics, which are the astrophysical reaction rates. The impact of using different models or data is always presented with respect to the possible modification of the reaction rates, not the cross sections.

At first, the basic equations through which nuclear processes enter astrophysical models are introduced in Sec.~\ref{sec:netrate}. The astrophysically relevant energy ranges are defined in Sec.~\ref{sec:energies}. The thorough discussion of the special stellar effects affecting reactions in Sec.~\ref{sec:stellareffects} is the heart of this review. Stellar and effective cross sections are derived in Sec.~\ref{sec:stellar}, then the relation between laboratory reactions and those occurring in a stellar plasma is investigated in Sec.~\ref{sec:stellarexp}. As it is important to realize when individual reactions are important and when they are not, reaction equlibria are introduced in Sec.~\ref{sec:equilibria}. Cross sections in a plasma are affected by the free electrons present which shield the nuclear charge. This is explained in Sec.~\ref{sec:electro}. Finally, astrophysically relevant reaction mechanisms are reviewed in Sec.~\ref{sec:mech} and reactions through isolated resonances (Sec.~\ref{sec:reso}), in systems with high nuclear level density (Sec.~\ref{sec:statmod}), and direct reactions (Sec.~\ref{sec:direct}) are discussed separately. This includes a detailed discussion of the sensitivities of the rates on the required input in Secs.~\ref{sec:relevance} and \ref{sec:directsensi}. Section \ref{sec:conclusion} presents a brief conclusion.

\section{Reaction networks and rates}
\label{sec:netrate}

Nuclear reactions are the engine of stellar evolution and
determine the overall production of the known chemical elements and their
isotopes in a variety of nucleosynthesis patterns. A detailed
understanding of the characteristic production and depletion rates of nuclei
within the framework of a nucleosynthesis process is crucial for
reliable model predictions and the interpretation of the observed
abundances.

Instead of the number $M$ of nuclei of a given species per volume $V$ (the number density $n=M/V$), it is advantageous
to use a quantity independent of density changes: the abundance $Y=n/(\rho_\mathrm{pla} N_\mathrm{A})$, where $\rho_\mathrm{pla}$ is
the plasma density and $N_\mathrm{A}$ denotes Avogadro's number.
The change of abundances $Y$ with time due to nuclear processes is
traced by coupled differential equations. Due to the nature of the involved reactions and the vastly different timescales
appearing, the equation system is non-linear and stiff. In addition, for complete solubility of the coupled equations
the number of equations $N$ has to equal the number of involved
nuclei acting as reaction partners and thus an equation matrix of size $N^2$ has
to be solved. Nucleosynthesis processes include thousands of nuclides and tens of thousands reactions.
This still makes it impossible to fully couple such a reaction
network to a full set of hydrodynamic equations as would be required for a complete modeling of nucleosynthesis in a given
astrophysical site.

A reaction network generally can be written as \cite{radiobook}
\begin{equation}
\label{eq:network}
 \dot{Y}_i=\frac{1}{\rho_\mathrm{pla} N_\mathrm{A}} \dot{n}_i = \frac{1}{\rho_\mathrm{pla} N_\mathrm{A}}
 \left\{ \sum_j {^{1}_{i}K_{j}} \; {_{i}\lambda_{j}} + \sum_{j} {^{2}_{i}K_{j}}\; {_{i}r_{j}} +
 \sum_j {^{3}_{i}K_{j}}\; {_{i}\hat{r}_{j}} + \dots \right\} \quad,
\end{equation}
where $1\leq i\leq N$ numbers the nucleus, $_i\lambda_j$ is the
$j$th rate for destruction or creation of the $i$th nucleus
without a nuclear projectile involved (this includes spontaneous
decay, lepton capture, photodisintegration), and $_ir_j$ is the
rate of the $j$th reaction involving a nuclear projectile and creating or
destroying nucleus $i$. Similarly, we have three-body reactions where nucleus $i$
is produced or destroyed together with two other (or similar) nuclei.
Reactions with more participants (denoted by $\dots$
above) are unlikely to occur at astrophysical conditions and are
usually neglected. The quantities $^1 _iK_j$, $^2_iK_j$, and $^3 _iK_{jk}$
are positive or negative integer numbers specifying the amount of
nuclei $i$ produced or destroyed, respectively, in the given process. As shown
below, the rates $\lambda$, $r$, and $\hat{r}$ contain the abundances of the
interacting nuclei. Rates of type $\lambda$ depend on one abundance (or number density),
rates $r$ depend on the abundances of two species, and rates $\hat{r}$ on three.

The rates $_i\lambda_j$ appearing in the first term of Eq.~(\ref{eq:network}) are reactions per time and volume, and only contain the
abundance $Y_j$. For example, ${_{i}\lambda_{j}}$ is
simply $n_j L_j=Y_j \rho_\mathrm{pla} N_\mathrm{A} L_j$ for $\beta$-decays. The factor
$L_j=(\ln 2) / ^jT_{1/2}$ is the usual decay constant (with the unit 1/time) and is related to the half-life $^jT_{1/2}$ of the
decaying nucleus $j$. It has to be noted that some decays depend on the plasma temperature and thus
$L_j$ is not always constant, even for decays (see Eq.~(\ref{eq:stelldecay}) in Sec.~\ref{sec:stellar}).

Two-body rates $r$ include the abundances of two interacting particles or nuclei.
In general, target $A$ and projectile $a$ follow specific thermal
momentum distributions $dn_A$  and $dn_a$ in an astrophysical plasma.
With the resulting relative velocities $\vec v_A -\vec v_a$,
the number of reactions per volume and time is given by
\begin{equation}
\label{eq:2body}
r_{Aa}=\int \hat{\sigma}(\vert \vec v_A -\vec v_a\vert)
 \vert \vec v_A -\vec v_a\vert dn_A dn_a \quad,
\end{equation}
and involves the reaction cross section $\hat{\sigma}$ as a function of velocity,
the relative velocity $\vec v_A -\vec v_a$ and the thermodynamic distributions
of target and projectile $dn_A$ and $dn_a$.
The evaluation of this integral depends on the type of particles
(fermions, bosons) and distributions which are involved.

However, many two-body reactions can be simplified and effectively expressed similarly to
one-body reactions, only depending on one abundance (or number density).
If reaction partner $a$ is a photon, the relative velocity is always $c$ and the quantities in the integral do not depend on $dn_a$.
This simplifies the rate expression to
\begin{equation}
\lambda_A=L_{\gamma}(T) n_A\quad,
\label{eq:lambda}
\end{equation}
where $L_{\gamma}(T)$ stems from an integration over a
Planck distribution for photons of temperature $T$.
This is similar to the decay rates introduced earlier and therefore we replaced $r$ by $\lambda$ in our notation
and can include this type of reaction in the first term of Eq.~(\ref{eq:network}).
A similar procedure is used for electron captures by protons and nuclei.
Because the electron is about 2000 times less massive than a nucleon, the
velocity of the nucleus is negligible in the center-of-mass system in
comparison to the electron velocity ($\vert \vec v_\mathrm{nucleus}- \vec v_\mathrm{electron} \vert
\approx \vert \vec v_\mathrm{electron} \vert$). The electron
capture cross section has to be integrated over a
Fermi distribution of electrons.
The electron capture rates are a function of the plasma temperature $T$ and the
electron number density $n_e=Y_e \rho_\mathrm{pla} N_A$. In a neutral, completely
ionized plasma, the electron abundance $Y_e$ is equal to the total proton
abundance $Y_e=\sum_i Z_i Y_i$ and thus
\begin{equation}
\lambda_\mathrm{nucleus,ec}=L_\mathrm{ec} (T,\rho_\mathrm{pla} Y_e)n_\mathrm{nucleus} \quad.
\end{equation}
Again, we have effectively a rate per target $L$ (with unit 1/time)
similar to the treatment of decays earlier and
a rate per volume including the number density of only one nucleus. We denote the latter
by $\lambda$ and use it in the first term of Eq.~(\ref{eq:network}).
This treatment can be applied also to the capture of positrons,
being in thermal equilibrium with photons, electrons, and nuclei.
Furthermore, at high densities ($\rho_\mathrm{pla} >10^{12}$gcm$^{-3}$) the size of the neutrino
scattering cross section on nucleons, nuclei, and electrons ensures that enough
scattering events occur to lead to a continuous neutrino energy distribution.
Then also the inverse
process to electron capture (neutrino capture) can occur as well as
other processes like, e.g., inelastic scattering, leaving a nucleus in an excited
state which can emit nucleons and $\alpha$ particles.
Such reactions can be expressed similarly to photon and electron
captures, integrating over the corresponding neutrino distribution.

In the following, we focus on the case of two interacting nuclei or nucleons as these
reactions will be extensively discussed in the following sections. (We mention in passing that Eq.~(\ref{eq:2body}) can be
generalized to three and more interacting nuclear species by integrating over the appropriate number of distributions,
leading to rates $\hat{r}$ and higher order terms in Eq.~(\ref{eq:network}).)

The velocity distributions in the rate definition in Eq.~(\ref{eq:2body}) can be replaced by energy distributions. Furthermore, it can be shown that
the two distributions can be replaced by a single one in the center-of-mass system.\cite{cauldrons,ilibook} Then the two-body rate $r$ is defined as an
interaction of two reaction partners with an energy distribution
$\phi(E)$ according to the plasma temperature $T$ and a reaction
cross section $\sigma' (E)$ specifying the probability of the
reaction in the plasma:
\begin{equation}
r_{Aa}=\frac{n_A n_a}{1+\delta_{Aa}} \int \limits_0^\infty \sigma'_{Aa} (E) \phi (E)\,dE \quad.
\label{eq:distrirate}
\end{equation}
The factor $1/(1+\delta_{Aa})$ with the Kronecker symbol $\delta$
is introduced to avoid double counting. The nuclear cross section $\sigma_{Aa}$
is defined as in standard scattering theory by
\begin{equation}
\sigma_{Aa} = \frac{\mathrm{number\ of\ reactions\ target^{-1} sec^{-1}}}{\mathrm{flux\ of\ incoming\ projectiles}}\quad.
\end{equation}
However, in an astrophysical plasma reactions not only proceed on the ground state of a nucleus but also from excited states.
This is implied in the notation for the modified cross section $\sigma'_{Aa}$, contrasting the usual laboratory cross section
(denoted by $\sigma_{Aa}$ or $\sigma_{Aa}^\mathrm{lab}$) for reactions acting only on the ground state of the target nucleus.
The implications of using such a cross section modified in the stellar plasma, instead of the usual laboratory one, will be discussed in Sec.\ \ref{sec:stellar}. It should be noted that $\sigma'$ may be a function not only of energy but also of plasma temperature.

The distribution of kinetic energies of nuclei in an astrophysical plasma with temperature $T$ follows a Maxwell-Boltzmann distribution (MBD) $\phi(E) = \phi_\mathrm{MB}(T)$ and we obtain finally:
\begin{align}
r_{Aa} &=  \frac{n_A n_a}{1+\delta_{Aa}} \langle \sigma v \rangle^*_{Aa}=Y_A Y_a \rho_\mathrm{pla}^2 N_A^2 \langle \sigma v \rangle^*_{Aa} \\
\langle \sigma v \rangle^*_{Aa} &= \left(\frac{8}{m_{Aa} \pi}\right)^{1/2}
(kT)^{-3/2} \int \limits_0 ^\infty {E \sigma'_{Aa} (E) e^{-\frac{E}{kT}}\,dE}\quad.\label{eq:rate}
\end{align}
Here, $m_{Aa}$ denotes the reduced mass of the two-particle system
and $\langle \sigma v \rangle^*_{Aa}$ is the reaction rate per particle pair or {\it
reactivity} under stellar conditions. The angle brackets stand for the appropriate averaging, i.e.\ integration, over the energy distribution. For the remainder of the paper we will be concerned with the determination of this reactivity and the involved cross sections, respectively.

\section{Relevant energies}
\label{sec:energies}

Before we proceed to the details of the determination of the reaction cross sections, it is instructive to further
investigate the rate equation and to derive the relevant energies at which the nuclear reaction cross sections have to be
known. Although the integral in Eq.~(\ref{eq:rate}) runs to infinity, the MBD folded with the cross section
selects a comparatively narrow energy range with non-negligible contributions to the total value of the integral. Historically,
this energy range is called the Gamow window because Gamow realized early on the astrophysical relevance of the fact
that -- if the energy dependence of the cross
section is dominated by the Coulomb barrier between the projectile and the target -- the integrand in Eq.~(\ref{eq:rate})
can be factorized as\cite{cauldrons,ilibook}
\begin{equation}
\mathcal{F} = E \sigma (E) e^{-\frac{E}{kT}}=S(E)e^{-\frac{E}{kT}}e^{-\frac{b}{\sqrt{E}}} \quad,\label{eq:penet}
\end{equation}
where $S$ is the astrophysical $S$-factor
\begin{equation}
S=\sigma Ee^{\frac{b}{\sqrt{E}}}\label{eq:sfactor}
\end{equation}
which is assumed to be only weakly dependent on the energy $E$ for
non-resonant reactions.
The second exponential in Eq.~(\ref{eq:penet}) is called the Gamow factor and
contains an approximation of the Coulomb penetration with the Sommerfeld parameter
\begin{equation}
\eta=\frac{\tilde{Z}_a \tilde{Z}_A e^2}{\hbar}\sqrt{\frac{m_{Aa}}{2E}}\quad,
\label{eq:sommerfeld}
\end{equation}
where $\tilde{Z}_a$,
$\tilde{Z}_A$ are the charges of projectile $a$ and target $A$, respectively, and $m_{Aa}$ is their reduced mass.
While the first exponential (the tail of the MBD)
decreases with increasing energy, the Gamow factor increases, leading
to a confined peak of the integrand, the so-called Gamow peak. The
location of the peak $E_{0}$ is shifted to higher energies with respect
to the maximum of the MBD at $E_{\mathrm{MB}}=kT$ ($kT=T_9/11.6045$ MeV when $T_9$ is the plasma temperature
in GK). The width of the peak gives the astrophysically relevant
energy range in which most of the reactions will take place.

In absence of a Coulomb barrier the energy dependence of the non-resonant cross section is roughly given by the one of the
wave number of the particle ($1/\sqrt{E}$) folded with the angular momentum barrier. This does not, however, lead to a
relevant shift of the peak of the integrand compared to the peak of the MBD. Thus, the effective energy window for
neutrons is simply the peak of the MBD.

\begin{figure}
\centerline{\includegraphics[angle=-90,width=0.5\textwidth]{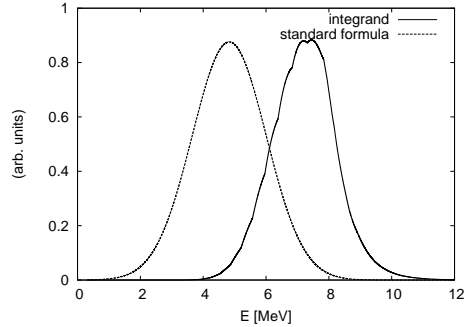}}
\vspace*{8pt}
\caption{\label{fig:sn112pa}Comparison of the actual reaction rate integrand
$\mathcal{F}$ and the Gaussian approximation of the Gamow window
for the reaction $^{112}$Sn(p,$\alpha$)$^{109}$In at $T=5$ GK.
The two curves have been arbitrarily scaled to yield similar maximal
values. The maximum of the integrand is shifted by several MeV to
energies higher than the maximum $E_{0}$ of the Gaussian. (Taken from Ref.~\protect\refcite{energywindows}, with kind permission.)}
\end{figure}

The above considerations concerning the location and size of the energy window have given rise to simple approximation
formulae extensively used by experimentalists to estimate the energies of interest. For instance, with a charged
projectile the location $E_0$ and width $\Delta$ of the Gamow window is often computed from
\begin{align}
E_{0} &=  0.12204\left(m_{Aa} \tilde{Z}_a^{2}\tilde{Z}_A^{2}T_{9}^{2}\right)^{\frac{1}{3}}\quad,\label{eq:e0approx}\\
\Delta &=  0.23682\left(m_{Aa} \tilde{Z}_a^{2}\tilde{Z}_A^{2}T_{9}^{5}\right)^{\frac{1}{6}}\quad,\label{eq:deltaapprox}
\end{align}
which is derived from Eq.~(\ref{eq:penet}) assuming a Gaussian shape of the peak and using appropriate numerical constants,
yielding $E_0$ and $\Delta$ in MeV.\cite{cauldrons,ilibook,rtk97}
This approximation of the Gamow window $E_0 \pm \left(\Delta / 2 \right)$ is valid for some but not all cases
because it is oversimplified. The above
factorization with the given Sommerfeld parameter including the charges of projectile $a$ and target $A$ implicitly assumes that the
energy dependence of the cross section is given by the Coulomb penetration in the entrance channel $a+A$ of the
reaction $a+A\rightarrow B+b$. It has been realized, however, that sometimes resonances below the Gamow window derived with the
above approximation significantly contribute to the reaction rate for certain capture reactions.\cite{ilibook,newton} In those
considered cases the energy dependence of the cross section is dominated by the energy dependence of the $\gamma$ width in the
exit channel instead of the charged particle width in the entrance channel. This can be generalized\cite{energywindows} and
leads to the important realization that the energy dependence of the integrand $\mathcal{F}$
has to be numerically examined in order to derive reliable energy windows. This has been generally and extensively studied in
Ref.\ \refcite{energywindows}. Here, only a few examples are shown. Figure \ref{fig:sn112pa} shows a comparison between the
actual $\mathcal{F}$ and the integrand assumed with the standard approximation. In this case, the relative shift of the energy
window is to higher energy because of the higher Coulomb barrier in the exit channel. In other cases, the shifts can also
be to much lower energy than predicted by the standard approximation.

For a detailed understanding of these differences one has to realize that both resonant and Hauser-Feshbach cross
sections (see Secs.~\ref{sec:reso},~\ref{sec:statmod}) can be expressed as\cite{descrau}
\begin{equation}
\sigma\propto\sum_{n}(2J_{n}+1)\frac{X_{\mathrm{in}}^{J_{n}}X_{\mathrm{fi}}^{J_{n}}}{X_{\mathrm{tot}}^{J_{n}}}\quad,\label{eq:cs}
\end{equation}
with $X$ being either Breit-Wigner widths or averaged Hauser-Feshbach
widths, depending on the context. The width of the entrance channel
is given by $X_{\mathrm{in}}^{J_{n}}$, the one of the exit channel
by $X_{\mathrm{fi}}^{J_{n}}$, and the total width including all possible
emission channels from a given resonance or compound state with spin
$J_{n}$ by $X_{\mathrm{tot}}^{J_{n}}=X_{\mathrm{in}}^{J_{n}}+X_{\mathrm{fi}}^{J_{n}}+\ldots$

It has become common knowledge that a cross section of the form shown
in Eq.~(\ref{eq:cs}) is determined by the properties of the smaller
width in the numerator if no other channels than the entrance and
exit channel contribute significantly to $X_{\mathrm{tot}}^{J_{n}}$ (see also Sec.~\ref{sec:sensigeneral}).
Then $X_{\mathrm{tot}}^{J_{n}}$ cancels with the larger width in
the numerator and the smaller width remains. (The effect is less pronounced
and requires a more detailed investigation when other channels are non-negligible
in $X_{\mathrm{tot}}^{J_{n}}$.) In consequence, the energy-dependence
of the cross section will then be governed by the energy dependence
of this smallest $X^{J}$. Only if this happens to be the charged-particle
(averaged) width in the entrance channel, the use of the standard
formula for the Gamow window (Eqs.~\ref{eq:e0approx}, \ref{eq:deltaapprox})
will be justified. Since $X_{\mathrm{in}}^{J}$ and $X_{\mathrm{fi}}^{J}$
have different energy dependences, it will depend on the specific
energy (weighted by the MBD) which of the widths is smaller.
Therefore, the above approximation should not be applied blindly but rather the
actual Gamow windows have to be determined from the true energy dependence by inspection of the integrand in
Eq.~(\ref{eq:rate}). Extensive tables of revised effective energy ranges for astrophysics from such a numerical inspection
are given in Ref.~\refcite{energywindows}. These ranges can be shifted by several MeV to higher or lower energy compared to the
ones obtained with the standard formula. Furthermore it is found that the assumption of a Gaussian shape of $\mathcal{F}$
is untenable for the majority of cases with intermediate and heavy target nuclei. Rather, the integrand $\mathcal{F}$ may
show a pronounced asymmetry around its maximum value. Therefore, the energy of the maximum alone is not sufficient to
determine the astrophysically relevant energy range.

Although derived from cross sections of a specific model prediction, the energy windows given in Ref.\ \refcite{energywindows}
are supposed to be robust. This can be understood by realizing that they mainly depend on the relative energy dependence of
the acting reaction channels and not the absolute value of the cross sections. This dependence is governed by the relative
energy and the Coulomb barrier seen in each reaction channel. Therefore, the limits of the energy windows are set by the
knowledge of the charges of the nuclei involved in the different reaction channels, and the reaction $Q$-values. Only the latter may be unknown for
nuclei far off stability and mass measurements may have an impact.

Another important consequence of using the correct energy dependence is that different reactions may not have necessarily the
same effective energy window, even when projectile and target nucleus are the same. This is immediately seen when considering the case of a reaction with a positive $Q$-value for
one channel but a negative one for another reaction channel. The astrophysically relevant energy window of the exothermic
reaction may lie below the threshold of the other reaction channel. Obviously, the relevant energy window for the endothermic
channel cannot open below the threshold energy and thus has to lie at higher energies than the one for the capture. A randomly
chosen example for such a reaction pair would be $^{104}$Pd(p,$\gamma$)$^{105}$Ag and $^{104}$Pd(p,n)$^{104}$Ag, with the energy
windows at $1.5-2.8$ MeV and $5.07-5.7$ MeV, respectively, for a plasma temperature of 2 GK.\cite{energywindows} Below the (p,n) threshold at 5.06 MeV,
the (p,n) cross section is zero and does not give a contribution to the integral in Eq.~(\ref{eq:rate}). This also further illustrates the limitation of the
standard approximation which yields identical energy windows for reactions with identical entrance channels but different exit channels. It has to be noted, though,
that the effective energy windows only point out the energy range contributing mostly to the rate integral at given stellar
temperature but do not make a statement on the size of the rate or its astrophysical relevance.

It has already been mentioned that Eq.~(\ref{eq:cs}) applies to reactions either exhibiting isolated resonances treatable
by a Breit-Wigner resonance formula or smooth cross sections stemming from an averaging over a large number of narrowly spaced
and unresolved resonances. Therefore the derived energy windows are also applicable to obtain the relevant energy ranges in
which narrow resonances have to be considered. They do not, however, specify the relative strengths of the resonances within a given
window.

Definition (\ref{eq:sfactor}) only makes sense when using the laboratory cross section $\sigma=\sigma^\mathrm{lab}$ (see Eq.~(\ref{eq:labcs}) in the following section).
Strictly speaking, with the laboratory cross section the energy windows apply to laboratory measurements only, i.e.\ to the determination of the ground state component of the actual stellar cross section. For low stellar temperatures and positive $Q$-values (see the discussion in Sec.~\ref{sec:stellar}), however, this will dominate the stellar cross section. Relevant energy windows can also be derived numerically for stellar cross sections (see next section), of course, in the same manner. Since these cannot be measured (yet), they may be of limited use, though.

A general scrutiny of the astrophysically relevant energy windows up to a plasma temperature of 5 GK (above this temperature, reaction
equilibria are established which do not require the knowledge of individual rates; see, e.g., Refs.~\refcite{radiobook,ilibook})
shows that the appearing interaction energies are small by nuclear physics standards. For neutron-induced reactions, the
encountered maximum energies are a few hundred keV, depending on the examined nucleosynthesis process
(e.g., in the $s$ process they are more like 8-60 keV).\cite{radiobook,ilibook,sprocess} The relevant energy windows are shifted to
higher energies for charged reactants, with a few MeV for reactions with protons and several MeV up to about 10 MeV for reactions
involving $\alpha$ particles. The formulae given in Eqs.\ (\ref{eq:e0approx}), (\ref{eq:deltaapprox}) are inadequate for the determination
of these energies and should not be used anymore.

\section{Stellar effects}
\label{sec:stellareffects}

\subsection{Stellar cross sections and reciprocity of stellar rates}
\label{sec:stellar}

In an astrophysical plasma, nuclei quickly (on the
timescale of nuclear reactions and scattering) reach thermal
equilibrium with all plasma components. This allows thermal
excitation of nuclei which follows a Boltzmann law and gives rise
to the {\it stellar} cross section
\begin{equation}
\label{eq:csstar}
\sigma^*_{Aa}(E,T)=\frac{\sum_\mu g_\mu \sigma^\mu(E_\mu) e^{-\frac{E_\mu^\mathrm{x}}{kT}}}{\sum_\mu g_\mu e^{-\frac{E_\mu^\mathrm{x}}{kT}}}
=\frac{\sum_\mu g_\mu \sigma^\mu(E_\mu) e^{-\frac{E_\mu^\mathrm{x}}{kT}}}{G(T)}=\sum_\mu \mathcal{P}_\mu \sigma^\mu \quad,
\end{equation}
where the sum runs over all excited states $\mu$ of the target nucleus $A$ (for
simplicity, here we assume the projectile $a$, i.e.\ the second
reaction partner, does not have excited states) with spin factor $g_\mu=2J_\mu+1$
and excitation energy $E_\mu^\mathrm{x}$. Thus, the stellar cross section is the sum of cross sections $\sigma^\mu$ (evaluated at their respective center-of-mass energies $E_\mu=E-E_\mu^\mathrm{x}$ with $\sigma^\mu=0$ when $E_\mu<0$) for
reactions on a nucleus in excited state $\mu$, weighted by the population coefficient $\mathcal{P}_\mu$. The above
relation can be derived from a Saha equation\cite{ilibook}. It is to be noted that the stellar cross section depends on energy \textit{and} temperature, contrary to the usual cross section which only is a function of energy.

The quantity $G$ is the partition
function of the nucleus. Often, the partition function normalized to the ground state
\begin{equation}
\begin{split}
G_0(T)=\frac{G(T)}{g_0}=&\frac{1}{g_0}\left[ \left\{\sum_\mu g_\mu e^{-\frac{E_\mu^\mathrm{x}}{kT}}\right\}+\right. \\
&\left.+\int\limits _{E_{\mu^\mathrm{last}}^\mathrm{x}}^{\infty}\sum _{J,
\pi} g_J
e^{-\epsilon/(kT)}\rho \left(\epsilon,J,\pi\right)\,d\epsilon \right]
\end{split}
\label{eq:partfuncint}
\end{equation}
is used (the ground state is labeled as $\mu=0$, the first excited state as $\mu=1$, \dots). Equation~(\ref{eq:partfuncint}) shows how the computation can be extended beyond the energy of the
highest known discrete energy level $\mu^\mathrm{last}$ by using an integration of a nuclear level density
$\rho$ over a range of excitation energies $\epsilon$. Likewise, the sums appearing in (\ref{eq:csstar}) can be amended with a supplemental integration over the level density above the last discrete level used. Although the product of the Boltzmann
factor $(2J+1)e^{-\epsilon/(kT)}$ and $\rho$ does not have a trivial energy dependence,
it has been shown that for the application of (\ref{eq:partfuncint})
at temperatures $T\leq 10$ GK it is sufficient to integrate only up to $E_\mathrm{max}=25$ MeV.\cite{rath,hipartfunc} Temperatures above 10 GK are encountered in some explosive astrophysical events, in
accretion disks, and in the formation of neutron stars and black holes. Nuclear transformations in such
environments are described in reaction equilibria between several or all possible reactions
(with the exception of reactions mediated via the weak interaction because they are too slow in most
cases), replacing full reaction networks by simplified abundance equations (see, e.g., Refs.~\refcite{radiobook,ilibook} for details) but still containing the partition functions. At such high temperatures, a straightforward application of
(\ref{eq:partfuncint}) would overestimate the partition function because continuum effects have to be
taken into account. These can be treated by approximated correction factors to $\rho$ and
extending the integration to $E_\mathrm{max} \gg 35$ MeV.\cite{hipartfunc} A more rigorous
treatment of the correction would be desireable but the sheer number of involved nuclei proves prohibitive for fully microscopic approaches.

\begin{figure}
\centerline{\includegraphics[width=0.8\textwidth]{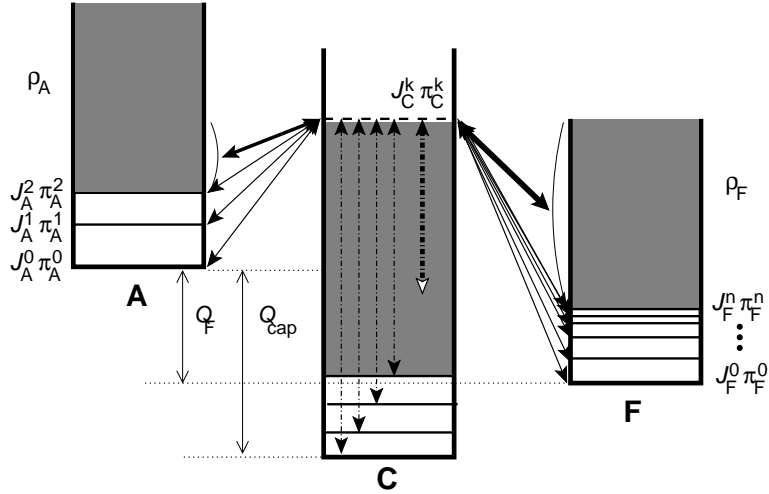}}
\vspace*{8pt}
\caption{Schematic view of the transitions (full arrows denote particle transitions, dashed arrows are $\gamma$ transitions) in a compound reaction involving the nuclei A and F, and proceeding via a compound state (horizontal dashed line) with spin $J_\mathrm{C}^k$ and parity $\pi_\mathrm{C}^k$ in the compound nucleus C. The reaction $Q$ values for the capture reaction (Q$_\mathrm{cap}$) and the reaction A$\rightarrow$F (Q$_\mathrm{F}$=Q$_{Aa}$) are given by the mass differences of the involved nuclei. Above the last state, transitions can be computed by integrating over nuclear level densities (shaded areas). \label{fig:CNscheme}}
\end{figure}

The use of stellar cross sections in the calculation of reaction rates assures an important property, the reciprocity of forward and reverse rate. A scheme of the energetics and the transitions between nuclear levels
in the involved nuclei is shown in Fig.~\ref{fig:CNscheme}. For a reaction $a+A\rightarrow b+F$ involving only one initial level $\mu$
in nucleus $A$ and one final level $\nu$ in nucleus $F$ (this can also be the compound nucleus $C$ if the ejectile $b$ is a photon) the well-known reciprocity relation between forward ($\sigma^{\mu \nu}_{Aa}$) and reverse ($\sigma^{\nu \mu}_{Fb}$) reaction cross section is\cite{ilibook,BW52}
\begin{equation}
\label{eq:reci_single}
\sigma^{\nu \mu}_{Fb}(E_F)=\frac{g_\mu^A g_a}{g_\nu^F g_b}\frac{m_A E_A}{m_F E_F}\sigma^{\mu \nu}_{Aa}(E_A) \quad,
\end{equation}
where $m_A$, $m_F$ are the reduced masses, $E_A$, $E_F$ the center-of-mass energies relative to the levels $\mu$, $\nu$, respectively, and $g$ the spin factors as before. This relation connects one initial with one final state and therefore is not applicable to the
regular laboratory reaction cross sections which connect one initial state (usually the ground state $\mu=0$) of the target nucleus with a number of possible final states
\begin{equation}
\label{eq:labcs}
\sigma=\sigma^\mathrm{lab}=\sigma^{\mu=0}=\sigma^0=\sum_\nu \sigma^{0\nu}_{Aa} \quad.
\end{equation}
In order to obtain a cross section obeying reciprocity one has to construct a theoretical quantity called
\textit{effective cross section} $\sigma^\mathrm{eff}$.
The effective cross section is a sum over all energetically possible transitions of initial levels to final levels (capture: in nuclei $A$ and $C$, otherwise: in nuclei $A$ and $F$; as indicated in Fig.~\ref{fig:CNscheme}), applying to compound reactions as well as to direct reactions (see Sec.~\ref{sec:mech}). It includes all the transitions shown by arrows in Fig.~\ref{fig:CNscheme} and therefore sums over all final levels $\nu$
\textit{and} initial levels $\mu$
\begin{equation}
\label{eq:effcs}
\sigma^\mathrm{eff}_{Aa}(E_0)=\sum_\mu \sum_\nu \frac{g_\mu}{g_0} \frac{E_\mu}{E_0}\sigma^{\mu\nu}(E_\mu)=\sum_\mu \frac{g_\mu}{g_0} \frac{E_\mu}{E_0} \sigma^\mu(E_\mu) \quad .
\end{equation}
As before, the relative center-of-mass energy of a transition proceeding from level $\mu$ is denoted by $E_\mu=E_0-E_\mu^\mathrm{x}$, and $\sigma^{\mu\nu}=\sigma^\mu=0$ for $E_\mu<0$.
The first summand ($\mu=0$) in the sum over $\mu$ is just the laboratory cross section $\sigma^0$. Note that the effective cross section is only a function of energy, like the usually defined cross section, and does not depend on temperature, contrary to the stellar cross section.

When interchanging the labels $\mu$ and $\nu$, a similar quantity $\sigma^\mathrm{eff}_{Fb}$ is obtained for the reverse direction commencing on
levels $\nu$. It is straightforward to show that the two effective cross sections obey the reciprocity relations
\begin{align}
\sigma_{Fb}^\mathrm{eff}&=\frac{g_0^A g_a}{g_0^F g_b}\frac{m_{Aa} E_0^A}{m_{Fb} E_0^F}\sigma^\mathrm{eff}_{Aa} \quad , \label{eq:reci_eff1} \\
\sigma_{Fb}^\mathrm{eff}E_0^F&=\frac{g_0^A g_a}{g_0^F g_b}\frac{m_{Aa}}{m_{Fb}}\sigma^\mathrm{eff}_{Aa}E_0^A \quad , \label{eq:reci_eff2}
\end{align}
which are identical to the one in (\ref{eq:reci_single}) for a single transition between two states in two nuclei. The relative energies of the transitions proceeding on the ground states of the two target nuclei for forward and reverse reaction are denoted by $E_0^A$ and $E_0^F$, respectively.

The effective cross section is, of course, unmeasureable. Its usefulness becomes apparent when we combine definition (\ref{eq:csstar}) of the stellar cross section with the definition (\ref{eq:rate}) for the astrophysical reaction rate. Since excited states of target nuclei are populated in a stellar plasma according to (\ref{eq:csstar}), we have to sum over the rates for reactions from each level and weight each summand with the population factor
\begin{equation}
\langle \sigma v \rangle^*_{Aa}=\left(\frac{8}{m_{Aa} \pi}\right)^{1/2}
(kT)^{-3/2} \sum_\mu {\left\{ \mathcal{P}_\mu \int \limits_0^\infty {\sigma^\mu_{Aa} E_\mu^A e^{-\frac{E_\mu^A}{kT}}\,dE_\mu^A}\right\} }\quad.\label{eq:stellrate1}
\end{equation}
This means that projectiles with MB distributed energies are acting on each level $\mu$ separately. Insertion of definition (\ref{eq:csstar}) for the population factor $\mathcal{P}_\mu$ leads to
\begin{align}
\sum_\mu {\left\{ \mathcal{P}_\mu \int \limits_0^\infty {\sigma^\mu_{Aa} E_\mu^A e^{-\frac{E_\mu^A}{kT}}\,dE_\mu^A}\right\} }
&= \sum_\mu \int \limits_0^\infty { \mathcal{P}_\mu \sigma^\mu_{Aa} E_\mu^A e^{-\frac{E_\mu^A}{kT}} \,dE_\mu^A} =  \nonumber \\
 = \sum_\mu \int \limits_0^\infty { \frac{g_\mu^Ae^{-E_\mu^\mathrm{x}/(kT)}}{g_0^A G^A_0} \sigma^\mu_{Aa} E_\mu^A e^{-\frac{E_\mu^A}{kT}} \,dE_\mu^A} &= \frac{1}{G^A_0} \sum_\mu \int \limits_0^\infty { \frac{g_\mu^A}{g_0^A} \sigma^\mu_{Aa} E_\mu^A e^{-\frac{E_\mu^A+E_\mu^\mathrm{x}}{kT}}}\,dE_\mu^A
\quad, \label{eq:stellrate2}
\end{align}
where $G^A_0$ is the normalized partition function as defined in (\ref{eq:partfuncint}).
In order to obtain an expression similar to the original single MBD, the integral can be transformed by replacing $dE_\mu^A \rightarrow dE_0^A$, with $E_0^A=E_\mu^A+E_\mu^\mathrm{x}$ and this yields
\begin{align}
&\frac{1}{G^A_0} \sum_\mu \int \limits_0^\infty { \frac{g_\mu^A}{g_0^A} \sigma^\mu_{Aa}(E_\mu^A) E_\mu^A e^{-\frac{E_\mu^A+E_\mu^\mathrm{x}}{kT}}}\,dE_\mu^A = \nonumber \\
&=\frac{1}{G^A_0} \sum_\mu { \int \limits_{E_\mu^\mathrm{x}}^\infty { \frac{g_\mu^A}{g_0^A} \sigma^\mu_{Aa}(E_0^A-E_\mu^\mathrm{x}) \left[E_0^A-E_\mu^\mathrm{x}\right] } e^{-\frac{E_0^A}{kT}}\,dE_0^A } = \nonumber \\
&=\frac{1}{G^A_0} \sum_\mu { \int \limits_{0}^\infty \frac{g_\mu^A}{g_0^A} \sigma^\mu_{Aa}(E_\mu^A) \left[E_0^A-E_\mu^\mathrm{x}\right] e^{-\frac{E_0^A}{kT}}\,dE_0^A }
\quad. \label{eq:stellrate3}
\end{align}
In the last line above, the lower limit of the integration was reset to Zero. This is allowed because cross sections at negative energies do not give any contribution to the integral.
It has been pointed out in Ref.~\refcite{fow74} that it is mathematically equivalent when sum and integral are exchanged, leading to\cite{fow74,hwfz}
\begin{align}
\langle \sigma v\rangle_{Aa} ^* &=\left(\frac{8}{m_{Aa} \pi}\right)^{1/2}
(kT)^{-3/2}\frac{1}{G^A_0} \sum_\mu { \int \limits_{0}^\infty \frac{g_\mu^A}{g_0^A} \sigma^\mu_{Aa}(E_\mu^A) \left[E_0^A-E_\mu^\mathrm{x}\right] e^{-\frac{E_0^A}{kT}}\,dE_0^A } = \nonumber \\
&=\left(\frac{8}{m_{Aa} \pi}\right)^{1/2}
(kT)^{-3/2}\frac{1}{G^A_0} \int \limits_{0}^\infty { \left\{ \sum_\mu { \frac{g_\mu^A}{g_0^A} \sigma^\mu_{Aa}(E_\mu^A) E_\mu^A } \right\} e^{-\frac{E_0^A}{kT}}\,dE_0^A } = \nonumber \\
&=\left(\frac{8}{m_{Aa} \pi}\right)^{1/2}
(kT)^{-3/2}\frac{1}{G^A_0} \int \limits_{0}^\infty { \sigma^\mathrm{eff}_{Aa} E_0^A e^{-\frac{E_0^A}{kT}}\,dE_0^A } = \frac{\langle \sigma^\mathrm{eff} v\rangle_{Aa}}{G^A_0} \quad.
\label{eq:effrate}
\end{align}
The last line was obtained by realizing that the expression in the curly brackets is identical to $\sigma^\mathrm{eff}_{Aa} E_0^A$, with the effective cross section from (\ref{eq:effcs}). Thus, the weighted sum over many
MBDs acting on the thermally populated excited states is reduced to a single MBD acting on an effective cross section and divided by the normalized partition function. In terms of relevant physics, this means that the
Boltzmann factor in the population probability is offset by shifting down each MBD to the same relative energy.\cite{fow74} Now $\sigma'$ in (\ref{eq:rate}) can be identified as $\sigma'=\sigma^\mathrm{eff}/G^A_0$, introducing a temperature dependence while $\sigma^\mathrm{eff}$ is conveniently independent of $T$.

It may be confusing that the stellar reactivity frequently is written as $\langle \sigma^* v \rangle$ instead of $\langle \sigma v \rangle^*$. This is not meant to imply that the stellar cross section as defined in (\ref{eq:csstar}) is inserted in a single integral over a MBD as shown in (\ref{eq:rate}). Rather, the angle brackets imply a separate integration for each populated state as performed in (\ref{eq:stellrate1}) and (\ref{eq:stellrate2}) in this case.

Equation (\ref{eq:effrate}) not only simplifies the numerical calculation of the stellar rate but also allows to better understand certain details. For instance, it can immediately be seen that stellar rates obey a reciprocity relation because the effective cross sections do. A similar expression has to hold for the reverse reactivity
$\langle \sigma v\rangle^*_{Fb}$ as for the forward reactivity $\langle \sigma v\rangle^*_{Aa}$, being
\begin{equation}
\langle \sigma v\rangle^*_{Fb} = \left(\frac{8}{m_{Fb} \pi}\right)^{1/2}
(kT)^{-3/2}\frac{1}{G^F_0} \int \limits_{0}^\infty { \sigma^\mathrm{eff}_{Fb} E_0^F e^{-\frac{E_0^F}{kT}}\,dE_0^F } \quad.\label{eq:revreac}
\end{equation}
This expression is derived in the same manner as (\ref{eq:effrate}) but by starting from thermally populated excited states in the final nucleus.
With the help of (\ref{eq:reci_eff2}) we can express the reverse reactivity of (\ref{eq:revreac}) in terms of the forward reactivity:\cite{ilibook,hwfz}
\begin{equation}
\label{eq:revrate}
\frac{\langle \sigma v\rangle^*_{Fb}}{\langle \sigma v \rangle^*_{Aa}}=\frac{g_0^A g_a}{g_0^F g_b} \frac{G^A_0}{G^F_0} \left( \frac{m_{Aa}}{m_{Bb}}\right) ^{3/2}e^{-Q_{Aa}/(kT)} \quad ,
\end{equation}
where $Q_{Aa}=E_0^F-E_0^A$ is the reaction $Q$-value of the forward reaction. The reciprocity relation (\ref{eq:reci_eff2}) applies to photodisintegration and captures as well. In relating the photodisintegration rate $\lambda$ to the capture rate, however, it has to be assumed that the denominator $\exp(E/(kT))-1$ of the Planck distribution for photons appearing in the
photodisintegration reactivity $L_\gamma=\langle \sigma v \rangle^*_{C\gamma}$ can be replaced by $\exp(E/(kT))$, similar to the one of a MBD with the same temperature $T$. With this approximation and realizing that $g_\gamma=2$, one obtains\cite{ilibook,rath,hwfz,fcz67}
\begin{equation}
\frac{L_\gamma}{\langle \sigma v \rangle^*_{Aa}}=\frac{g_0^A g_a}{g_0^C} \frac{G^A_0}{G^C_0}
\left( \frac{m_{Aa}kT}{2\pi \hbar^2}\right)^{3/2} e^{-Q_{Aa}/(kT)} \label{eq:revphoto}
\end{equation}
in the same manner as (\ref{eq:revrate}).
Using the approximation of the denominator resulting in (\ref{eq:revphoto}) is very important for the application in reaction networks. Employing the expressions (\ref{eq:revreac}) and (\ref{eq:revphoto}) avoids numerical inconsistencies in network calculations which may arise when forward and reverse rates are calculated separately (or even from different sources). The proper balance between the two reaction directions can only be achieved in such a treatment. Furthermore, simplified equations for reaction equilibria (see Sec.~\ref{sec:equilibria}) can be derived which prove important in the modeling and understanding of nucleosynthesis at high temperature.

\begin{figure}
\centerline{\includegraphics[angle=-90,width=0.63\columnwidth,clip]{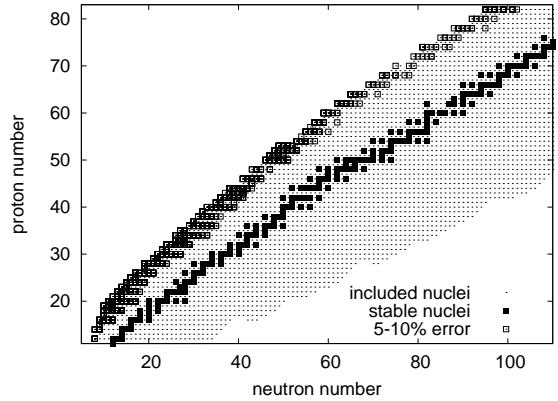}}
\vspace*{8pt}
\caption{\label{fig:mbcomp_proton}Target nuclei for which an error of 5--10\% is introduced in the relation (p,$\gamma$)/($\gamma$,p) at $T=1$ GK with the standard approximation of the Planck distribution.}
\end{figure}

\begin{figure}
\centerline{\includegraphics[angle=-90,width=0.63\columnwidth,clip]{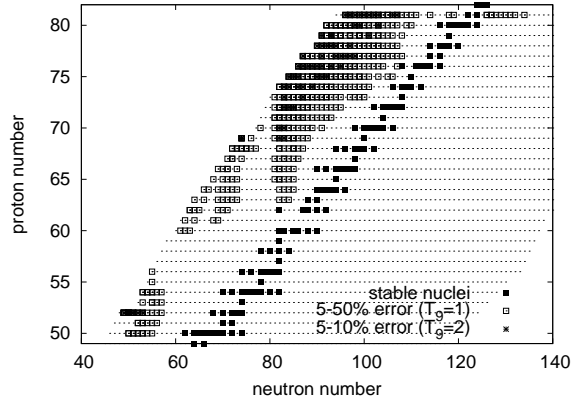}}
\vspace*{8pt}
\caption{\label{fig:mbcomp_alpha}Target nuclei for which an error of 5--50\% is introduced in the relation ($\alpha$,$\gamma$)/($\gamma$,$\alpha$) at $T=1$ and $T=2$ GK with the standard approximation of the Planck distribution. Maximum errors only reach 10\% at $T=2$.}
\end{figure}

\begin{figure}
\centerline{\includegraphics[angle=-90,width=0.63\columnwidth,clip]{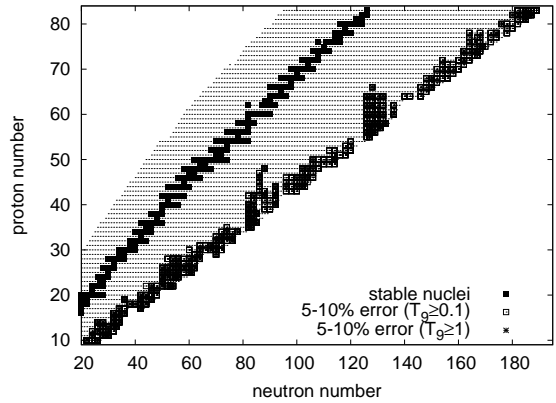}}
\vspace*{8pt}
\caption{\label{fig:mbcomp_neut}Target nuclei for which an error of 5--10\% is introduced in the relation (n,$\gamma$)/($\gamma$,n) for $T\geq0.1$ and $T\geq1$ GK, respectively, with the standard approximation of the Planck distribution.}
\end{figure}

How large is the error stemming from the approximation involved in the derivation of (\ref{eq:revphoto})?
Although mathematically unsound, it turns out that setting $\exp(E/(kT))-1\approx \exp(E/(kT))$ is a good approximation for the calculation of the rate integrals and introduces an error of less than a few percent for astrophysically relevant temperatures and rate values.\cite{ilibook,hwfz,rau95,mathewsreci}
In other words, the contributions to the integral in (\ref{eq:lambda}) are negligible at the low energies where Planck and MBDs differ considerably (see also Fig.~3.5 in Ref.~\refcite{ilibook}, and Ref.~\refcite{rau95,mathewsreci}).
This is assured by either a sufficiently large and positive $Q_{Aa}$, which causes the integration over the Planck distribution to start not at Zero energy but rather at a sufficiently large threshold energy, or by vanishing effective cross sections at low energy due to, e.g., a Coulomb barrier. The assumption may not be valid for s-wave neutron captures with very small (of the order of $Q\lesssim kT$) or negative $Q$-values, but the required correction still is only a few \% as can be shown in numerical comparisons between photodisintegration rates calculated with the two versions of the denominator. Such a comparison was performed with the code SMARAGD (version 0.8s; see Sec.~\ref{sec:codes}) and the results are shown in Figs.~\ref{fig:mbcomp_proton}, \ref{fig:mbcomp_alpha}, and \ref{fig:mbcomp_neut}. Generally, larger errors appear at lower temperature. This results in astrophysical irrelevance of the errors in many cases because either the rates are too slow (especially for rates involving charged projectiles) or the target nuclei in question are so short-lived that they will never be produced at low plasma temperature. The largest error found was between 50 and 100\% for a few heavy nuclei at the driplines for proton- or $\alpha$-capture at $T<0.3$ GK. For neutron captures, the errors when applying the standard approximation for the reverse rate were never larger than 10\% at any investigated temperature, even at the driplines. In the figures, errors of 5\% and smaller are not presented in detail because they are assumed to be negligible, especially given the remaining uncertainties in the prediction of the rates far from stability.

Figure \ref{fig:mbcomp_proton} shows target nuclei for proton capture where errors reach 5--10\% at $T=1$ GK when computing the photodisintegration rate from (\ref{eq:revphoto}). As expected, this occurs close to the dripline where the reaction $Q$-value is small or negative. At $T\geq2$ GK, however, the errors for all nuclei with mass number $\tilde{A}>56$ are below 5\% already and thus negligible.
Figure \ref{fig:mbcomp_alpha} shows target nuclei for $\alpha$ capture where errors maximally reach 5--50\% at $T=1$ and 5--10\% at $T=2$ GK, respectively, when computing the photodisintegration rate from (\ref{eq:revphoto}). Again, this occurs for some (but not all, depending on the energy-dependence of the effective cross section) $\alpha$ captures with small or negative $Q$-values. At $T=2$ GK even fewer rates are affected and the maximal error is below 10\%. The shown rates have little or no astrophysical impact because those nuclei can only be reached at higher temperatures. For instance, the $\gamma$-process (p-process) significantly photodisintegrates nuclei close to stability at $T\geq2.5$ GK and the rp-process also requires such high temperatures and probably does not proceed beyond $\tilde{A} \approx 110$.\cite{tombranch,schatzend}

Figure \ref{fig:mbcomp_alpha} shows target nuclei for neutron capture where errors reach 5--10\% at any plasma temperature. As expected, captures with low $Q$-value, either at the dripline or in the vicinity of closed shells, exhibit the largest errors but those do not exceed 10\%. Nuclei so far from stability are expected to be synthesized only at higher temperature. For $T\geq 1$ GK a smaller number of nuclei shows errors above 4\%. In fact, for most reactions across the chart the error introduced in the reverse rate due to the approximation of the Planck distribution is less than 1\%.

Having assured that the approximation of the Planck distribution in the derivation of (\ref{eq:revphoto}) does not introduce a considerable error in the obtained rates, it is important to realize that further conditions have to be fulfilled to allow the application of the relations discussed above.
We started from introducing a thermal population $\mathcal{P}_\mu$ in (\ref{eq:csstar}) above and consequently all derivations up to here depend on the assumption that the excited states in all participating nuclei are occupied according to this population factor. This is a valid assumption for most astrophysical plasmas and most nuclei reach thermal equilibrium very rapidly through collisions and interactions with photons and other plasma components. However, there are some nuclei which exhibit long-lived isomeric states (well-known examples are $^{26}$Al, $^{176}$Lu, and $^{180}$Ta)\cite{wardfowler,meyeral26,rhhw02,moka07,mobi09} with such spins that they cannot be easily excited or de-excited through electromagnetic transitions. At sufficiently high temperature they may still get into equilibrium, sometimes through couplings to intermediate states, but the relevant transitions have to be carefully studied. This can be achieved through an internal reaction network, not connecting different nuclei but rather including the different levels within one nucleus.\cite{wardfowler,meyeral26} Levels not being in thermal equilibrium can be included in regular networks in such a manner as if they were a different nucleus. Also in this case, however, the populating and depopulating reactions have to be known explicitly. This also applies when ensembles of excited states are in equilibrium but the different ensembles within a nucleus are not. Then each ensemble can be treated as a separate species in a reaction network and the reactions connecting the ensembles have to be included explicitly.

Although not discussed in further detail here, it is worth mentioning that also weak interactions are affected by the thermal population of excited states. For example, the $\beta$-decay half-life $T_{1/2}$ of a nucleus will be changed relative to its ground state half-life when the ground state becomes depopulated and excited states with different decay half-lives are populated. Thus, the decay ``constant'' $L_\beta=\ln(2)/T_{1/2}$ is actually temperature-dependent
\begin{equation}
L^*_\beta (T)=\sum_\mu \mathcal{P}_\mu L_\beta^\mu \quad,
\label{eq:stelldecay}
\end{equation}
where $L_\beta^\mu=1/\tau_\beta^\mu$ and $\tau_\beta^\mu$ is the decay lifetime of the excited state. Similar considerations also apply to other processes, such as electron capture and neutrino-induced reactions.

\subsection{Reaction equilibria}
\label{sec:equilibria}

Using the reciprocity relations derived above allows to simplify the full reaction networks as defined in (\ref{eq:network}). Such simplifications are instructive because they enable us to study nucleosynthesis properties which are independent of details in the hydrodynamic evolution of the system or even, as shown below, independent of individual reaction rates. They usually go along with a restriction to only types of reactions in the network which are actually necessary instead of blindly evolving a large system of differential equations. Such an approach is not always feasible but considerable understanding of nucleosynthesis has been gained in the past through such means by circumventing the necessity of computationally intensive calculations. Such simplifications remain important today because it is still impossible to couple multi-dimensional hydrodynamic simulations to full reaction networks. Furthermore, restriction to the essential often provides a much better insight into the physical processes than a brute-force full network calculation. Here we are concerned with the simplifications because it has to be understood when it is necessary to know astrophysical reaction rates and when not.

Setting the abundance change $\dot{Y}_i=0$ on the left-hand side of (\ref{eq:network}) implies that the sum of all rates destroying the nuclear species $i$ is exactly balanced by all production rates and the net change in abundance is therefore Zero, leaving the abundance constant. This is called \textit{steady flow equilibrium}. It is especially useful with reaction chains where most rates (perhaps except one) are in steady flow equilibrium. Then the slowest reaction sets the timescale of the reaction flow and all other reactions adjust. As long as steady flow is upheld, no full reaction network has to be solved. Rather, the ratios of the steady flow abundances of the involved nuclei are related by the ratios of their net destruction rates (or, equivalently, production rates as these have to be the same). For illustration, let us assume a chain of reactions $A \rightarrow B \rightarrow C \rightarrow D \rightarrow \dots$ connecting nuclei through reactions with the same projectile, where all net reactions are in steady flow and therefore the same, except for the one starting at nucleus $A$. Then
\begin{equation}
\frac{Y_B}{Y_C}=\frac{\langle \sigma v \rangle^*_{C\rightarrow D}}{\langle \sigma v \rangle^*_{B\rightarrow C}}
\end{equation}
and
\begin{equation}
\frac{Y_C}{Y_D}=\frac{\langle \sigma v \rangle^*_{D\rightarrow \dots}}{\langle \sigma v \rangle^*_{C\rightarrow D}} \quad .
\end{equation}
The slowest rate sets the abundance of $B$ through $\dot{Y}_A=-\dot{Y}_B=-r_{A\rightarrow B}/(\rho_\mathrm{pla} N_A)$. The use of a complete set of
coupled differential equations is not required anymore but the important rates still have to be known. Steady state considerations are helpful when
investigating hydrostatic hydrogen burning of stars through the pp-chains and the CNO cycles.\cite{ilibook,clayton} In the past they have also been used for sequences of neutron captures in the s-process on nuclei in between magic numbers.\cite{clayton} The fact that separate steady flows can be assigned to each mass region between closed shells has been termed \textit{local approximation} in s-process studies.

A slightly different concept is to assume equilibrium between a forward and its reverse rate. This is the case when the two rates are equal or, in practice, very close. Since stellar rates obey the simple reciprocity relations
(\ref{eq:revrate}) and (\ref{eq:revphoto}), respectively, it is trivial to show that
\begin{equation}
\frac{Y_AY_a}{Y_FY_b}=\frac{g_0^A g_a}{g_0^F g_b} \frac{G^A_0}{G^F_0} \left( \frac{m_{Aa}}{m_{Bb}}\right) ^{3/2}e^{-Q_{Aa}/(kT)}
\end{equation}
for a reaction $A+a \leftrightarrow F+b$ and
\begin{equation}
\frac{Y_AY_a}{Y_C}=\frac{g_0^A g_a}{g_0^C} \frac{G^A_0}{G^C_0} \left( \frac{m_{Aa}kT}{2\pi \hbar^2}\right)^{3/2} e^{-Q_{Aa}/(kT)}
\end{equation}
for a reaction $A+a \leftrightarrow C+\gamma$. The individual rates do not appear anymore in the relation between the abundances. Note that this does not imply that the abundances remain constant, they still depend on $T$ which may vary with time as well as on $Y_a$ and $Y_b$.

Depending on the plasma density\footnote{The rate of the triple-$\alpha$ reaction $\alpha + \alpha + \alpha \rightarrow ^{12}$C is very sensitive to the density and will not get into equilibrium for $\rho_\mathrm{pla}\lesssim 10^5$ g/cm$^3$.}, above $T\approx 4-5$ GK all reactions (with the exception of the weak interaction) achieve equilibrium. It can be shown that the equilibrium abundance of a nucleus $A$ can be calculated from a set of three equations\cite{arnett}
\begin{align}
Y_A &=G_A \left(\rho_\mathrm{pla}N_A\right)^{\tilde{A}-1}
\frac{\tilde{A}^{3/2}}{2^{\tilde{A}}}
\left( \frac{2\pi\hbar^2}{m_u kT} \right)^{3(\tilde{A}-1)/2} e^{B_A/(kT)}Y_\mathrm{n}^{\tilde{N}}Y_\mathrm{p}^{\tilde{Z}} \quad, \label{eq:nse} \\
1 &= \sum_i \tilde{A}_i Y_i \quad, \label{eq:masscons} \\
Y_e &= \sum_i \tilde{Z}_i Y_i \quad, \label{eq:chargeconv}
\end{align}
where $\tilde{A}=\tilde{N}+\tilde{Z}$ is the mass number, $Y_\mathrm{n}$, $Y_\mathrm{p}$ are the abundances of the free neutrons and protons, respectively, and $m_u$ the nuclear mass unit. The binding energy of the nucleus with neutron number $\tilde{N}$ and proton number $\tilde{Z}$ is denoted by $B_A$. The sums run over all species of nuclei in the plasma, including neutrons and protons. Equation (\ref{eq:masscons}) expresses mass conservation and (\ref{eq:chargeconv}) is the charge conservation. The unknown abundances $Y_A$, $Y_\mathrm{n}$, and $Y_\mathrm{p}$ are obtained with the above equation set. Note that reactions mediated by the weak interaction are not included in the equilibrium and $Y_e$ may be time-dependent. Again, individual rates are not required to determine the abundances.

When all abundances in the network obey the above relations, full \textit{nuclear statistical equilibrium} (NSE) is achieved. In this case, no reaction rates have to be known. In realistic cases, more or less extended groups of nuclei are in statistical equilibrium and the relative abundances within a group can be described by equations similar to (\ref{eq:nse}). The different groups are connected by comparatively slow reactions not being in equilibrium, which determine the abundance level of one group with respect to another group similar to what was shown above for the steady state equilibrium. The rates of these slow, connecting reactions have to be known explicitly. This is called \textit{quasi-statistical equilibrium} (QSE). It appears in various kinds of high-temperature burning, such as hydrostatic oxygen and silicon burning in massive stars and different explosive scenarios.

A special kind of QSE is the (n,$\gamma$)$-$($\gamma$,n) equilibrium or \textit{waiting point approximation}, often used in r-process calculations.\cite{ctt,freiburg} This is nothing else than a QSE within an isotopic chain, where neutron captures and ($\gamma$,n) reactions are in equilibrium under very neutron-rich conditions ($n_\mathrm{n} \geq 10^{20}$ cm$^{-3}$) and $T\approx 1-2$ GK. For the r-process the network is reduced to neutron captures and their inverse reactions, and to $\beta^-$ decays (with possible subsequent neutron emission). The decays are not in equilibrium and determine the timescale with which matter is processed from small $\tilde{Z}$ to the heaviest nuclei. When (n,$\gamma$)$-$($\gamma$,n) equilibrium is achieved, the abundances of nuclei within an isotopic chain are connected by
\begin{equation}
\frac{Y_{A'}}{Y_A}=n_\mathrm{n} \frac{G_{A'}}{2G_A} \left( \frac{\tilde{A}+1}{\tilde{A}} \right)^{3/2}\left( \frac{2\pi\hbar^2}{m_u kT} \right)^{3/2} e^{S_\mathrm{n}^{A'}/(kT)}\quad,
\label{eq:nggn}
\end{equation}
which connects the abundance of nucleus $A$ (mass number $\tilde{A}$) with the one of nucleus $A'$ (mass number $\tilde{A}+1$). There is an exponential dependence on the neutron separation energy $S_\mathrm{n}^{A'}=Q_{\mathrm{n}\gamma}^A$ of $A'$. Also in this type of equilibrium there is no dependence on the individual capture or photodisintegration rates. The r-process flow to higher elements, however, depends on the $\beta^-$-decay rates which connect the isotopic chains and are not in equilibrium. They are very slow compared to the rates in equilibrium and that is why ``waiting points'' are established, which are just the nuclei (usually only one or two within a chain) with the highest abundances according to (\ref{eq:nggn}). The r-process cannot proceed until they decay and their decay rates have to be known.\footnote{The waiting point(s) in two neighboring isotopic chains do not have to be contiguous and therefore the notion of an r-process ``path'' similar to the s-process path is not valid.}

A similar equilibrium, but between proton captures and ($\gamma$,p), is reached in the late phase of the rp-process on the surface of mass accreting neutron stars.\cite{schatz} There, the waiting points are established close to the proton dripline.

Nucleosynthesis under extreme conditions, such as encountered in some explosive scenarios, involves exotic nuclei far from stability. According to the above, reaction rates are not needed for all of them because reaction equilibria are established at such extreme conditions. Required are nuclear masses (to determine $Q$-values or binding energies) as well as spectroscopic information and nuclear level densities (entering the calculation of the partition functions $G$).

But although NSE, QSE, and the waiting point approximation do not contain the rates explicitly, they implicitly depend on them because they determine whether nuclei are participating in the equilibrium or not. The higher the rates, the lower the temperature at which equilibrium is reached. With a time-dependent $T$ evolution, this means that the rates determine whether equilibrium is reached earlier with increasing $T$ or the freeze-out happens later with decreasing $T$.

\subsection{Stellar cross sections and experiments}
\label{sec:stellarexp}

In principle, stellar cross sections $\sigma^*$ as defined in (\ref{eq:csstar}) correspond to physically measurable quantities, contrary to the purely theoretical effective cross sections introduced in (\ref{eq:effcs}). In practice, cross sections $\sigma=\sigma^{\mu=0}$ measured in terrestrial laboratories do not include thermal
effects. Therefore the rates derived from them do not obey reciprocity relations. To derive astrophysical rates appropriate for the utilization in reaction networks, laboratory cross sections almost always have to be supplemented by theory to account for the additional transitions not included in the measurement.

The energy- and temperature-dependent stellar enhancement factor for the cross sections (again obtained via theory)
\begin{equation}
\label{eq:sefsig}
f_\mathrm{SEF}^\mathrm{c.s.}(E,T)=\frac{\sigma^*(E,T)}{\sigma(E)}
\end{equation}
compares the stellar cross section including reactions from thermally populated excited states to the cross section obtained with reactions proceeding from the ground state of the target nucleus only. Another definition of a stellar enhancement factor involves the rates or reactivities,
\begin{equation}
\label{eq:sefrate}
f_\mathrm{SEF}^\mathrm{rate}(T)=\frac{\langle \sigma v\rangle^*}{\langle \sigma v\rangle} = \frac{\int_0^\infty \sigma^\mathrm{eff} E e^{-E/(kT)}\,dE}{G_0(T) \int_0^\infty \sigma E e^{-E/(kT)}\,dE}\quad.
\end{equation}
In fact, $f_\mathrm{SEF}^\mathrm{rate}$ is the astrophysically interesting quantity because it shows the deviation introduced in the \textit{rate} at a chosen stellar temperature $T$ when using the laboratory cross section instead of the stellar cross section. Then the fraction of the ground state contribution to the stellar rate is $\mathcal{X}=1/(G_0 f_\mathrm{SEF}^\mathrm{rate})$.

On the other hand, $f_\mathrm{SEF}^\mathrm{c.s.}$ is supposed to provide information for the experimentalist on how much off the measured $\sigma$ is in comparison to $\sigma^*$ at each energy for a given stellar temperature. To this end it was quoted in literature occasionally. However, a much more useful definition is
\begin{equation}
\label{eq:sefeff}
f_\mathrm{SEF}^\mathrm{eff}(E,T)=\frac{\sigma^\mathrm{eff}(E)}{G_0(T) \sigma(E)}
\end{equation}
which weights the excited states appropriately for a straightforward comparison.
This is the same as $f_\mathrm{SEF}^\mathrm{rate}$ only if $f_\mathrm{SEF}^\mathrm{eff}$ is independent of $E$. This is not necessarily fulfilled because the energy-dependence of laboratory and effective cross section may be different and this would lead to a different evaluation of the reaction rate integral. (It turns out that in practice $f_\mathrm{SEF}^\mathrm{eff}$ is often more slowly varying with energy than the cross section across the relevant energy window (see Sec.\ \ref{sec:energies}) and that it can be approximated by an energy-independent factor in this case.) It is recommended to use only either $f_\mathrm{SEF}^\mathrm{eff}$ or $f_\mathrm{SEF}^\mathrm{rate}$, depending on whether an energy-dependent measure is desired or one independent of interaction energy. As has become clear from the derivation of the stellar rate in Sec.~\ref{sec:stellar} there is no simple relation between $f_\mathrm{SEF}^\mathrm{rate}$ and $f_\mathrm{SEF}^\mathrm{c.s.}$.

\begin{figure}
\centerline{\includegraphics[angle=-90,width=0.64\textwidth]{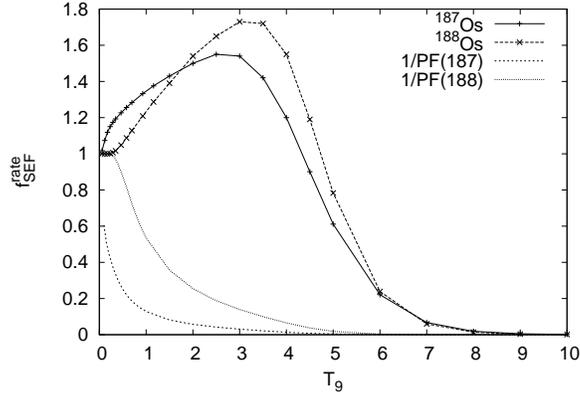}}
\vspace*{8pt}
\caption{Stellar enhancement factors $f_\mathrm{SEF}^\mathrm{rate}$ for neutron capture on $^{187}$Os and $^{188}$Os as a function of stellar temperature ($T_9$ is in GK). The reciprocals of the normalized partition functions for $^{187,188}$Os are also shown. \label{fig:ossef1}}
\end{figure}

It should be noted that while the stellar population factors $\mathcal{P}_\mu$ are normalized to Unity and the normalized partition functions $G_0$ cannot become smaller than Unity, the three types of stellar enhancement factors defined above can assume any positive value, \textit{larger or smaller} than Unity.

In which case do we have to expect large deviations
\begin{equation}
\mathcal{D} =\exp \left( \lvert\ln \left( \langle \sigma v\rangle^* \right) - \ln \left( \langle \sigma v\rangle \right) \rvert  \right) = \exp \left( \lvert \ln \left( f_\mathrm{SEF}^\mathrm{rate} \right) \rvert \right)  \label{eq:logdev}
\end{equation}
from the laboratory value? (The above definition assures $\mathcal{D} \geq 1$; without thermal effects $\mathcal{D}=1$.)
Again, a scrutiny of the effective cross section appearing in (\ref{eq:sefrate}) helps to understand the various dependences.

\begin{figure}
\centerline{\includegraphics[angle=-90,width=0.5\textwidth]{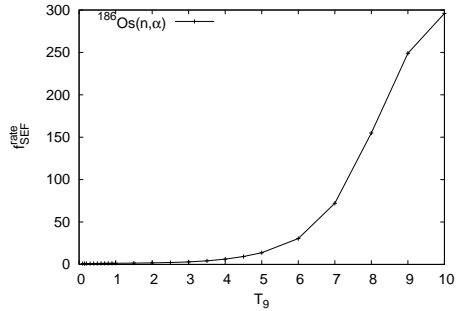}}
\vspace*{8pt}
\caption{Stellar enhancement factors $f_\mathrm{SEF}^\mathrm{rate}$ for $^{186}$Os(n,$\alpha$)$^{183}$W as a function of stellar temperature ($T_9$ is in GK).\label{fig:ossef2}}
\end{figure}

Let us start with the dependence on stellar temperature. Naively, one would assume that the higher the temperature, the larger the stellar enhancement will become. Indeed, this is often the case but there are further intricacies. There are two dependences on $T$ appearing in (\ref{eq:sefrate}), one in the rate in the numerator and the other in the normalized partition function in the denominator. The latter is monotonically increasing with increasing $T$. The integration of the product of the effective cross section and the MBD may show a different $T$-dependence, however, which even may not prove to be monotonic. This can be understood by combining the knowledge of the relevant energy windows from Sec.~\ref{sec:energies} with definition (\ref{eq:effcs}) of the effective cross section. Larger $T$ shifts the energy window to higher relative energy, both in the entrance channel $A+a$ with the relevant $E_0^A$ becoming larger and in the exit channel $F+b$ (or $C+\gamma$) with the
relevant $E_0^F=E_0^A+Q_{Aa}$ becoming larger (see Fig.~\ref{fig:CNscheme}). Since the effective cross section sums over transitions with relative energies $0<E^A\leq E_0^A$ and $0<E^F\leq E_0^F$, it becomes obvious that the higher the relevant energy window, the more transitions are included. This does not necessarily result in an increased rate although it often will. It is also conceivable that the additionally included transitions at large $T$ have small cross sections and do not provide a considerable increase in the rate. Depending on the type of reaction, cross sections for transitions already included at small $T$ may also decrease with increasing relative energy. In these cases, $f_\mathrm{SEF}^\mathrm{rate}$ will be reduced at larger $T$ because $G_0$ is always increasing. The stellar ``enhancement'' may even become smaller than Unity and not live up to its name anymore. This behavior should not be viewed as monotonic, either, because as additional transitions become accessible at even larger $T$ the stellar rate again may increase faster than $G_0$.

It is to be expected, however, that $f_\mathrm{SEF}^\mathrm{rate}$ is decreasing to very small values for very large $T$ after having reached a maximum value, i.e., for very high (but not necessarily astrophysically important) temperatures $1/G_0 \leq f_\mathrm{SEF}^\mathrm{rate} \ll 1$ will always be achieved. This is because with increasing relevant energy $E_0^{A,F}$ other reaction channels become increasingly important, reducing the cross sections of the individual transitions. Additionally, more unbound levels will be included in the effective cross section. These may lose particles to other reaction channels (e.g., through pre-equilibrium emission; see also Sec.~\ref{sec:statmod}) and also not contribute to the effective cross section anymore. It was suggested in Ref.~\refcite{hwfz,fcz75} to include only bound states in the definition of the effective cross section. Although this may be a good approximation, it may neglect some transitions which still can contribute to the effective cross section at high temperature. Note, however, that unbound states may not be in thermal equilibrium although the timescale for reaching equilibrium under stellar burning conditions is short compared to the one of a nuclear reaction.\cite{fcz75} A more suitable cutoff, if required, would be the energy $E_\mathrm{max}$ appearing as cutoff in the calculation of the partition function in (\ref{eq:partfuncint}), although this may include already too many transitions with negligible cross sections.
Examples for the above considerations are shown in Figs.~\ref{fig:ossef1}, \ref{fig:ossef2} for neutron-induced reactions on Os isotopes. Neutron capture on $^{187}$Os is important to understand the Re-Os cosmochronometer which can be used for age determinations in our Galaxy.\cite{clayton,reos} Modern measurements of neutron capture in the astrophysically relevant energy range have reached a precision that requires the inclusion of the  $f_\mathrm{SEF}^\mathrm{rate}$ correction, even when it is only a few tens of percent.\cite{marita1,marita2} Due to low-lying levels ($J^\pi=3/2^-$ at 9.75 and 74.3 keV, $5/2^-$ at 75 keV, and $7/2^-$ at 100 keV) in $^{187}$Os, $f_\mathrm{SEF}^\mathrm{rate}$ is higher for $^{187}$Os(n,$\gamma$)$^{188}$Os at low temperature than for neutron captures on neighboring isotopes. The enhancement factors of the reactions $^{187}$Os(n,$\gamma$) and $^{188}$Os(n,$\gamma$) both rise to a maximum and decline from there, soon reaching values below Unity and thus not being ``enhancements'' anymore. While the $f_\mathrm{SEF}^\mathrm{rate}$ for both capture reactions stay below a factor of two (and reach only a few tens of percent at the s-process temperature relevant for the cosmochronometer), the $f_\mathrm{SEF}^\mathrm{rate}$ of $^{186}$Os(n,$\alpha$)$^{183}$W is also shown here to give an example for larger values (see below for the even larger factors encountered for photodisintegration reactions).

Having realized the importance of constraining $E_0^A$ by the relevant energy window at a given temperature and the definition of the range of energies $0\leq E_\mu^\mathrm{x} < E_0^A$ of relevant excited states, we can arrive at a more general understanding of when $\mathcal{D}$ will considerably differ from Unity. In the assessment of which excited states in the target are contributing to the stellar rate, it is incorrect to directly use the Boltzmann population factors $\mathcal{P}_\mu$ from Eq.~(\ref{eq:csstar})! These are only appropriate when each state is bombarded by its own MBD of projectiles. They cannot be used when we calculate the rate as usual, integrating over just one MBD with the energy scale being relative to the ground state.

\begin{figure}
\centerline{\includegraphics[width=0.64\textwidth]{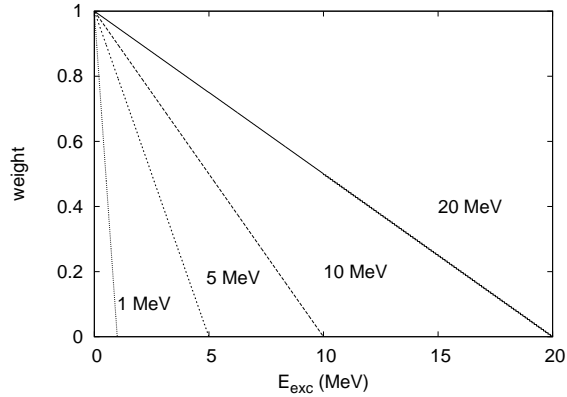}}
\vspace*{8pt}
\caption{Example for relative effective weights $W$ for levels at excitation energy $E_\mathrm{exc}$ and zero spin. The weights are independent of stellar temperature but their slopes depend on the maximal transition energy $E_0$ as indicated by the labels.\label{fig:excweights1}}
\end{figure}

As shown in Eqs.~(\ref{eq:stellrate2})--(\ref{eq:effrate}) the transformation to a single MBD is possible because the Boltzmann weight $\exp(-E_\mu^x/(kT))$ in the population factor offsets the exponentials in the MBD for each state. This offset is not a complete cancellation but results in a transformation of the population $\mathcal{P}_\mu$. The individual weights are transformed to linearly declining functions of the excitation energy
\begin{align}
\mathcal{P}_\mu &\longrightarrow \mathcal{W}_\mu \quad, \nonumber \\
\mathcal{P}_\mu &=\frac{g_\mu}{g_0 G_0(T)}e^{-\frac{E_\mu^\mathrm{x}}{kT}}=\mathcal{N}_0 g_\mu e^{-\frac{E_\mu^\mathrm{x}}{kT}} = \mathcal{N}_0 P_\mu \quad, \nonumber \\
\mathcal{W}_\mu &=\frac{g_\mu}{g_0 G_0(T)} \left( 1-\frac{E_\mu^\mathrm{x}}{E_0} \right) = \mathcal{N}_0 g_\mu \left( 1-\frac{E_\mu^\mathrm{x}}{E_0} \right) = \mathcal{N}_0 W_\mu \quad. \label{eq:effweights}
\end{align}
This is also readily seen in the weighting of the reactions from excited states in the definition of the effective cross section by Eq.~(\ref{eq:effcs}). These relative \textit{effective weights} $W_\mu$ are the ones to be employed when using the standard definition of the rate as in (\ref{eq:rate}), instead of the relative weights $P_\mu$. Contrary to the $P_\mu$, the relative effective weights $W_\mu$ are not explicitly temperature dependent anymore. However, they depend on the energy $E_0$ of the transitions to the ground state which are the transitions with the highest possible relative energy. In the application to astrophysical reaction rates the range of $E_0$ is given by the relevant energy windows as discussed in Sec.~\ref{sec:energies}. It is a relatively narrow range, depending on stellar temperature and the type of reaction (and thus introducing an implicit temperature dependence). Due to the linearity in excitation energy $E_\mu^\mathrm{x}$ -- contrary to the $P_\mu$ which show an exponential decrease -- almost all states up to $E_0$ contribute. Neglecting the spin weights $g_\mu$, one has to consider levels up to $E_\mu^\mathrm{x}\approx (2/3) E_0$ to include 90\% of the levels with non-negligible weight. Depending on the nucleus and considered reactions, some levels with large spin values and/or large reaction cross sections may require the inclusion of all levels up to very close to $E_\mu \approx E_0$. The relative effective weights $W_\mu$ as the product of the population factors with their respective MBDs are shown in Fig.~\ref{fig:excweights1}.

Understanding the $W_\mu$ we are ready to make some general statements on the magnitude of $\mathcal{D}$ in various reactions in a range of astrophysically relevant temperatures.
For example, neutron captures have their relevant energies around the maximum of the MBD, $E_0\approx kT$. This is between a few keV up to several tens of keV for the s-process\cite{radiobook,sprocess} (see also Sec.~\ref{sec:energies}). Therefore only (exothermic) captures on nuclei with low-lying excited states below several tens of keV will exhibit $\mathcal{D}>1$. For light and intermediate mass nuclei, the average level spacings typically are larger than $\approx 100$ keV and thus $\mathcal{D}$ remains close to Unity, with a few notable exceptions. The r-process\cite{ctt,arngorr} involves neutron captures at temperatures of $1-2$ GK which translates to a location of the MBD peak at $\approx 80-160$ keV and we will expect slightly larger $\mathcal{D}$ than for the s-process on average.\footnote{Since the s-process proceeds along the line of stability, the level structure of the involved nuclei, including isomers, is known quite well. This is problematic in the r-process as the experimental nuclear structure information far off stability is limited.}

\begin{figure}
\centerline{\includegraphics[width=0.8\textwidth]{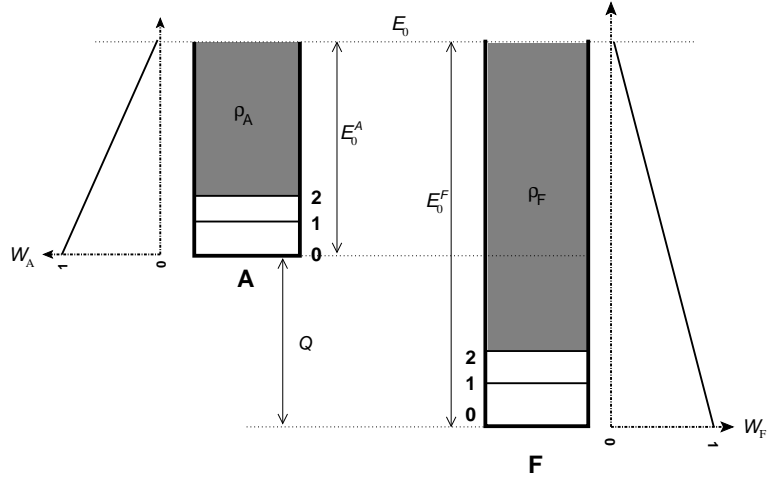}}
\vspace*{8pt}
\caption{Schematic view of the relative effective weights $W_A$ and $W_F$ in the initial and final nucleus, respectively, under the assumption of a positive reaction $Q$-value $Q=Q_{Aa}$. In both nuclei a few low-lying states are explicitly numbered, above them nuclear level densities are indicated (shaded areas). \label{fig:excweights2}}
\end{figure}

Reactions having their relevant energy window determined by charged particle widths have considerably larger $E_0$, in the range of about $0.5-13$ MeV, depending on the stellar temperature and the charges of the interacting particles. Accordingly, $\mathcal{D}$ can already become large at rather low temperature because transitions from many excited states have to be considered. Only in light, strongly bound nuclei with level spacings of several MeV above the ground state, $\mathcal{D}$ may still remain close to Unity.

The relative effective weights $W$ derived in (\ref{eq:effweights}) also allow to understand a rule-of-thumb which has been known, but never quantified, for quite some time (see, e.g., Ref.~\refcite{hwfz}). The rule states that exothermic reactions ($Q=Q_{Aa}>0$) usually have smaller $\mathcal{D}$ than their endothermic inverses. Defining the forward reactions by the exothermic reaction and the reverse reaction by its endothermic counterpart, this means that $\mathcal{D}^\mathrm{rev}>\mathcal{D}^\mathrm{forw}$ and frequently $\mathcal{D}^\mathrm{rev}\gg \mathcal{D}^\mathrm{forw}$. This is not immediately comprehensible upon inspection of the Boltzmann weights $P_\mu$ as these seem to act similarly in the initial and final nuclei. As stated above, however, it is a mistake to straightforwardly use the $P_\mu$ together with a single MBD. The rule becomes obvious when using the appropriate $W_\mu$. The situation is sketched in Fig.~\ref{fig:excweights2} where the behavior of the weights is shown for two nuclei $A$ and $F$ being the target and the final nucleus of a reaction, respectively. For the relevant interaction energy set to $E_0$ the maximum relative transition energy is $E_0^A$ in the target nucleus and $E_0^F=E_0^A+Q$ in the final nucleus. Provided that $Q>0$ the energy range of possible transitions $0<E_\nu \leq E_0^F$ in the final nucleus is larger than in the target nucleus. In consequence, the relative effective weights $W_\nu$ in nucleus $F$ decline slower with increasing excitation energy and thus a larger range of levels contributes. This assumes similar spin structure in the two nuclei, of course, and the rule may not work at small $|Q|$ when $W_\mu$, $W_\nu$ are similar and the spin factors $g_\mu$, $g_\nu$ dominate.

\begin{table}[t]
\tbl{Stellar enhancement factors  in ($\gamma$,n) reactions. Values are from a NON-SMOKER calculation,\protect\cite{nonsmokerphoto} as reported in Ref.~\protect\refcite{utso}.\label{tab:photosefs}}
{\begin{tabular}{rrrrrrrr} \toprule
Nucleus: & $f_\mathrm{SEF}^\mathrm{rate}$ & Nucleus: & $f_\mathrm{SEF}^\mathrm{rate}$ & Nucleus: & $f_\mathrm{SEF}^\mathrm{rate}$ & Nucleus: & $f_\mathrm{SEF}^\mathrm{rate}$ \\
\colrule
$^{186}$W: & 400 &$^{190}$Pt: & 5500 &$^{197}$Au:&1100 &$^{204}$Hg:&43 \\
$^{185}$Re: & 1300& $^{192}$Pt: & 3300 &$^{196}$Hg:&1700 &$^{204}$Pb:&160\\
$^{187}$Re: & 1200 &$^{198}$Pt: & 310 &$^{198}$Hg:&750 &&\\
\botrule
\end{tabular} }
\end{table}

The largest reaction $Q$-values are encountered in capture reactions. For instance, neutron captures close to stability exhibit $Q$-values of the order of $5-13$ MeV, for highly proton-rich nuclei they can reach $20-29$ MeV. Similar values are found for proton captures around stability and on the neutron-rich side of the nuclear chart. In the light of the above it is not surprising that endothermic photodisintegration reactions exhibit very large $f_\mathrm{SEF}^\mathrm{rate}$ of the order of several hundreds to thousands as illustrated in Table~\ref{tab:photosefs}.

An exception to the above rule $\mathcal{D}^\mathrm{rev}>\mathcal{D}^\mathrm{forw}$, however, has recently been discovered.\cite{sefkiss,seftom} Although the effective weight $\mathcal{W}_\mu$ may be slowly decreasing with increasing excitation energy $E_\mu^\mathrm{x}$ of a level, the corresponding cross section $\sigma^\mu$ may decrease much faster, even exponentially. This is because with increasing $E_\mu^\mathrm{x}$ the relative interaction energy in that channel $E_\mu=E_0-E_\mu^\mathrm{x}$ is reduced. If the cross section $\sigma^\mu$ is strongly decreasing with decreasing energy -- as it is the case in the presence of a Coulomb barrier or at high relative angular momentum -- transitions on excited states will cease to importantly contribute to the effective cross section even when being strongly weighted. This is the reason why charged particle reactions show only moderate values of $\mathcal{D}$ at low $T$ even though their relevant energy window may be high above the reaction threshold. This, of course, acts in the entrance channel of a reaction as well as in its exit channel.

The point, however, is to realize that the barriers may be different in the two channels, leading to a different suppression of contributions. For instance, a (n,p) or (n,$\alpha$) reaction has a Coulomb barrier only in the exit channel, (p,$\alpha$) reactions have different barriers in entrance and exit channel. This has the consequence that low-energy transitions in the exit channel of an exothermic reaction may be suppressed in such a manner as to yield $\mathcal{D}^\mathrm{rev}<\mathcal{D}^\mathrm{forw}$. Whether this is the case will strongly depend on the $Q$-value because it determines the range of transition energies to be efficiently suppressed. The higher the barrier, the larger $Q$-value is allowed while still permitting to suppress most transitions from excited states. The stellar temperature also plays a role but its impact is smaller.

\begin{figure}
\centerline{\includegraphics[angle=-90,width=0.7\textwidth]{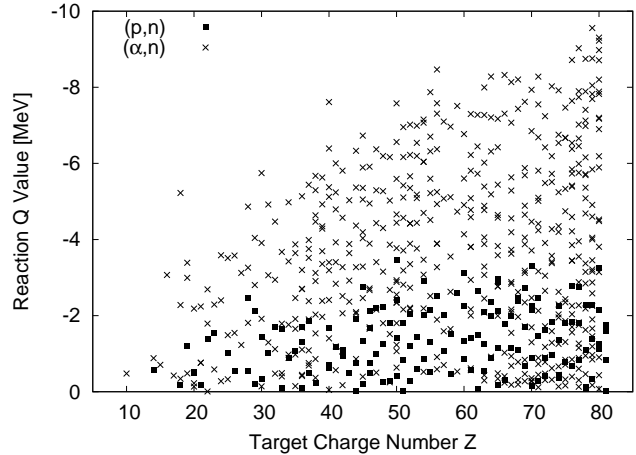}}
\vspace*{8pt}
\caption{Reaction $Q$-values of endothermic (p,n) and ($\alpha$,n) reactions which still permit a considerable Coulomb suppression of the stellar enhancement factors. \label{fig:qsupp}}
\end{figure}

A global search across the chart of nuclides for reactions exhibiting $\mathcal{D}^\mathrm{rev}<\mathcal{D}^\mathrm{forw}$ was performed in Refs.~\refcite{sefkiss,seftom}. The study focused on identifying cases which are interesting for experiments by requiring $\mathcal{D}^\mathrm{rev}<1.5$ and the stellar enhancement factors for forward and reverse reaction to differ by more than 10\%. At stellar temperatures $T \leq 4.5$ GK about 1200 reactions were found but not all of them are astrophysically interesting.\cite{seftom} Figure~\ref{fig:qsupp} shows the dependence of the $Q$-values on the charge of the target nucleus of two types of reactions selected from the total set. The envelope from the maximal $|Q|$ appearing for each type at each charge $Z$ illustrates the action of the increasing Coulomb barrier. It can be clearly seen that
larger maximal $\lvert Q\rvert$ is allowed with increasing charge $Z$. The increase with increasing charge is steeper for ($\alpha$,n) reactions than for (p,n) reactions due to the higher Coulomb barrier for $\alpha$ particles. The scatter at a fixed charge number $Z$ is due to the range of $Q$-values found within an isotopic chain.

Summarizing the above, comparing the position of the relevant energy window with the average level spacing in a nucleus already gives an estimate of the magnitude of the stellar enhancement to be expected.
When measuring reaction cross sections for astrophysics it is desireable to perform the experiment in the reaction direction showing smallest $\mathcal{D}$ in order to stay as close as possible to the actually required stellar cross section. This favored direction turns out to be the exothermic one in the vast majority of cases, i.e., a reaction with positive $Q$-value, with comparatively few exceptions (compared to the total number of conceivable reactions) as discussed above. It should be remembered that the reverse rate can always be calculated using the reciprocity relations (\ref{eq:revrate}) and (\ref{eq:revphoto}).\footnote{There is another reason why the knowledge of the exothermic rate is favored which is not connected to the stellar enhancement. Reaction network calculations implement rates either through tables or by using fits of the rates as functions of temperature. In the latter case it is highly advantageous to be able to perform the fit of the exothermic rate. This is because any small deviation of the fit from the actual rate will be strongly enhanced by the factor $\exp\left(-Q/(kT)\right)$, appearing in (\ref{eq:revrate}), (\ref{eq:revphoto}), when $Q<0$.}

Finally, it is worth mentioning that similar considerations for the transformation of effective weights and deviations $\mathcal{D}$ from the ground state values also apply to other types of reactions, not only the ones discussed here. For instance, also for electron captures and other reactions mediated by the weak interaction an effective cross section, effective weights, and relevant energy windows can be derived, applying the appropriate energy distributions. They may become more complicated, however, than the ones obtained for the MBD because an explicit dependence on the chemical potential is required in the general case. Moreover, neutrinos are not in thermal equilibrium with nuclei and thus their temperature will enter as an additional parameter.

\subsection{Electronic plasma effects}
\label{sec:electro}

The plasma temperature in the astrophysical sites where nuclear reactions occur is so high that the nuclei are fully ionized and embedded in a cloud of free electrons. This situation affects reactions and decays and has to be considered when preparing a rate to be used in astrophysical reaction networks. Similar to the treatment of stellar enhancement factors, these corrections also have to be modeled theoretically and experimental rates have to be corrected for them. For completeness, some of the effects are discussed briefly without going into detail here.

Decays (and electron captures) are affected not only by the thermal excitation of the nucei as shown in (\ref{eq:stelldecay}) but also by the electrons surrounding the nuclei. A nucleus in an atom can be converted by, e.g., capture of an electron from the atomic K-shell. In the plasma, electrons are captured from the electron cloud and this may alter the half-life of a nucleus considerably. A well-known example for this is the decay of $^7$Be. Its lifetime under central solar conditions ($\tau_\mathrm{ec}=140$ d) is almost double the one under terrestrial, non-ionized conditions ($\tau_\mathrm{ec}=77$ d). Also
the $\beta^-$ decay lifetimes are modified by a change in the electron emission probability. The plasma electrons reduce the phasespace available for emission to the continuum and thus increase $\tau_\beta^-$. On the other hand, bound-state decay, i.e.\ the placement of the emitted electron into a low-lying atomic shell, becomes possible even when it were forbidden in an atom because of the occupation of available electron shells. Similar considerations apply to charged-current reactions with electron neutrinos. These effects act in addition to the alteration of the lifetimes through thermal excitation of the nucleus and depend on the plasma temperature, density, and composition, which may all affect the distribution of the electrons throughout the plasma.

Nuclear reactions with charged nuclei are affected by \textit{electron screening}. Electrons in the vicinity of the nucleus shield part of its charge and thus effectively lower its Coulomb barrier. Theoretical predictions of cross sections and rates always assume bare nuclei without any electrons and therefore have to be corrected for screening. The magnitude of the screening is strongly dependent on the temperature $T$, density $\rho_\mathrm{pla}$, and composition $Y=\sum_i Y_i$ of the plasma. At high density, electron screening can increase the rate by several orders of magnitude and lead to \textit{pycnonuclear} burning at much lower temperatures than otherwise needed to ignite nuclear burning. Moreover, dynamical effects due to the fast movement of electrons may also play a role, although there is discussion on whether this is important when all reactants are in thermodynamic equilibrium.\cite{BrSa97} It is apparent that the theoretical treatment of screening is complicated and we are far from complete microscopic descriptions.\cite{Shaviv,Gasques05,Sawyer} The proper inclusion of screening effects remains a challenging problem in plasma physics.

Under most conditions, the screened reactivity can be decomposed into the regular stellar reactivity and a screening factor\cite{salpeter}
\begin{equation}
\langle \sigma v \rangle^*_\mathrm{screened} = \mathcal{C}(T,\rho_\mathrm{pla},Y) \langle \sigma v \rangle^* \quad.
\end{equation}
This implies that the Coulomb potential $V_c$ seen by the reaction partners can be described by the bare Coulomb potential and an effective screening potential $U$
\begin{equation}
V_c(r,T,\rho_\mathrm{pla},Y)=V_c^\mathrm{bare}(r)+U(r,T,\rho_\mathrm{pla},Y)=\frac{Z_A Z_a e^2}{r}+U(r,T,\rho_\mathrm{pla},Y) \quad.
\end{equation}
Then the screening factor acquires the form $\mathcal{C}=\exp(-\tilde{U}/(kT))$.
The challenge is in the determination of $\tilde{U}$ depending on the plasma conditions.

An often used static approximation, being appropriate for early burning stages of stars, is \textit{weak screening} in the Debye-H\"uckel model.\cite{salpeter,debhu} Weak screening assumes that the average Coulomb energy of each nucleus is much smaller than the thermal energy, i.e.\ $ZeV_c\ll kT$. Then a nucleus will be surrounded by a polarized sphere of charges, with a radius
\begin{equation}
R_\mathrm{D} =\sqrt{\frac{kT}{4\pi e^2 \rho_\mathrm{pla}N_A\zeta}}
\end{equation}
with
\begin{equation}
\zeta=\sum_i {\left( Z_i^2+Z_i\right) ^2 Y_i} \quad,
\end{equation}
where the sum runs over all charged plasma components. The screening factor $\mathcal{C}$ is transformed to
$\mathcal{C}_\mathrm{D}=1-\tilde{U}_\mathrm{D}/(kT)$ with $\tilde{U}_\mathrm{D}=-e^2 Z_A Z_a/R_\mathrm{D}$.

For \textit{strong screening} in high density plasmas it is more appropriate to use the ion-sphere model instead of the Debye-H\"uckel approximation.\cite{salpeter} The ion-sphere model is equivalent to the Wigner-Seitz model used in condensed matter theory.

Another type of screening is observed in nuclear experiments in the laboratory. There, nuclei are present in atoms, molecules, or metals, each with specific electron charge distributions around the nucleus. Although completely different from plasma screening, this type of screening has to be understood because it is especially important at the low interaction energies of astrophysical relevance. The \textit{measured} reaction cross sections have to be corrected to obtain the bare cross section which can be compared to theory or used to determine the rate. Atomic screening can be treated in the adiabatic approximation, leading to
\begin{equation}
\frac{\sigma_\mathrm{screened}(E)}{\sigma(E)}=\frac{E}{E+U_\mathrm{e}} \exp \left( \frac{\pi\eta
U_\mathrm{e}}{E} \right)
\end{equation}
with the Sommerfeld parameter $\eta$ from (\ref{eq:sommerfeld}). The screening potential $U_\mathrm{e}$ in this approximation is given by the difference in the electron binding energy of the target atom and the atom made from target atom plus projectile $U_\mathrm{e}=B_\mathrm{e}^{A+a}-B_\mathrm{e}^A$. In light systems, the velocity of the atomic electrons is comparable to the relative motion between the nuclei. Therefore a dynamical model is more appropriate.\cite{shoppa} However, the adiabatic approximation provides an upper limit on the expected screening effect on the cross section.

There seem to be discrepancies between theory and laboratory determinations of $U_\mathrm{e}$, the latter often yielding much larger values of $U_\mathrm{e}$. Some of them have been resolved through improved stopping powers used in the determination of the experimental cross sections,\cite{shoppa2,bang,bertu} while others remain puzzling, especially regarding cross sections of nuclei implanted in metals.\cite{formicola,czer,rai1,rai2,rai3} Thus, the laboratory screening seems to be less understood than the stellar screening.

\section{Reaction mechanisms}
\label{sec:mech}

\subsection{General considerations}
\label{sec:mechintro}

Having discussed the special requirements of astrophysics for the determination of the stellar rates in the preceding sections, we turn to the question of how to obtain the cross sections $\sigma^{\mu\nu}$ required in Eqs.~(\ref{eq:rate}), (\ref{eq:csstar}) and (\ref{eq:effcs}) at the energies of astrophysical relevance. It has become apparent that astrophysical rates include more transitions than usually obtained in straightforward laboratory reaction cross section measurements. Cleverly designed experiments may study some of them but especially at larger stellar temperatures, as they are typical for explosive burning, theory will be indispensable for providing the appropriate \textit{stellar} reaction rates. Even at stability, the small cross sections of charged-particle reactions at astrophysically relevant energies (see Sec.~\ref{sec:energies}) pose a considerable challenge for current measurements and future experiments have to employ novel techniques or new facilities to address this problem. Moreover, hot astrophysical environments produce highly unstable nuclei which cannot be studied in the laboratory, yet. Reaction cross sections of nuclei far from stability at astrophysical energies probably will never be experimentally determined. Therefore reaction networks for explosive nucleosynthesis have to include the majority of their reaction rates from theoretical predictions although experiments may help to determine nuclear properties and some of the transitions required for the calculation of effective cross sections.

Although reaction theory dates back as far as the 1950s, the special requirements of astrophysics and the need for cross section predictions of nuclei far from stability provide an interesting and stimulating environment for the application and further developments of different approaches. The challenges are manyfold. On one hand, astrophysical energies are very low and, as we will see below,  different reaction mechanisms may contribute or interfere. On the other hand, even if the reaction mechanism is unique and well understood, nuclear properties entering the reaction model have to be predicted for nuclei far off stability. This proves challenging even for modern nuclear structure calculations. Although a fully microscopic treatment is preferrable, good parameterizations and averaged quantities are still necessary in many cases due to the sheer number of reactions and involved nuclei, especially for intermediate and heavy nuclei consisting of more than $30-40$ nucleons and thus not allowing the application of few-body models. Finally, the interpretation of experiments has to be supported by theory. This latter case may involve different methods than the one dealing with the prediction of astrophysical rates because experiments may be conducted at higher energies and theory is needed to extract the information to be included in the rates.\cite{Bertu2010} For example, the properties of excited states and their spectroscopic factors can be studied by (d,p) reactions at comparatively high energy which are not directly relevant in astrophysics. In the following I focus on theory for the prediction of astrophysical reaction rates.

Theoretical models can be roughly classified in three categories:\cite{descrau,De03}
\begin{enumerate}
\item Models involving
adjustable parameters, such as the $R$-matrix\cite{LT58} or the $K$-matrix\cite{Hu72} methods;
parameters are fitted to the available experimental data and the cross sections are extrapolated
down to astrophysical energies. These fitting procedures, of course, require the knowledge of
data, which are sometimes too scarce for a reliable extrapolation.
\item ``Ab initio" models, where the cross sections are determined from the wave functions
of the system. The potential model\cite{BD85}, the Distorted Wave Born Approximation (DWBA)\cite{OS91,Bertu2010}, and microscopic models\cite{WT77,La94,Bertu2010} are, in principle, independent of experimental data. More realistically, these models depend on some physical parameters, such as a nucleus-nucleus or a nucleon-nucleon interaction which can be
reasonably determined from experiment only. The microscopic Generator Coordinate Method (GCM)
provides a ``basic" description of a nucleonic system, since the whole information is obtained
from a nucleon-nucleon interaction. Since this interaction is nearly the same
for all light nuclei, the predictive power of the GCM is high for such nuclei.
\item The above models can be used for low level-density nuclei only. This condition
is fulfilled in most of the reactions involving light nuclei ($A \leq 20$). However
when the level density near threshold is large (i.e. more than a few levels per MeV),
statistical models, using averaged optical transmission coefficients, are
more suitable (see Sec.\ \ref{sec:statmod}).
\end{enumerate}
The nuclear level density (NLD) at the compound nucleus excitation energy corresponding to the astrophysical energy window determines which reaction mechanism is applicable and which model to choose. The compound formation energy $E_\mathrm{form}=E_0+E_\mathrm{sep}$ is given by the astrophysical energy $E_0$ relative to the ground state of the target nucleus and the separation energy of the projectile $E_\mathrm{sep}$. Around stability, $E_\mathrm{sep}$ usually is high and dominates $E_\mathrm{form}$. Statistical models will then be applicable for intermediate and heavy nuclei with sufficiently high NLD. Even at stability, however, the NLD may not be high enough at nuclei with closed shells (see Sec.~\ref{sec:statmod}). Approaching the driplines, the neutron- or proton-separation energies strongly decrease, resulting in low $E_\mathrm{form}$ for neutron- or proton-induced reactions, respectively. This leads to low NLD at $E_\mathrm{form}$ even for intermediate and heavy nuclei.\cite{rtk97} Isolated resonances but also direct reactions will become important.\cite{MaMe83}

In the following, reactions at intermediate (Sec.~\ref{sec:reso}), high (Sec.~\ref{sec:highnld}), and low (Sec.~\ref{sec:direct}) compound NLD are discussed separately although there may be contributions from several reaction mechanisms simultaneously, especially in systems with low and intermediate NLD. The discussion will focus on models more or less applicable for large-scale predictions across the nuclear chart. Further models are presented in, e.g., Refs.\ \refcite{descrau,De03,La94,Bertu2010}.

\subsection{Resonant reactions}
\label{sec:reso}

Resonances in reaction cross sections are important for the majority of nuclei.
Depending on the number of nucleons in the target nucleus resonances appear in the reaction cross section at lower or higher energy and their average spacing also depends on the structure of the nucleus. Astrophysical energy windows cover regions of widely spaced, isolated resonances to regions of a large number of overlapping, unresolved resonances. Accordingly, different approaches have to be combined. The latter region is more important for reactions between charged reactants because the Coulomb barrier shifts the relevant energy window to higher energy compared to reactions where neutrons determining the location of the window (see Sec.~\ref{sec:energies}).

Isolated resonances in the low and intermediate NLD regimes can be treated in the $R$-matrix\cite{LT58} or the $K$-matrix\cite{Hu72} approaches or by applying simple single-level or multi-level Breit-Wigner formulae. In all these methods, the resonance properties (resonance energy, spin, partial and total widths) have to be known. Often, an inverse approach is used and the resonance properties are derived from experimental data by, e.g., $R$-matrix fits. Where this is impossible, nuclear theory has to be invoked to predict the required quantities. This remains problematic, however, because the reaction cross sections are very sensitive to the resonance properties. The astrophysical rates are also sensitive but since their calculation involves an integration over an energy range, only strong resonances truly contribute and others may be averaged out. Nevertheless, there are large uncertainties in reaction rates off stability due to the unknown resonance contributions (see also Sec.~\ref{sec:sensi}). Cluster models (see Sec.~\ref{sec:optmod} and Ref.~\refcite{De03}) have been successful in describing resonant cross sections in light nuclei but cannot be easily applied to nuclei at intermediate and heavy mass.

Although resonances with the same spin $J$ interfere and single resonances may also show interference with a direct reaction (Sec.~\ref{sec:direct}), the single-level Breit-Wigner formula (BWF) is often used:\cite{ilibook,BW52}
\begin{equation}
\label{eq:breit}
\sigma^{Aa\rightarrow Fb}_{\rm BW}(E_\mu)=\sigma^\mu_{\rm BW}=\frac{\pi}{k^2_\mu}
\frac{1+\delta_{Aa}}{g_\mu g_a}\sum_{i=1}^n g_i
\frac{\Gamma_i^\mu(E_\mu)
\hat{\Gamma}_i^{Fb}(E_\mu)}{(E_\mu-E_{i,\mu}^\mathrm{res})^2+(\hat{\Gamma}^{\rm tot}_i(E_\mu)/2)^2} \quad .
\end{equation}
It is quoted here despite its restrictions because it allows to demonstrate some important principles important for the relation between stellar and laboratory rates. Equation (\ref{eq:breit}) gives the BWF for $n$ non-interfering resonances. The wave number is denoted by $k_\mu$. The total
width $\hat{\Gamma}^{\rm tot}_i$ of a resonant state $i$ in the compound nucleus is the sum over the widths of the individual decay channels $\hat{\Gamma}^{\rm tot}_i =\hat{\Gamma}_i^{Aa}+\hat{\Gamma}_i^{Fb}+\dots$, also including transitions to other reaction channels beyond the exit channel $F+b$. The widths of the individual decay channels are summed over transitions to all possible final states in the channel. Thus, $\hat{\Gamma}_i^{Fb}=\sum_\nu \Gamma_i^\nu$. Figure \ref{fig:CNscheme} shows the energy scheme and the contributing transitions in each channel. When the resonance energy $E_i^\mathrm{res}$ is known, the widths $\Gamma$ and $\hat{\Gamma}$ can be calculated from the transmission coefficients obtained by solution of a Schr\"odinger equation in the optical model (see Sec.~\ref{sec:optmod}) and a spectroscopic factor (see Sec.~\ref{sec:direct}). Both resonance energies and spectroscopic factors should, in principle, be predictable in the shell model (see, e.g., Refs.~\refcite{kwas94,herbro97,fisker}) or other microscopic theories but this is currently not feasible for all nuclei across the nuclear chart. Different approaches yield results which differ more than it is tolerable in the calculations of astrophysical reaction rates.\cite{microdirect}

According to (\ref{eq:labcs}), we take $\mu=0$ for the usual laboratory cross section. It is interesting to note that also resonances located below the reaction threshold may contribute due to their finite width reaching above the threshold. These are called \textit{sub-threshold resonances}.\cite{cauldrons,ilibook} In reactions with a large, positive $Q$-value the energy dependence of the partial width in the exit channel $\Gamma_i^{Fb}$ can be neglected. This is not true when the $Q$-value is small or negative.

As has become obvious in the discussion of the stellar cross section in Sec.~\ref{sec:stellar}, for the astrophysical reaction rate, the effective cross section has to be employed in the integration for the reaction rate and thus a weighted sum over excited target states $\mu$ has to be performed and we obtain
\begin{equation}
\langle \sigma v \rangle_\mathrm{BW}^* = \left(\frac{8}{m_{Aa} \pi}\right)^{1/2}
(kT)^{-3/2} \int \limits_{0}^\infty { \left\{ \sum_\mu \mathcal{W}_\mu \sigma_\mathrm{BW}^\mu (E_\mu) \right\} E_0^A e^{-\frac{E_0^A}{kT}}\,dE_0^A } \quad, \label{eq:stellBW}
\end{equation}
with the effective weights $\mathcal{W}$ taken from (\ref{eq:effweights}) and $E_\mu=E_0-E_\mu^\mathrm{x}$. This can be simply achieved by replacing $\Gamma_i^\mu$ in (\ref{eq:breit}) by $\hat{\Gamma}_i^{Aa}$ (summing over all possible transitions in the entrance channel) and dividing the resulting cross section by the normalized partition function of the target nucleus $G_0^A$.
This can easily be shown when combining definitions (\ref{eq:effcs}) and (\ref{eq:breit}) for a single resonance with spin $J$,
\begin{align}
\sigma^\mathrm{eff}_\mathrm{BW}&=\sum_\mu \frac{g_\mu}{g_0} \frac{E_\mu}{E_0} \sigma^\mu_\mathrm{BW} = \frac{\pi g_J}{g_0 g_a E_0} \sum_\mu \frac{E_\mu}{k_\mu^2} \frac{\Gamma^\mu \hat{\Gamma}^{Fb}}{(E_\mu-E_\mu^\mathrm{res})^2+(\hat{\Gamma}^\mathrm{tot}/2)^2} = \nonumber \\
&= \frac{\pi g_J (1+\delta_{Aa})}{g_0 g_a E_0} \sum_\mu \frac{\hbar^2}{2m_{Aa}}\frac{\Gamma^\mu \hat{\Gamma}^{Fb}}{(E_0-E_0^\mathrm{res})^2+(\hat{\Gamma}^\mathrm{tot}/2)^2} = \nonumber \\
&= \frac{\pi}{k_0^2} \frac{g_J (1+\delta_{Aa})}{g_0 g_a} \frac{\sum_\mu \Gamma^\mu \hat{\Gamma}^{Fb}}{(E_0-E_0^\mathrm{res})^2+(\hat{\Gamma}^\mathrm{tot}/2)^2} = \nonumber \\
&= \frac{\pi}{k_0^2} \frac{g_J (1+\delta_{Aa})}{g_0 g_a} \frac{\hat{\Gamma}^{Aa} \hat{\Gamma}^{Fb}}{(E_0-E_0^\mathrm{res})^2+(\hat{\Gamma}^\mathrm{tot}/2)^2} \quad.
\end{align}
This results in
\begin{align}
\langle \sigma v \rangle_\mathrm{BW}^* &= \sum_{i=1}^n \langle \sigma v \rangle_{\mathrm{BW},i}^* = \nonumber \\
&= \left(\frac{8}{m_{Aa} \pi}\right)^{1/2}
(kT)^{-3/2} \frac{1}{G_0^A} \int \limits_{0}^\infty { \left\{ \sum_{i=1}^n \sigma_\mathrm{{BW},i}^\mathrm{eff} \right\} E_0^A e^{-\frac{E_0^A}{kT}}\,dE_0^A } = \nonumber \\
&= \left(\frac{8\pi}{m_{Aa}}\right)^{1/2} \frac{1}{G_0^A (kT)^{(3/2)}}
\frac{1+\delta_{Aa}}{k_0^2 g_a g_0} \times \nonumber \\
&\times \int \limits_{0}^\infty { \left\{ \sum_{i=1}^n g_i
\frac{\hat{\Gamma}_i^{Aa}(E_0^A)
\hat{\Gamma}_i^{Fb}(E_0^A)}{(E_0^A-E_{i}^\mathrm{res})^2+(\hat{\Gamma}^{\rm tot}_i(E_0^A)/2)^2} \right\} E_0^A e^{-\frac{E_0^A}{kT}}\,dE_0^A } \label{eq:sumgamma}
\end{align}
for the stellar rate.

Some simplifications can be made depending on the resonance widths. For simplicity, the derivations are given for a single resonance with spin $J$, for $n>1$ the contributions can be added. A frequently used simplification is the one for narrow resonances, assuming that the widths $\hat{\Gamma}$ in the numerator in the integral and the Boltzmann factor $\exp (E_0^A/(kT))$ do not change across the width of the resonance.\cite{cauldrons,ilibook,fcz75} Then their values can be taken at the resonance energy and the integration can be performed analytically (see also Eq.~(\ref{eq:hfaverage})), yielding
\begin{equation}
\langle \sigma v \rangle_\mathrm{narrow}^* = \left(\frac{2\pi}{m_{Aa} kT}\right)^{3/2} \hbar^2 e^{-E_0^\mathrm{res}/(kT)}\frac{1}{G_0^A} \frac{(1+\delta_{Aa})g_J}{g_a g_0} \frac{\hat{\Gamma}^{Aa}(E_0^\mathrm{res})
\hat{\Gamma}^{Fb}(E_0^\mathrm{res})}{\hat{\Gamma}^{\rm tot}(E_0^\mathrm{res})} \quad,
\label{eq:narrowres}
\end{equation}
where the resonance energy $E_0^\mathrm{res}$ is given relative to the ground state of the target nucleus. How does this compare to the ground state rate usually measured in the laboratory? The stellar rate can be recast as
\begin{align}
\langle \sigma v \rangle_\mathrm{narrow}^* &= \langle \sigma v \rangle_\mathrm{narrow}^\mathrm{g.s.} \frac{1}{G_0^A} \left(1+\sum_\mu \frac{\Gamma(E_\mu)\hat{\Gamma}^{Fb}(E_\mu)\hat{\Gamma}^\mathrm{tot}(E_0)}{\Gamma(E_0)\hat{\Gamma}^{Fb}(E_0)\hat{\Gamma}^\mathrm{tot}(E_\mu)} \right) \approx \nonumber \\
&\approx \langle \sigma v \rangle_\mathrm{narrow}^\mathrm{g.s.} \frac{1}{G_0^A} \left(1+\sum_\mu \frac{\Gamma(E_\mu)}{\Gamma(E_0)} \right) \quad,
\label{eq:narrowcontrib}
\end{align}
with
\begin{equation}
\langle \sigma v \rangle_\mathrm{narrow}^\mathrm{g.s.} = \left(\frac{2\pi}{m_{Aa} kT}\right)^{3/2} \hbar^2 e^{-E_0^\mathrm{res}/(kT)} \frac{(1+\delta_{Aa})g_J}{g_a g_0} \frac{\Gamma^0(E_0^\mathrm{res})
\hat{\Gamma}^{Fb}(E_0^\mathrm{res})}{\hat{\Gamma}^{\rm tot}(E_0^\mathrm{res})} \quad.
\label{eq:narrowgs}
\end{equation}
The second line in (\ref{eq:narrowcontrib}) was obtained by neglecting the energy dependence of $\hat{\Gamma}^{Fb}$ and $\hat{\Gamma}^\mathrm{tot}$. This is a valid assumption provided the reaction has a sizeable, positive $Q$-value.
As can be seen from (\ref{eq:narrowcontrib}), the contributions from excited states vanish quickly because often $\Gamma$ is strongly energy dependent and vanishes fast with decreasing energy (remember that $E_\mu=E_0-E_\mu^\mathrm{x}$). This is certainly true for reactions between charged particles and low resonance energies. Nevertheless, the resonant transitions from excited states may dominate a resonant stellar rate.\cite{ilibook,schatz2005}

For broader resonances the above approximation cannot be used and the rate has to be determined by numerical integration of Eq.~(\ref{eq:rate}). The wings of broad resonances can also contribute significantly to the rate even when the resonance energy is outside the relevant energy window for the rate. Sometimes the values for $\Gamma^0$, $\hat{\Gamma}^{Fb}$, and $\hat{\Gamma}^{\rm tot}$ are known experimentally at the resonance energy. Then a frequently used approach in experimental nuclear physics is to assume that $\hat{\Gamma}^{Fb}$ and $\hat{\Gamma}^{\rm tot}$ are approximated by energy-independent values and the energy-dependence of $\Gamma^0$ is only due to a barrier penetration factor derived from an optical model. Even if the energy dependence of $\hat{\Gamma}^{Fb}$, $\hat{\Gamma}^{\rm tot}$ is accounted for explicitly, this type of extrapolation does not include the stellar enhancement and is only valid for laboratory cross sections. The stellar rate must be calculated from a weighted sum of resonant contributions, as shown in (\ref{eq:stellBW}), both for the value at the resonance energy and in the extrapolation. It follows from (\ref{eq:sumgamma}) that the only difference in the energy dependence, however, stems from the width in the entrance channel where $\Gamma^0$ has to be replaced by $\hat{\Gamma}^{Aa}$. The additional transitions to excited states in the target nucleus can be measured in principle. If they are not available, $\hat{\Gamma}^{Aa}$ at the resonance energy has to be predicted from theory. Also the extrapolation is more involved because $\hat{\Gamma}^{Aa}$ will have a different energy dependence than $\Gamma^0$. The same methods can be used as in the extrapolation of $\Gamma^0$ but they have to be applied to all contributing transitions separately.

The discussion of electron screening in the stellar plasma in Sec.~\ref{sec:electro} applied to nonresonant rates. It can be shown that the same screening corrections can be applied for resonant rates when $\hat{\Gamma}^{Aa}\gg \hat{\Gamma}^{Fb}$.\cite{ilibook} A more complicated form arises for $\hat{\Gamma}^{Aa}\ll \hat{\Gamma}^{Fb}$, see Refs.~\refcite{SalHorn69} and \refcite{mit77} for further details.

\subsection{Resonant reactions at large compound level density}
\label{sec:highnld}

\subsubsection{Optical model}
\label{sec:optmod}

Microscopic models are based on basic principles of quantum mechanics, such as the treatment of all nucleons, with exact antisymmetrization of the wave functions. The hamiltonian of an $\tilde{A}$-nucleon system is
\begin{equation}
H = \sum^{\tilde{A}}_{i=1} \mathcal{T}_i \ + \ \sum^{\tilde{A}}_{i<j=1} \mathcal{V}_{ij},
\label{eq:mic1}
\end{equation}
where $\mathcal{T}_i$ is the kinetic energy and $\mathcal{V}_{ij}$ a nucleon-nucleon interaction.\cite{WT77,La94}
The Schr\"odinger equation associated with this hamiltonian can not be solved exactly when $\tilde{A}>3$. The Quantum Monte Carlo method  represents a significant breakthrough in this direction, but is currently limited to $\tilde{A}=10$.\cite{PW01} In addition its application to continuum states is not feasible for the moment (it has been applied to the d($\alpha$,$\gamma$)$^6$Li reaction but the $\alpha$+d relative motion is described by a nucleus-nucleus potential).\cite{NWS01}

In cluster models, it is assumed that the nucleons are grouped in clusters and internal wave functions describing the relative cluster motions are generated.\cite{descrau} The main advantage of cluster models with respect to other microscopic theories is its ability to deal
with reactions, as well as with nuclear spectroscopy. Over the past years, much work has been devoted to the improvement of the internal wave functions: multicluster descriptions\cite{DB94,DD96}, large-basis shell model extensions\cite{De96}, or monopolar distortion\cite{BK92}. The main limitation arises from the number of
channels included in the wave function, which reduces the validity of the model at low energies.
Also large NLDs require many channels in the wave functions. Therefore the application of cluster models is limited to light nuclei.

Due to the complexity of the nucleon-nucleon (NN) interaction, one often
resorts to working with effective interactions instead of
solving microscopic models based on NN potentials. Widely used in calculating different reaction
mechanisms is the optical model. In that model, the complicated
many-body problem posed by the interaction of two nuclei is replaced by
the much simpler problem of two particles interacting through an
effective potential, the so-called optical potential.\cite{sat83,glen83,gh92}
Such an approach is usually feasible only with few
contributing channels. Always included is
elastic scattering. That is why optical potentials can be derived from
elastic scattering data.

The time-independent radial Schr\"odinger equation is numerically solved
with an optical potential which provides a mean interaction potential,
averaging over individual NN interactions. In consequence,
single-particle resonances cannot be described in such a model. However,
resonances stemming from potential scattering can still be found.
Elementary scattering theory yields expressions for the elastic cross
section and the reaction cross section. The latter includes all
reactions and inelastic processes which cause loss of flux from the
elastic channel. With the diagonal element $S^{\alpha \alpha}$ of the
S-matrix (sometimes also called scattering
matrix or collision matrix),
the reaction cross section for spinless particles
is then given by\cite{gh92}
\begin{equation}
\label{eq:csreac}
\sigma^{\alpha \alpha}_\mathrm{r} = \frac{\pi}{k^2}
\sum_\ell \left(2\ell+1\right) \left( 1-\left| S^{\alpha \alpha}_\ell\right|^2
\right) \quad .
\end{equation}
This can be generalized to other outgoing channels $\beta$, not just the elastic one.
The elements of the S-matrix are complex, in general, and related to the
scattering amplitude $f$ of the outgoing wave function
\begin{equation}
f^{\alpha \beta}=\frac{1}{2ik}\sum_\ell (2\ell+1) (S^{\alpha \beta}_\ell
-1)P_\ell(\cos \theta) \quad ,
\end{equation}
which, in turn, is nothing else than the transition amplitude
$t^{\alpha \beta}=-(2\pi \hbar)/m_\beta f^{\alpha \beta}$ connecting entrance channel $\alpha$ and exit channel $\beta$, with $m$ being the reduced mass in the entrance channel and $P_\ell$ a Legendre polynomial.
The imaginary part of the
optical potential gives rise to an absorption term in the solution of
the Schr\"odinger equation, thus removing flux from the considered
channels. Therefore, the matrix element $S^{\alpha \alpha}$ is also
related to the transmission coefficient
\begin{equation}
T_\ell = \left( 1-\left| \mathrm{e}^{2i\delta_\ell}\right|^2 \right)
=1-\left| S^{\alpha \alpha}_\ell\right|^2
\end{equation}
which describes the absorption of the projectile by the nucleus.
Important for practical application is that the phase shifts $\delta_\ell$
can be derived from elastic data.

The optical model is well suited for describing transitions between states of intermediate and heavy nuclei. It has been and is still used also to treat reactions with light nuclei although other methods exist for these.

\subsubsection{Statistical model}
\label{sec:statmod}

The optical model can be used to compute the widths ($\Gamma$, $\hat{\Gamma}$, $\hat{\Gamma}^\mathrm{tot}$) appearing in the BWF for resonances, see (\ref{eq:breit}) above. As mentioned before, the relevant energy windows for astrophysics also include compound nucleus excitation energies with such high NLD that individual resonances cannot be separated because the average resonance width $\langle \Gamma \rangle$ becomes larger than the average level spacing $D=1/\rho$. In fact, this is the case for the majority of reactions included in astrophysical reaction networks. Instead of explicitly dealing with a large number of unknown resonances, one moves to averaged resonance properties.

Starting with the BWF given in (\ref{eq:breit}), the sum of individual resonances can be replaced by an average over an energy interval $\Delta E$ using the mathematical relation\cite{descrau}
\begin{align}
\left< \frac{\Gamma_i^\mu \hat{\Gamma}_i^{Fb} }{(E_\mu-E_{i,\mu}^\mathrm{res})^2+(\hat{\Gamma}^{\rm
tot}_i/2)^2}\right>
&=\frac{1}{\Delta E}\int\frac{\Gamma_i^\mu \hat{\Gamma}_i^{Fb}}{(E_\mu-E_{i,\mu}^\mathrm{res})^2+
(\hat{\Gamma}^{\rm tot}_i/2)^2}\,dE_\mu \nonumber \\
&\approx \frac{2\pi}{\Delta E}\frac{\Gamma_i^\mu \hat{\Gamma}_i^{Fb}}{\hat{\Gamma}^{\rm
tot}_i} \quad.
\label{eq:hfaverage}
\end{align}
Here, the angle brackets denote the average as defined by the above equation. Note that the approximation for narrow resonances, as also used in (\ref{eq:narrowres}) and (\ref{eq:narrowgs}), was applied to arrive at the last line of (\ref{eq:hfaverage}). This is obviously allowed because of the assumption of a large number of narrowly spaced resonances.
With this we rewrite the sum over resonances in the BWF as
\begin{align}
\left<\sum_i g_i \frac{\Gamma_i^\mu \hat{\Gamma}_i^{Fb} }{(E_\mu-E_{i,\mu})^2+
(\hat{\Gamma}^{\rm tot}_i/2)^2} \right>
&=\sum_{J\pi} g_J 2\pi\frac{\Delta n_\mathrm{nuc}(J\pi)}{\Delta
E}\left<\frac{\Gamma_{J\pi}^\mu \hat{\Gamma}_{J\pi}^{Fb}}{\hat{\Gamma}^{\rm
tot}_{J\pi}}\right> \nonumber \\
&=\sum_{J\pi} g_J 2\pi\rho(J,\pi)
\frac{\left<\Gamma_{J\pi}^\mu \right>\left<\hat{\Gamma}_{J\pi}^{Fb}
\right>}{\left<\hat{\Gamma}^{\rm tot}_{J\pi}\right>} \tilde{W}(J,\pi)&\quad .
\end{align}
The number of resonances $\Delta n_\mathrm{nuc}$ within an energy interval $\Delta E$ was replaced by the NLD $\rho$ in the last line.
The averaged widths, the NLD, and the $\tilde{W}$ are energy-dependent, of course.
The width fluctuation coefficients $\tilde{W}$ account for the different averaging in the last line
\begin{equation}
\tilde{W}(E,J,\pi)=\left<\frac{\Gamma^\mu_{J\pi}(E)\hat{\Gamma}^{Fb}_{J\pi}(E)}{\hat{\Gamma}^{\rm tot}_{J\pi}(E)}\right>
\frac{\left<\hat{\Gamma}^{\rm tot}_{J\pi}(E)\right>}{
\left<\Gamma^\mu_{J\pi}(E)\right>\left<\hat{\Gamma}^{Fb}_{J\pi}(E)\right>}
\quad.
\label{eq:wfc}
\end{equation}
In terms of physics, they
describe non-statistical correlations between the widths in the channels
$A+a$ and $F+b$. In practice, they differ from Unity only close to channel openings.\cite{gh92,sar82}

Making use of the relation between transmission coefficients obtained from the solution of the Schr\"odinger equation with an optical potential and the averaged widths
$T^\mu=2\pi \rho \left< \Gamma^\mu \right>,\dots$ the cross section for the statistical model of compound reactions can be written as
\begin{equation}
\label{eq:hf}
\sigma_\mathrm{HF}^{\mu}=\frac{\pi}{k_\mu^2}
\frac{1+\delta_{Aa}}{g_\mu g_a}\sum_{J\pi j\ell j'\ell'}
g_J \frac{T^\mu_{Jlj} \hat{T}^{Fb}_{
Jl'j'}}{\sum_{c lj} \hat{T}^c_{Jlj}}
W^{Aa\rightarrow Fb} \quad.
\end{equation}
The summation in the denominator runs over all channels $c$ leading to the same compound nucleus, not only (but including) $A+a$ and $F+b$. Thus, this sum is equivalent to $\hat{T}^\mathrm{tot}_{J\pi}$. Also, the sums over channel spins $j$ and partial waves $\ell$ are explicitly written to emphasize that the transmission coefficients must include these quantum numbers. Each transmission coefficient includes transitions from states at the compound energy $E_\mathrm{form}=E+E_\mathrm {sep}$ ($E_\mathrm {sep}$ being the separation energy of the projectile in the compound nucleus). While $T^\mu$ only includes those to the state $\mu$ in the target nucleus, the $\hat{T}$ include all transitions allowed by energetics and quantum selection rules.

Comparing (\ref{eq:hf}) with (\ref{eq:breit}) and (\ref{eq:narrowgs}) it is readily seen that the statistical model cross section is an averaged Breit-Wigner cross
section for narrow resonances, when $\tilde{W}=1$. Completely equivalently to the Breit-Wigner case, it can be shown that for the calculation of the stellar rate it is sufficient to replace $T^\mu$ by $\hat{T}^{Aa}$ in (\ref{eq:hf}) and divide the resulting integral by the normalized partition function $G_0^A$, giving\cite{hwfz}
\begin{align}
\langle \sigma v \rangle^*_\mathrm{HF} &= \left(\frac{8\pi}{m_{Aa}}\right)^{1/2} \frac{1}{G_0^A (kT)^{(3/2)}}
\frac{1+\delta_{Aa}}{k_0^2 g_a g_0} \times \nonumber \\
&\times \int \limits_{0}^\infty { \left\{ \sum_{J\pi} g_J
\frac{\hat{T}_{J\pi}^{Aa}(E_0^A)
\hat{T}_{J\pi}^{Fb}(E_0^A)}{\hat{T}^{\rm tot}_{J\pi}(E_0^A)} \right\} E_0^A e^{-\frac{E_0^A}{kT}}\,dE_0^A } \quad.
\label{eq:hfrate}
\end{align}
The total transmission coefficients $\hat{T}^z$ in each channel $z=$$Aa$, $Fb$, \dots include a sum over final states $\xi$ in that channel. Similar to the treatment of the partition functions in (\ref{eq:partfuncint}), the sum over discrete states can be extended by an integration over a level density above the energy $E^\mathrm{x}_{\xi^\mathrm{last}}$ of the last discrete state included,
\begin{align}
\hat{T}^z(E,J,\pi ) & = \left\{ \sum _{\xi} T^{\xi }(E,J,\pi ,
E^{\xi },J^{\xi },\pi ^{\xi }) \right\} + \nonumber \\
 &+\int _{E^\mathrm{x}_{\xi^\mathrm{last}}}^{E_\mathrm{form}} \sum _{J^z\pi^z}T^z(E,J,
\pi ,E^z,J^z,\pi^z)\rho^z (E^z,J^z,\pi^z)\,dE^z\quad .\label{eq:tottrans}
\end{align}
The integration is over the NLD $\rho^z$ in the channel $z$, i.e.\ in the target nucleus $A$ for channel $Aa$, in the final nucleus $F$ for channel $Fb$, and so on. The transmission $T^z$ is the same as $T^\xi$, only that it is a transition to an artifical state with given $(E^z,J^z,\pi^z)$ and weighted by the NLD $\rho^z (E^z,J^z,\pi^z)$. The relative transition energy in channel $z$
is $E^c_\xi=E-S_\mathrm{sep}^z-E^\mathrm{x}_\xi=E_\mathrm{form}-E^\mathrm{x}_\xi$, where $S_\mathrm{sep}^z$ is
the channel separation energy. The reader is advised to consult Fig.~\ref{fig:CNscheme} to get an overview of the included transitions and their relative energies.

Particle transmission coefficients have to obey spin selection rules and thus
\begin{equation}
\label{eq:parttrans}
T^{\xi }(E,J,\pi ,E^{\xi },J^{\xi },\pi ^{\xi })=\sum _{\ell=\left|j-s\right|}
^{j+s}\sum _{j=\left|J^{\xi }-J\right|}^{J^{\xi }+J}T_{j\ell}
(E^c_\xi) \quad.
\end{equation}
Here the angular momentum $\vec{\ell}$ and the channel spin
$\vec{j}=\vec{J}+\vec{J}^{\xi }$ are connected by $\vec{j}=\vec{\ell}+\vec{s}$ including the particle spin $s$. Each $T_{j\ell}$ can be directly obtained from the solution of the (time-independent, radial) Schr\"odinger equation at the energy $E^c_\xi$ with an appropriate optical potential.

The calculation of radiative transmission coefficients proceeds equivalently to (\ref{eq:parttrans}) but electromagnetic selection rules (see, e.g., Appendix B of Ref.~\refcite{ilibook}) have to be obeyed. The parities $\pi$, $\pi^\xi$ and the angular momentum $\ell$ select the type of allowed electromagnetic transition (E1, E2, M1, M2, etc.) and accordingly the appropriate description has to be invoked to calculate the transition strength $T_{j\ell}$. To phenomenologically account for pre-equilibrium particle emission at higher compound excitation energy (see Sec.~\ref{sec:statmodmod}), sometimes the integration in (\ref{eq:tottrans}) is only carried out to a cut-off energy $\min(E_\mathrm{cut},E_\mathrm{form})$, with an appropriately chosen $E_\mathrm{cut}$ (e.g., the energy at which the $\gamma$ transmission exceeds a certain fraction of the total transmission) for the $\gamma$ transmission coefficient appearing in the numerator of (\ref{eq:hfrate}). The total transmission coefficient $\hat{T}^{\rm tot}_{J\pi}$ in the denominator, however, always has to include the full integration up to $E_\mathrm{form}$.

The statistical model of compound reactions was initially developed by Bohr, who conceived the independence hypothesis.\cite{boh36} It states that the projectile forms a compound system
with the target, shares its energy among all of the nucleons, and finally the compound nucleus decays by emitting photons or particles independently of the formation process.
This implicitly requires long reaction timescales as the compound
nucleus has to live long enough to establish complete statistical
equilibrium among the nucleons. Compared to the direct mechanism (Sec.\
\ref{sec:direct}) the timescale is about $5-6$ orders of magnitude longer and
includes many degrees of freedom. In the independence hypothesis, the (laboratory) cross section can
be factorized into two terms
\begin{equation}
\sigma^{Aa \rightarrow Fb} _\mathrm{HF}=\sigma^{Aa}_\mathrm{form}b_\mathrm{dec}
=\sigma^{Aa}_\mathrm{form}\frac{\left< \Gamma^{Fb}
\right>}{\left< \Gamma_\mathrm{tot}\right>}=\sigma^{Aa}_\mathrm{form}\frac{\hat{T}^{Fb}}{\hat{T}^\mathrm{tot}}\quad ,
\label{eq:independence}
\end{equation}
the formation cross section $\sigma_\mathrm{form}^{Aa}$ and a
branching ratio describing the probability for decay to the
observed channel $Fb$.
An early implementation of this was the Weisskopf-Ewing theory.\cite{wei40}
Since then, the Hauser-Feshbach approach has been widely used, which also
incorporates conservation of angular momentum partially lifting the independence assumption but thus being more realistic.\cite{hf52}  Equation (\ref{eq:hf}) is the cross section from the full Hauser-Feshbach formalism. Nevertheless, although too simplified, Eq.~(\ref{eq:independence}) is sometimes useful when estimating the relative feeding of different reaction channels.

Although it might seem tempting to conclude that the cross section
of a reaction proceeding through the compound mechanism should be
smooth because it is formed from the superposition of amplitudes
from a very large number of states with random phases, this is a
wrong assumption.\cite{gh92} It was first shown in Ref.~\refcite{eri60}
that the cross sections can
continue to show large fluctuations. The usual Hauser-Feshbach
equations do not account for these fluctuations. Therefore, a
meaningful comparison to experimental data is only possible after
averaging the data over a sufficiently wide energy range,
comparable to the average resonance widths. When using the
statistical model to compute astrophysical reaction rates (or when
deriving rates experimentally directly) this is taken care of
automatically. However, when using beams with a very narrow energy
spread it should be noted that the results cannot be directly
compared to calculations.\cite{gh92,koe04}

It is worthwhile to point out that the reaction rate is rather ``forgiving'' to deviations around a ``true'' cross section value, provided the deviations go both ways and can cancel within the integration in (\ref{eq:rate}). Therefore, the statistical model approach may even be applicable in the presence of small but isolated resonances as long as their average contribution is correctly accounted for.

This is closely connected to the question of the applicability of the statistical model. This does not necessarily mean the actual validity of the model but rather its suitability to calculate astrophysical reaction rates. As pointed out above, the model cross sections may deviate from the actual ones with large consequences for the rate. The point is that the cross section \textit{averaged} within the relevant energy window is described correctly so that the evaluation of the integral yields the ``true'' value. This is a different and less stringent criterion than one asking for a close reproduction of experimental cross sections. The rule-of-thumb\cite{hwfz,whfz} of 10 levels within the relevant energy window has been found quite sound on average in simple numerical tests.\cite{rtk97} It has to be emphasized that this is an average value. Sometimes even one or a few resonances, if broad enough to almost fill the energy window, may be sufficient. On the other hand, it is always desireable to get as much experimental information as possible for any nucleus but especially for those with low NLD in the relevant energy window.

\begin{figure}
\centerline{\includegraphics[width=0.8\textwidth]{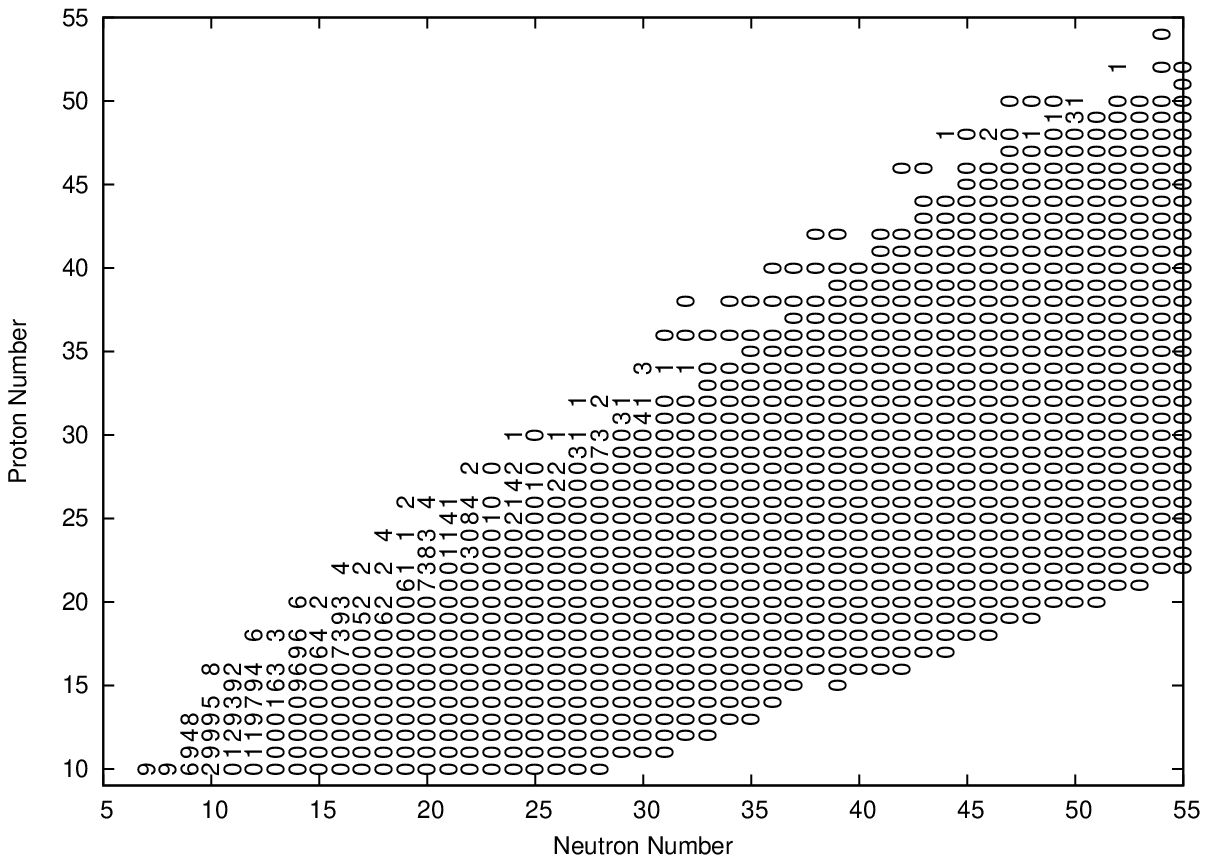}}
\vspace*{8pt}
\caption{Estimated applicability of the statistical model for proton captures. For each nucleus the number labels the temperature region where the statistical model becomes applicable to calculate the astrophysical reaction rate ($T_9$ is in GK): $T_9\leq 0.5$ (0), $0.5<T_9\leq 1.0$ (1), $1.0<T_9\leq 1.5$ (2), $1.5<T_9\leq2.0$ (3), $2.0<T_9\leq 2.5$ (4), $2.5<T_9\leq 3.0$ (5), $3.0<T_9\leq 3.5$ (6), $3.5<T_9\leq 4.0$ (7), $4.0<T_9\leq 5$ (8), $T_9>5$ (9). \label{fig:pgapplic}}
\end{figure}

Using the criterion of 10 levels per energy window, applicability maps for neutron-, proton-, and $\alpha$-induced reactions were shown in Ref.~\refcite{rtk97}, showing the lowest stellar temperature at which the statistical model is applicable to predict the rate. It is important to note that those maps were derived with the approximations (\ref{eq:e0approx}) and (\ref{eq:deltaapprox}) for the Gamow windows and not with the correct energy windows as explained in Sec.~\ref{sec:energies}. Within the accuracy with which the plots can be read, this may not impact capture reactions too much, however, especially the energy windows for (n,$\gamma$) reactions have not been changed.\cite{energywindows} For comparison, Fig.~\ref{fig:pgapplic} shows the minimal temperatures required for (p,$\gamma$) (although on a coarser temperature grid) for a region of the nuclear chart when applying the correct energy windows. It has to be kept in mind, though, that reactions with different exit channels do not exhibit the same relevant energy window, as explained in Sec.~\ref{sec:energies}. The general picture arising is the same as described in Ref.~\refcite{rtk97}. The statistical model can be applied to the majority of neutron capture reactions, with exceptions close to magic neutron numbers or low neutron separation energy, both leading to a low NLD at the compound formation energy $E_\mathrm{form}$. Since the relevant energy windows for charged-particle capture are shifted to higher energy with respect to the ones for (n,$\gamma$), the applicability is even broader. This may also apply to reactions with a charged particle in entrance or exit channel but no general statement can be made because it depends on which width determines the location of the energy window. Finally, endothermic reactions always require higher temperature to have appreciable rates but that does not necessarily mean that the compound nucleus is formed at high excitation energy. Therefore, they may require even higher temperature until the statistical model becomes applicable.

\subsubsection{Modifications of the standard statistical model}
\label{sec:statmodmod}

In the preceding section, the applicability of the statistical model (Hauser-Feshbach model, HFM) due to the required average NLD has been discussed. Modern reaction theory knows a multitude of models, each suited to a particular reaction type and mechanism. These all imply certain approximations. Of course, Nature is continuous, many types of reactions occur simultaneously and we have to choose an approximation suited to describe the dominant effects. Astrophysical energy windows prefer low projectile energies but still may include transitional regions between several reaction mechanisms.

Discussed below are two types of modifications to the rate calculations: 1) Accounting for additional reaction mechanisms, and 2) modifying the HFM itself to provide a smooth transition to low NLD regimes. The inclusion of isospin conservation is a further modification which is discussed in a separate subsection of Sec.~\ref{sec:relevance}.

Direct reactions are known to be relevant at high projectile energies but were also found to significantly contribute at low energies in nuclei with low NLD. They are further discussed in Sec.~\ref{sec:direct}.

Semi-direct\cite{bois72} and multistep reactions\cite{gh92} also occur on a faster timescale than equilibrated compound reactions. They become important at several to several tens of MeV projectile energy and are thus outside the astrophysically relevant energy range. The capture of $\alpha$-particles on intermediate and heavy nuclei may barely reach such energies but only at high plasma temperature. It was shown that semi-direct capture is negligible even for very neutron-rich nuclei at r-process conditions.\cite{chiba08} When assessing the importance of additional mechanisms it is essential to consider the change they bring about for the astrophysical rates. Even if the ground state transitions (as usually explored in theoretical and experimental nuclear physics) may barely reach the required high energy for additional processes, such as semi-direct reactions, this may not affect the reaction rates much. This is because they also include transitions from excited states according to (\ref{eq:effrate}) and (\ref{eq:effweights}), which proceed at lower relative energy and thus remain unaffected by modifications of the cross sections at higher energy.

The HFM assumes that the compound nucleus has sufficient time to distribute the energy gained through the interaction with the projectile among all nucleons of the compound system before it decays. At a high compound formation energy $E_\mathrm{form}$ of several tens of MeV transitions occurring before this energy-equilibration lead to multistep reactions and pre-equilibrium emission of particles and photons.\cite{gh92} The required compound excitation energy is too high to be of astrophysical relevance.

The second type of modifications introduced above is discussed in the following, focussing on two ideas. The first is that subsequent emission of several particles may also occur when the final nucleus is produced at an energy above a further particle emission threshold. Astrophysically this can be relevant in nuclei close to the driplines with low separation energies. It can be treated approximately by applying an iterative application of (\ref{eq:hf}), (\ref{eq:tottrans}), and (\ref{eq:parttrans}) to transitions to states in each system formed in each emission process. This has been used, for example, for calculating neutrino-induced particle emission which is relevant to the construction of neutrino detectors.\cite{kolbe,boydneutrino} In such calculations, the formation transmission coefficients $T^\mu_{J}$ in (\ref{eq:hf}) are replaced by neutrino transmission coefficients $^{\nu_\mathrm{e},\bar{\nu}_\mathrm{e}}T_J^\mu$ describing the population of compound states by neutrino interactions (using, e.g., the random phase approximation). Neutrino reactions select high excitation energies and thus multi-particle emission will also be important at stability. The same approach can be used to determine $\beta$-delayed particle emission when $\beta$-decay produces a daughter nucleus in an excited state above a particle-separation energy. It can also be used to study $\beta$-delayed fission. Both processes are important in the r-process.\cite{ctt,arngorr}

The second idea revolves around the fact that regular HFM calculations assume a compound formation probability independent of the compound NLD at the compound formation energy. Therefore the sum in (\ref{eq:cs}) runs over all $J^\pi$ pairs (a high-spin cutoff is introduced in practical application of the model because spin values far removed from the spins appearing in the initial and final nuclei do not contribute significantly to the transmission coefficients). The availability of compound states and doorway states defines the applicability of the HFM.\cite{rtk97,gh92} Relying on an average over resonances, the HFM is not applicable with a low NLD at compound formation. Furthermore, not all spins and parities will be available with equal probability at each $E_\mathrm{form}$, especially at low NLD. On average the HFM will then overpredict the resonant cross section (unless single resonances dominate) because it will overestimate the compound formation probability. This can be treated by introducing a modification of the formation cross section which includes the compound NLD dependence. The summands of (\ref{eq:cs}) will then be weighted according to the available number of states with the given $J^\pi$. (Formally this is the same as assuming $J^\pi$ dependent potentials for particle channels.) In the most general case this will require a dependence on a $J^\pi$-dependent compound nucleus NLD at $E_\mathrm{form}$. Note that the standard HFM only includes the NLD of the compound nucleus in the photon transmission coefficients (see Sec.~\ref{sec:relevance}) to determine the endpoints of the $\gamma$-transitions.

The assumption that all spins and parities are available can be lifted in several steps. A parity-dependent, global NLD $\rho^z$ was used in the calculation of the transmission coefficients for nuclei without experimentally determined excited states (see Sec.~\ref{sec:relevance}).\cite{darko03,darko05,darko07} Shortly thereafter, the parity-dependence of the compound formation was implemented in a modified HFM in version 4.0w of the code NON-SMOKER$^\mathrm{WEB}$ (see also Sec.~\ref{sec:codes}).\cite{websmoker4p0} A discussion of the implications of a parity-dependent compound formation for astrophysical neutron capture is given in Ref.~\refcite{loens}. Although the parities are not equidistributed up to sizeable excitation energies, the impact on \textit{stellar} rates remains small (in comparison to other uncertainties) because the effective cross section (\ref{eq:effcs}) also includes transitions from excited states which washes out the selectivity on parity. This gives rise to factors of two modifications close to the neutron dripline (but see Sec.~\ref{sec:equilibria}).

Additionally, NON-SMOKER$^\mathrm{WEB}$ offered the option of weighting the HFM cross section by a function depending on the total NLD since version 4.0w.\cite{websmoker4p0} An improved implementation with $J^\pi$ dependent weighting of the summands, thus implicitly accounting for a low NLD at the compound formation energy, is introduced in the SMARAGD code and will be used for a future update of large-scale reaction rate predictions.\cite{smaragd,osaka,cyburt,rauprep} Preliminary results with this modification are shown in Fig.~\ref{fig:avdc} in Sec.~\ref{sec:avdc}.

Obviously, these modifications of the HFM depend on the NLD treatment to obtain the spin- and parity distribution. See the paragraphs on the NLD in Sec.~\ref{sec:relevance} for further details.

\subsection{Sensitivities of HFM rates to nuclear properties and other input}
\label{sec:sensi}

\subsubsection{General considerations}
\label{sec:sensigeneral}

There are different ingredients required to calculate the HFM cross section with the formulas given in the preceding sections.
Which ingredients impact
different parts of the calculation in what manner is discussed below, but how a change in the transmission coefficients
(of certain or all included transitions) affects the resulting cross section can
only be understood with the help of (\ref{eq:cs}). Similar to the determination of the energy-dependence of the cross section (which is crucial in the derivation of the relevant energy windows in Sec.~\ref{sec:energies}), the sensitivity of the cross section and rate to a change in the nuclear properties of the participating nuclei depends on which transmission coefficient (or averaged width) actually affects the cross section while the others cancel from (\ref{eq:cs}). As already pointed out in Sec.~\ref{sec:energies}, the discussion applies to both the BWF and the HFM. Thus, the sensitivities are rather well known in certain parts of the experimental community studying resonances and in the field of nuclear data evaluation. Since they do not seem to be so well known in Nuclear Astrophysics and their implications also have to be interpreted in terms of stellar cross sections and astrophysical reaction rates, it is helpful to outline briefly the main points here.

By comparison to (\ref{eq:hf}) and (\ref{eq:hfrate}) we find that (\ref{eq:cs}) transforms to
\begin{equation}
\mathcal{R}_\mathrm{lab}=\frac{T_{J\pi}^\mu
\hat{T}_{J\pi}^{Fb}}{\hat{T}^{\rm tot}_{J\pi}}\quad \mathrm{and} \quad
\mathcal{R}_\mathrm{rate}=\frac{\hat{T}_{J\pi}^{Aa}
\hat{T}_{J\pi}^{Fb}}{\hat{T}^{\rm tot}_{J\pi}}
\label{eq:fraction}
\end{equation}
for each $J,\pi$-dependent summand. In laboratory cross sections (usually with $\mu=0$) only few $J,\pi$-summands contribute due to the spin selection rules, and so the application of $\mathcal{R}_\mathrm{lab}$ to determine the sensitivities to changes in the widths is straightforward. The situation is different for stellar rates and ratios $\mathcal{R}_\mathrm{rate}$ because transitions from excited target states additionally contribute and more terms in the sum may be relevant, depending on the nucleus and the plasma temperature. It is interesting to note that $\hat{T}^{\rm tot}_{J\pi}$ includes $\hat{T}_{J\pi}^{Aa}$ in both cases. Depending on whether the average entrance transmission $T_{J\pi}^\mu$ significantly defines the size of $\hat{T}_{J\pi}^{Aa}$,
a variation of the entrance transmission will affect the total transmission more
or less. This can lead to a different sensitivity of laboratory cross sections to the entrance
channel than to the transmission in the exit channel, even if everything else
is comparable. The interpretation of experimental results concerning the impact on stellar rates has also to proceed carefully. For example, if a strong dependence is found on
the entrance transmission coefficient and some deficiencies when comparing the model
to experimental data, this does not necessarily mean that this is of
relevance to the astrophysical rate.\footnote{In this discussion, the term ``transmission coefficient'' can also be replaced by ``width'' which may be more familiar to some readers. However, if averaged widths $\langle \hat{\Gamma} \rangle=\hat{T}/(2\pi\rho)$ or strength functions $S_f=\hat{T}/(2\pi)$ can be determined experimentally, it is more useful to obtain the latter because they are directly proportional to the transmission coefficients $\hat{T}$ without an additional dependence on the compound NLD.\cite{koe04}}
Since the astrophysical rate involves $\hat{T}_{J\pi}^{Aa}$ also in the numerator, it may more often cancel with the denominator, even if it would not for $T_{J\pi}^\mu$ in the numerator.
Only for rates with low SEF $\mathcal{R}_\mathrm{lab}\approx \mathcal{R}_\mathrm{rate}$.
This shows again that stellar rates and laboratory cross sections do not have a one-to-one correspondence and additional, mostly theoretical considerations have to be included.

Several special cases can appear in (\ref{eq:fraction}): (i) The larger of the two transmissions in the denominator also dominates the numerator; then the cross section or rate will change similarly to a change of the smaller transmission coefficient in the numerator and be oblivious to any others; (ii) Three channels will be important when neither of the transmission coefficients in the numerator dominates the total transmission; then any change in the two transmissions in the numerator will be translated into an equal change in the cross section but a change in the one determining the denominator will result in an inverse proportional change in the cross section.
In the general case when $\hat{T}^{\rm tot}_{J\pi}$ is not dominated by a single channel the situation is more complicated as any change in a transmission coefficient will not fully affect the cross section at a similar level.

\begin{figure}
  \centerline{\includegraphics[angle=-90,width=0.6\columnwidth]{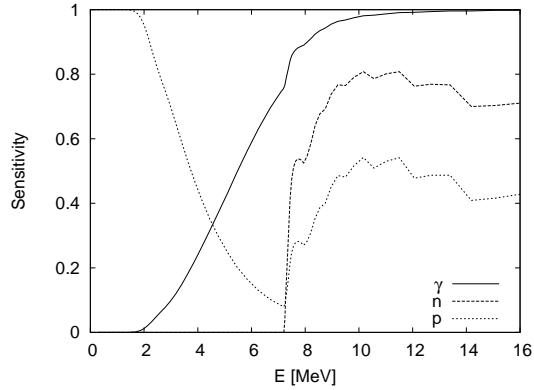}}
  \vspace*{8pt}
  \caption{Sensitivities $\hat{s}$ for $^{96}$Ru(p,$\gamma$)$^{97}$Rh when multiplying the transmission coefficients (averaged widths) by a factor of 2.\label{fig:sensi}}
\end{figure}

A helpful visual aid to estimate the relative importance of the different channels is a sensitivity plot. I define the sensitivity $\hat{s}$ as a measure of a change in the cross section $f_\sigma=\sigma_\mathrm{new}/\sigma_\mathrm{old}$ as the result of a change in a transmission coefficient (or width) by the factor $f_\omega$, with $\hat{s}=0$ when no change occurs and $\hat{s}=1$ when the cross section changes by the same factor as the transmission coefficient (or width):
\begin{equation}
\hat{s}=\left\{ \begin{array}{cl}
\frac{f_\sigma-1}{f_\omega-1} & \mathrm{if}\, f_\sigma >1, f_\omega>1\, \mathrm{ or }\, f_\sigma<1, f_\omega<1\,, \\
\frac{1-f_\sigma}{(f_\omega-1)f_\sigma} & \mathrm{if }\, f_\sigma <1, f_\omega>1\, \mathrm{ or }\, f_\sigma>1, f_\omega<1\,.
\end{array} \right.
\end{equation}
Plotting $\hat{s}$ as a function of the c.m.\ energy yields a plot like the example shown in Fig.\ \ref{fig:sensi} for the reaction $^{96}$Ru(p,$\gamma$)$^{97}$Rh. Its astrophysically relevant energy window is $1.63 \leq E \leq 3.42$ MeV for the typical p-process temperature $T=2.5$ GK.\cite{energywindows} It can clearly be seen in Fig.\ \ref{fig:sensi} that the sensitivities are very different at lower and higher energies. For example, a measurement closely below the neutron threshold would be in a region where $\hat{s}$ is largest for the $\gamma$ transmission coefficient (or width) but smallest for the proton transmission, just the opposite of what is found in the astrophysically relevant energy region. Above the neutron threshold the situation is even more complicated because there is additional sensitivity to the neutron channel (dominating $\hat{T}^\mathrm{tot}$), although not as large as to $\hat{T}^{^{97}\mathrm{Rh},\gamma}$. If any discrepancy between measured and predicted cross sections was found, it would be hard to disentangle the different contributions. In this example, no information on the astrophysically important proton transmission coefficient can be extracted from a measurement at higher energies.

The above example also shows that some of
the standard assumptions usually used in nuclear physics experiments
do not apply. For instance, it is usually
assumed that the $\gamma-$width is always smaller than particle widths
and therefore a capture reaction will always only be sensitive to
the $\gamma-$width. Many, if not most, astrophysical
reactions with charged particles, however, proceed close to or below the Coulomb
barrier. This leads to very narrow charged particle widths and they
may well become smaller than the $\gamma-$widths. Even neutron widths
close to a neutron threshold become smaller than photon widths (which
are relatively independent of energy compared to the particle widths).
Therefore it is always important to closely inspect the widths and
to perform a thorough sensitivity study when investigating the astrophysical
impacts of changes in the HFM inputs.

From the general considerations above it also follows that it is advantageous to use reactions with neutrons in one channel for investigating the sensitivity to the charged-particle optical potential, i.e., using ($\alpha$,n), (n,$\alpha$), (p,n), or (n,p) reactions. Except within a few keV above the neutron threshold, the neutron width will always be much larger than the charged-particle width, even at higher than astrophysical energies. Therefore it will cancel with the denominator in Eq.\ \eqref{eq:fraction} and leave the pure energy dependence of the charged-particle width. On the other hand, the neutron channel may not be open at the energies required to study the astrophysically important charged particle width and extrapolations have to be performed. Furthermore, this shows that it is difficult to obtain information on the neutron potential from reactions. But this also implies that the sensitivity of astrophysical rates to the neutron optical potential is not high.

\subsubsection{Relevance of nuclear input}
\label{sec:relevance}

Nuclear properties and how they affect the calculation of transmission coefficients (or averaged widths) are briefly discussed in the following. This is by no means meant to be an exhaustive listing and discussion but rather some exemplary points are taken to explain the special requirements in the calculation of astrophysical rates and to point out the challenges. Further information on the nuclear input used in statistical model calculations for nuclear physics and nuclear astrophysics can be found, e.g., in Refs.~\refcite{rtk97,rath,hwfz,whfz,ctt,arngorr,sar82,ripl-2,ripl-3,krbecrev,arngorp,talys} and references therein.

While the HFM has been used extensively to study reaction data, its application in Nuclear Astrophysics has a slightly different focus.  Standard nuclear physics investigations use the HFM by including measured or known properties of nuclei. Then reaction cross sections can be reproduced with high accuracy. Stellar cross sections not only require the modification of the HFM shown above but also the inclusion of mostly unknown further input because the reactions proceed at much lower energy and/or involve unstable nuclei. The required properties have to be predicted \textit{globally} for a large number of nuclei and it has to be realized which properties are important in which astrophysical process. This is where the challenge lies and where basic research is necessary, going beyond a mere application of a seasoned model. Although intellectually the most satisfying, modern microscopic models still are not applicable for large-scale predictions and do not produce all the necessary data with sufficient reliability. This calls for clever combinations of microscopic models and parameterizations (which can also use dependences derived from microscopic calculations for a limited range of nuclei). Furthermore, additional information from experiments is required to constrain such models.

\paragraph{Masses:}

Nuclear masses or, rather, mass differences determine separation energies and reaction $Q$-values. In this way, they also determine the range of transition energies to be considered in each reaction channel and through this the relative importance of a channel. Except in the NSE equations (\ref{eq:nse}), masses always appear in mass differences. This poses a potential difficulty when a certain mass region is not fully explored experimentally yet. Care has to be taken to avoid artifical breaks and structures in the mass surface when calculating mass differences from a mix of experimental and theoretical masses. This is usually considered in codes especially written for astrophysical applications. On the other hand, it is expected that mass differences can be measured as well as predicted with higher accuracy than single masses. This seems reassuring for the calculation of astrophysical rates far from stability.

A change in the mass of a nucleus impacts the rates in two different manners. First, the separation energies are altered in the reaction channels including the nucleus. This leads to a change in the transition energies in these channels (see Fig.~\ref{fig:CNscheme}). If the change is large, also more or fewer transitions may become possible. Although charged-particle transitions sensitively depend on the interaction energy, it has to be realized that even in this case the change in $Q$-value \textit{plus} ejectile energy has to be considerable to have a sizeable impact on the rate. The entrance channel is not affected.

Secondly, a change in the $Q$-value changes the relation between forward and reverse rate as shown in (\ref{eq:revrate}) and (\ref{eq:revphoto}). It is highly sensitive to a small change in $Q$-value due to the exponential dependence. This impacts the temperature at which the two rates become comparable and at which equilibria are reached. On the other hand, if forward and reverse rates are different by many orders of magnitude (i.e.\ for large $|Q|$), there may not be much astrophysical impact, nevertheless.

\paragraph{Properties of ground and excited states:}

It is standard procedure to include spin, parity, and excitation energy of low-lying discrete levels when calculating the transmission coefficients. Information about discrete states comes from experiments or from nuclear theory (single particle states and shell model states). Close to stability, a large number of excited states are well known. However, often excited states of nuclides not produced in reactions on stable target nuclei are only partially known experimentally, even if the nuclide itself is stable. Obviously, the situation worsens further off stability. Therefore, nuclear spectroscopy is important to provide the database for reaction modelling. Discrete states are not only important in the HFM but also for resonant reactions treated in an $R$-matrix or BWF approach (unbound states) and even more for the treatment of direct reactions (see Sec.~\ref{sec:direct}) (bound states).

It is very important to have a complete set of excited states because a large number of missing levels at and below a given excitation energy would lead to an incorrect prediction. Therefore it is important to set a cut-off in the excitation energy above which no states are included (even if there are some data) and below which the assumption of a complete level set holds. Above the cut-off, a NLD is employed (see below). It is difficult to define a useful cut-off energy but blindly including all existing data leads to worse results than neglecting too many, provided a reliable NLD description is used. Usually, an educated guess has to be made by comparing the level data to NLD predictions.

\paragraph{Nuclear level density:}

It should be noted that the level density $\rho$ used throughout this paper follows the ``experimental'' definition of number of
observed levels per energy interval $\rho(E)=\sum_{J,\pi} \rho(E,J,\pi)$, where $\rho(E,J,\pi)$ is the observed number of levels with spin $J$ and parity $\pi$ in a small energy interval
around an excitation energy $E$. It is not to be confused with a state density $\hat{\omega}(E)=\sum_{J,\pi} (2J+1) \rho(E,J,\pi)$ appearing in microscopic
nuclear theory, such as the shell model. The terms "level density" and "state density" are used inconsistently in the literature. For the relation of
the two types of density, see, e.g., Refs.~\refcite{haq80,gh92}.

In the regular HFM, the nuclear level density only enters in the calculation of the
transmission coefficients (Eq. \ref{eq:tottrans}) when there are no
or not enough discrete states known and therefore its importance depends on how many low-lying, discrete states were included (see above).

Because of the energetics connected
to particle emission (see Fig. \ref{fig:CNscheme}), usually only a small
fraction of the particle width is due to transitions calculated with
a level density at astrophysically relevant projectile energies as long as discrete excited states are known. This is especially true for neutron-capture reactions due to their low interaction energies. Reactions with charged particles in entrance or exit channel prefer somewhat
higher projectile energies due to the location of the relevant energy window (Sec.~\ref{sec:energies}
and thus may show slightly larger sensitivity to the level density
in the target and final nucleus. (These sensitivities are different from those encountered in usual nuclear reaction studies where projectile energies are much larger.)

On the other hand, there is a larger range of energies available
for $\gamma$-transitions (see Fig.~\ref{fig:CNscheme};
grey areas signify transitions calculated by integration over a level
density, as shown in the second term on the right hand side of Eq.~\ref{eq:tottrans}) and thus the impact of the level density in the compound nucleus will be
largest. In most cases, it is accurate to assume that a variation of the level
density will mostly (or only) affect the $\gamma$-widths.

Generally, the impact of a change in the NLD will be larger for nuclides with fewer known discrete states, i.e.\ far from stability. On the other hand, the $Q$-values of the astrophysically relevant reactions become lower and thus ground state transitions to the final nucleus may again dominate.

It is instructive to remember which excitation energies are the most important ones. For particle transmission coefficients the transitions with highest relative energy are most important, i.e.\ to the ground state and the lowest excitation energies in both target and final nucleus (similar considerations apply to direct reactions). Discrete levels or a NLD have to be known there. For electromagnetic transitions, the relevant excitation energy is around $E_\mathrm{form}/2$ in the compound nucleus (see the discussion of electromagnetic transmission coefficients below and Ref.~\refcite{tomgam} for a detailed explanation).

Due to the low excitation energies involved it is important to a) include the correct ground state spin and parity, and b) to account for a possible parity dependence of the NLD at low excitation. Both is automatically ensured when experimental information is available up to sufficiently high excitation energies. Otherwise, ground state properties and NLDs have to be predicted. This introduces additional uncertainties in rates far from stability.

Until recently, astrophysical rate predictions made use of equally distributed parities $\rho_+ = \rho_- = \rho/2$.\cite{rtk97,konnldcomparison} Modern rate predictions include parity-dependent NLDs in various manners. Either, microscopic NLDs are directly utilized in calculations\cite{konnldcomparison,hilgor06,gorhilkon08} or a parity-dependence is applied to a total NLD.\cite{darko03,darko05,darko07} The advantage of the latter approach is that it can conveniently be applied to total NLDs from different sources and for a large number of nuclei. The total NLD $\rho$ is not changed but the parities are redistributed according to excitation energy. This approach was used to study the impact of a parity-dependence across the nuclear chart. Since capture reactions mainly populate higher lying states, for which an equipartition of parities already is a good assumption, the impact of a parity-dependent level density is small, unless very low $E_\mathrm{form}$ is encountered due to low projectile separation energies and low plasma temperatures.\cite{darko07} According to the discussion above, however, the impact is larger in particle emission channels and for direct reactions.

A modified HFM is introduced in Sec.~\ref{sec:statmodmod} above, which accounts for the relative level distribution at the compound formation energy $E_\mathrm{form}$. Again, the impact of a parity-distribution is small for sufficiently large $E_\mathrm{form}$ and only becomes important far from stability at low separation energies (in the r-process).\cite{loens} However, it remains doubtful whether this actually is of astrophysical relevance as it is not clear whether the involved nuclei can be produced outside of equilibria (see Sec.~\ref{sec:equilibria}) and whether the uncertainty introduced by using the statistical model for nuclei with such low NLD at $E_\mathrm{form}$ is not much larger than the impact of the parity-dependence. On the other hand, a dependence on the total NLD and on the $J$ distribution in the further modified HFM may be more important at a larger range of excitation energies and thus also for nuclei closer to stability (see Sec.~\ref{sec:statmodmod}).

In any case, the uncertainties introduced by the predicted NLDs in rates far from stability are overall much smaller than the uncertainties stemming from other input to the HFM calculations, such as optical potentials and photon transmission coefficients.

\paragraph{Optical potentials:}

Optical potentials are required in the solution of the radial Schr\"odinger equations to determine the particle transmission coefficients as shown in (\ref{eq:parttrans}). Together with the electromagnetic transitions strengths (see below), the unknown optical potentials at low energy give rise to the largest uncertainties in astrophysical reaction rate predictions. There is a combination of two problems involved in the determination of appropriate optical potentials for astrophysical applications: the prediction of optical potentials for highly unstable nuclei and the extension to the astrophysically relevant energies. There is a large amount of reaction and scattering data along the line of stability and many parameterizations (usually of the Saxon-Woods shape), partly mass- and/or energy-dependent, are available. However, most scattering experiments to derive optical potential parameters have been performed at several tens of MeV, far above the astrophysically relevant energy window (see Sec.~\ref{sec:energies}). Even at stability, there are almost no data (not even reaction data) for charged-particle reactions at astrophysical energies.
Measuring low-energy cross sections for charged-particle reactions is especially problematic due to the Coulomb barrier causing the astrophysically relevant cross sections to be tiny. Even more problematic is the standard way to obtain information on optical potentials through elastic scattering experiments. The scattering cross section at low energy becomes indistinguishable from Rutherford scattering.

In the optical model, an imaginary part of the potential appears whenever there is loss of flux from the elastic channel due to any kind of inelastic process. The reaction cross section is sensitive to both real and imaginary part of the optical potential because they determine the relation of real and imaginary part of the nuclear wavefunction and thus the transmission coefficient.\cite{koe04,takiput79,barli74,hui62} Microscopic approaches to derive the optical potential are preferrable over parameterizations when predicting rates far from stability, especially because the available parameterizations were derived at far too high projectile energies. Especially the imaginary part of the potential may vary strongly with energy, due to the energy-dependence in the available reaction channels included in the imaginary part. Nevertheless, any more sophisticated theoretical approach also includes some parameters which have to be constrained by comparison with experiment and so even in this case there may be uncertainties at low energy (see below and the further discussion of how additional reaction mechanisms impact the optical potential in Sec.~\ref{sec:directintro}).

\begin{figure}
\centerline{\includegraphics[angle=-90,width=0.64\columnwidth,clip]{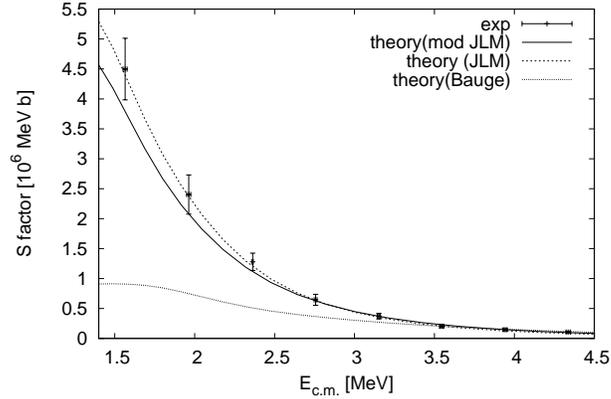}}
\vspace*{8pt}
\caption{\label{fig:pg1}Astrophysical S-factors for (p,$\gamma$) reactions on $^{70}$Ge (data are from Ref.~\protect\refcite{prc76kiss}) compared with theoretical values obtained with
different optical potentials: "standard" potential (JLM),\protect\cite{jlmlow}, modified standard potential (mod JLM),\protect\cite{sefkiss} and the potential from Ref.~\protect\refcite{bauge} (Bauge).}
\end{figure}

\begin{figure}
\centerline{\includegraphics[angle=-90,width=0.64\columnwidth,clip=true]{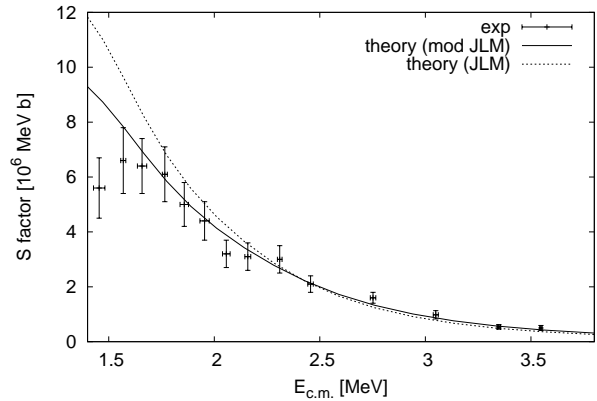}}
\vspace*{8pt}
\caption{\label{fig:pg2}Same as Fig.~\ref{fig:pg1} for $^{74}$Se(p,$\gamma$)}
\end{figure}

An optical potential widely used for interactions of nuclei with neutrons and protons uses the Br\"uckner-Hartree-Fock approximation with Reid's hard core
nucleon-nucleon interaction and adopts a local density approximation.\cite{jlm} A low-energy modification of this potential was provided specially for astrophysical applications.\cite{jlmlow} The latter has become the standard potential in predictions of astrophysical rates and is generally very successful when compared to the scarce experimental data at low energy. As stated in Sec.~\ref{sec:sensigeneral}, astrophysical rates are rather insensitive to the neutron potential. Regarding rates involving charged particles in entrance or exit channel, an important difference to cross sections at higher energies is in the fact that astrophysical cross sections are mostly sensitive to the charged particle widths instead of the $\gamma$- or neutron widths. A series of (p,$\gamma$) and (p,n) reactions was measured close to astrophysical energies recently (see, e.g., Refs.~\refcite{sefkiss,prc76kiss,gy03}, and references therein). The latter reactions are especially useful for testing the proton potential because the neutron width will (almost always) be larger than the proton width at all energies.

\begin{figure}
\centerline{\includegraphics[angle=-90,width=0.64\columnwidth,clip]{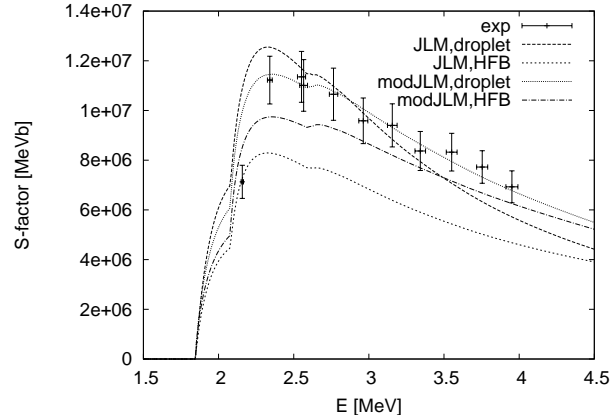}}
\vspace*{8pt}
\caption{\label{fig:dens1}Astrophysical S-factors of $^{85}$Rb(p,n)$^{85}$Sr (exp.\ data from \protect\refcite{sefkiss}) compared with theory using different
proton potentials and nuclear densities. Shown are results with nuclear density from a droplet model\protect\cite{droplet} and from a HFB model with Skyrme interaction (HFB-02)\protect\cite{hfb-02}, applied in the calculation of the ``standard'' potential\protect\cite{jlmlow} (JLM) and a modified version of this potential (modJLM) with increased absorption.}
\end{figure}

\begin{figure}
\centerline{\includegraphics[angle=-90,bb=50 100 554 770,width=0.64\columnwidth,clip=true]{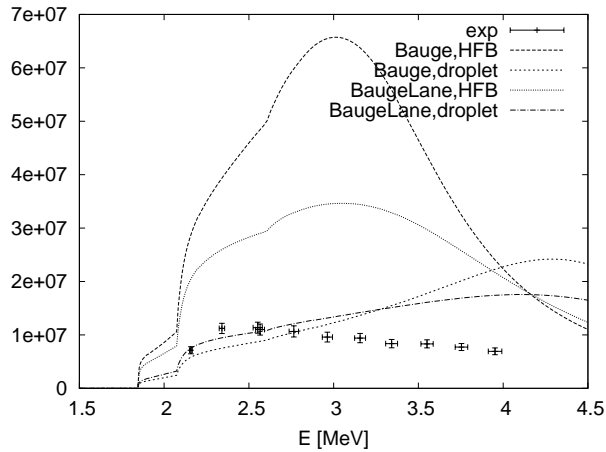}}
\vspace*{8pt}
\caption{\label{fig:dens2}Astrophysical S-factors of $^{85}$Rb(p,n)$^{85}$Sr (exp.\ data from \protect\cite{sefkiss}) compared with theory using different
proton potentials and nuclear densities. Shown are results with nuclear density from a droplet model\protect\cite{droplet} and from a HFB model with Skyrme interaction (HFB-02)\protect\cite{hfb-02}, applied in the calculation of the potential of Ref.~\protect\refcite{bauge98} (Bauge) and Ref.~\protect\refcite{bauge} (BaugeLane).}
\end{figure}

Despite of the overall good agreement when using the standard potential, certain systematic deviations at low energy were found recently (see, e.g., Figs.~\ref{fig:pg1}, \ref{fig:pg2}, more examples are shown in Ref.~\refcite{prc76kiss}). It was found that an increase in the strength of the imaginary part at low energies
considerably improves the reproduction of the data (denoted by ``mod JLM'' in Figs.~\ref{fig:pg1}, \ref{fig:pg2}, \ref{fig:dens1}).\cite{sefkiss,seftom,raujpconf} An increased absorption is permitted within the previous parameterization because the isoscalar and especially the isovector component of the imaginary part is not well constrained at low energies.\cite{jlmlow,bauge98,bauge,isovec} Therefore, the change has to be energy-dependent (i.e.\ acting only at low energy). More low-energy data is required to obtain a better picture.

Figures \ref{fig:pg1}, \ref{fig:pg2} also shows results obtained with another recent, Lane-consistent new parameterization of the JLM potential.\cite{bauge} Although it showed improved performance at higher energy, it yields worse agreement at astrophysically low energy. This is understandable as neither does it include the additional modifications of Ref.~\refcite{jlmlow}, nor can it constrain well the low energy part because it was fitted to data at higher energy.\cite{bauge,isovec} Similar considerations apply to another recent reevaluation of the standard potential.\cite{bauge98}

Required input to the calculation of this type of optical potentials is the nuclear density distribution $\rho^\mathrm{matt}$. Figure \ref{fig:dens1} shows results when employing a droplet model density\cite{droplet} and one from a Hartree-Fock-Bogolyubov model (HFB-2).\cite{hfb-02} For the reactions considered here, the droplet description yields better agreement to the data in both absolute scale and energy dependence of the theoretical S-factor. For comparison, Fig.~\ref{fig:dens2} also shows the results when employing the optical potentials of Refs.~\refcite{bauge98,bauge} with both densities. In the original work, HFB densities were employed.\cite{bauge98,bauge} For further information on how nuclear density distributions affect the astrophysical rates see the subsection on density distributions below, including Figs.~\ref{fig:pgdens1}$-$\ref{fig:ngdens2}.

The standard potential for neutrons and protons seems to work very well compared to the situation encountered when exploring the adequacy of $\alpha$+nucleus optical potentials for astrophysics. Global parameterizations describing scattering, reaction, and decay data have been notoriously hard to find for $\alpha$-particle potentials. Somewhat surprisingly, a mass- and energy-independent potential of Saxon-Woods type has been quite successful and is widely used to evaluate reaction data and also for astrophysical applications.\cite{mcf66} The potential was fitted to scattering data at 26.7 MeV $\alpha$-energy for a wide range of nuclei. However, it became obvious early on that it may be impossible to find a global potential with a predictive power comparable to those for nucleonic projectiles, especially at low energy.\cite{mcf66,iaea-report} The number of optical potential parameters can be reduced by using folding potentials $U_\mathrm{F}$ for the real part,\cite{satlove}
\begin{align}
U_\mathrm{F}(r)=\lambda\int d^3r_\mathrm{A}\int d^3r_\mathrm{a}
&\rho_\mathrm{A}^\mathrm{matt}(r_\mathrm{A})\rho_\mathrm{a}^\mathrm{matt}(r_\mathrm{a}) \times \nonumber \\
&v_\mathrm{eff}\left(E,\rho^\mathrm{matt}=\rho_{A}^\mathrm{matt}+\rho_{a}^\mathrm{matt},\tilde{s}=\left|\vec{r}+\vec{r}_\mathrm{a}-\vec{r}_\mathrm{A}\right|\right).
\end{align}
In this expression $r$ is the separation of the centers of mass of the two interacting
nuclei, $\rho_{a}^\mathrm{matt}$ and $\rho_{A}^\mathrm{matt}$
are their respective nucleon densities and $\lambda$ is an adjustable
strength factor. The factor $\lambda$ may differ slightly from Unity because it accounts for the
effects of antisymmetrization and the Pauli principle.
The effective nucleon-nucleon interaction $v_\mathrm{eff}$ for the
folding procedure is usually of the DDM3Y type.\cite{kob84} Density
distributions have to be taken from experiment or theory (see the section on matter density distributions below).
A global parameterization of the real part with such folding potentials was found based on extensive scattering data.\cite{atz} Unfortunately, there is no simple description for the imaginary part, for which shape and strength have to be energy-dependent. Correlations with the compound NLD, different parametrizations for the energy-dependence of the strength (Fermi-function, Brown-Rho dependence), and an energy-dependence in the relative strength of volume and surface imaginary parts have been suggested.\cite{rauring,rauhirsch,mohrrau,somorjai,deme,kumar,avri10} Again, those extrapolations to low energy are only loosely constrained due to the lack of scattering data. Based on apparently different potentials required for the description of $\alpha$-particles in entrance and exit channel and the fact that a potential fitted to reaction data is able to describe a number of reactions but does not reproduce scattering data, it was suggested that there may be some dependence on nuclear temperature and that absorption and emission potentials may be different from scattering potentials.\cite{kumar,avri89,avri94,avri06} Moreover, at energies close to or below the Coulomb barrier, so-called ``threshold anomalies'' have been observed, a rapid variation of optical potential parameters with energy.\cite{satanom} It has been shown that the dispersion relation connecting real and imaginary part of the optical potential is essential to describe these.\cite{satanom} There is a large literature on different local and global parameterizations of $\alpha$+nucleus potentials, mostly at high energy, underlining the lack of a coherent nuclear physics treatment. A complete review cannot be provided here but see, e.g., Refs.~\refcite{descrau,ripl-2,ripl-3,iaea-report,kumar} and references therein for further details.

\begin{figure}
\centerline{\includegraphics[angle=-90,width=0.64\columnwidth,clip]{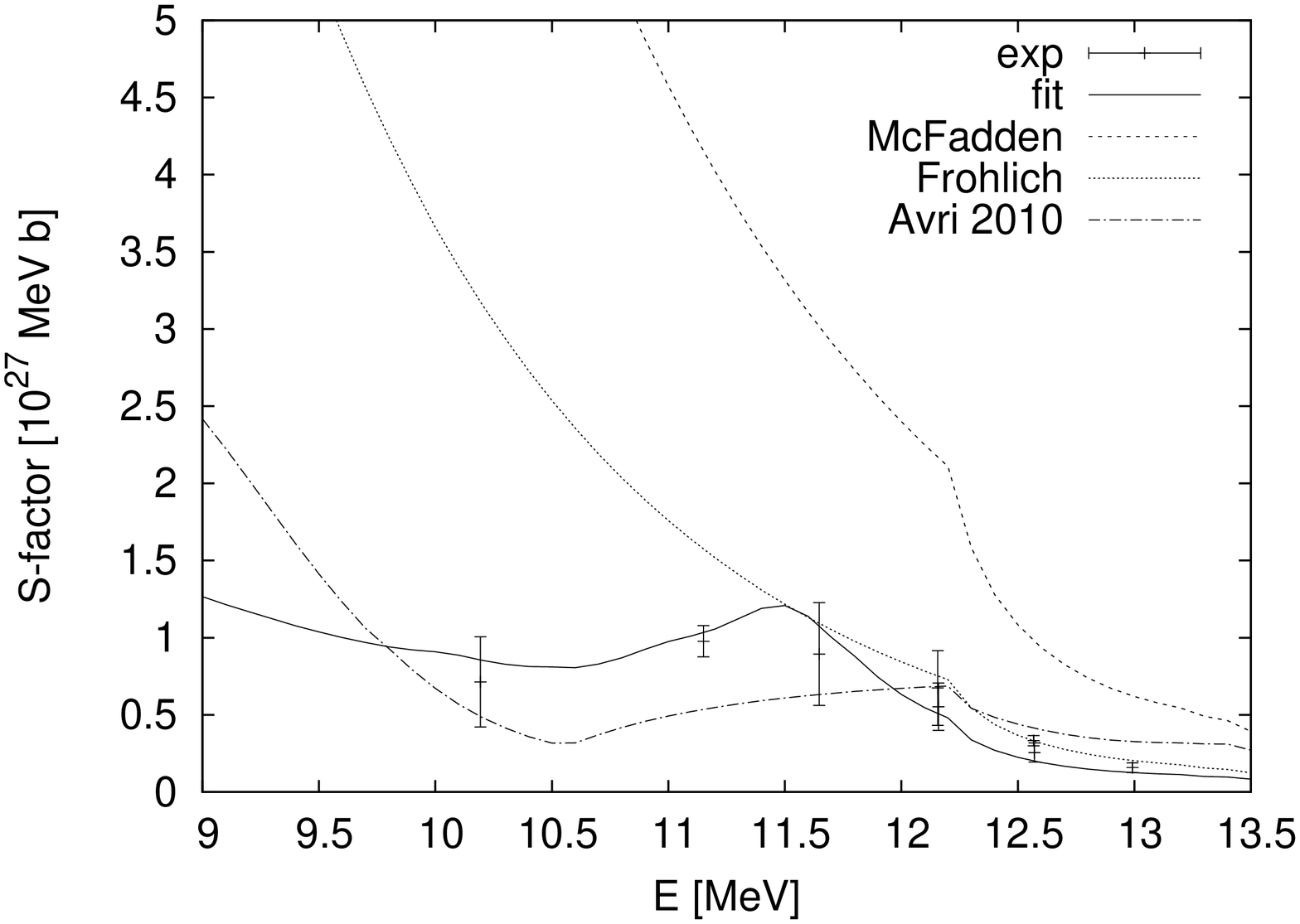}}
\vspace*{8pt}
\caption{\label{fig:sm144a}Astrophysical S-factors of $^{144}$Sm($\alpha$,$\gamma$)$^{148}$Gd (exp.\ data from Ref.~\protect\refcite{somorjai}) compared to HFM calculations with different optical $\alpha$ potentials, from Refs.~\protect\refcite{somorjai} (fit), \protect\refcite{mcf66} (McFadden), \protect\refcite{frohdip} (Froh), \protect\refcite{avri10} (Avri 2010). Transmission coefficients were computed with the routine by Refs.~\protect\refcite{smithroutine,arnouldthiel}.}
\end{figure}

\begin{figure}
\centerline{\includegraphics[angle=-90,width=0.64\columnwidth,clip=true]{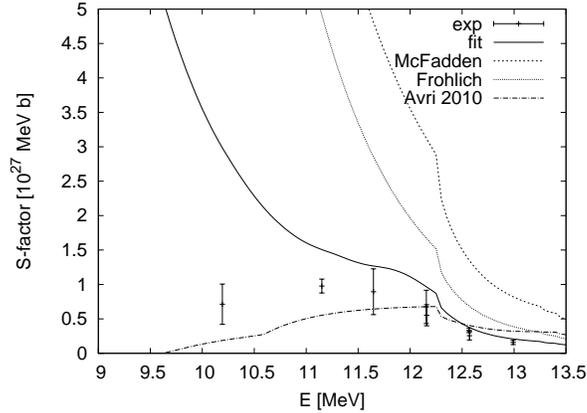}}
\vspace*{8pt}
\caption{\label{fig:sm144b}Same as Fig.~\protect\ref{fig:sm144a} but transmission coefficients were computed with a modern routine for the solution of the Schr\"odinger equation at energies close to the Coulomb barrier (Refs.~\protect\refcite{tedca,smaragd}).}
\end{figure}

For astrophysical applications, early reaction rate predictions made use of an Equivalent Square Well potential.\cite{hwfz,whfz,A72} Later on, the potential of Ref.~\refcite{mcf66} was used. With accumulating evidence that low-energy data deviates from the predicted cross sections more complicated parameterizations were tried, based on various combinations of scattering and reaction data, but these did not lead to a consistent picture. On the other hand, the potential by Ref.~\refcite{frohdip} (see also Refs.~\refcite{koe04,raufroh,raufrogu}) was fitted to simultaneously reproduce low-energy reaction cross sections of $^{143}$Nd(n,$\alpha$)$^{140}$Ce, $^{147}$Sm(n,$\alpha$)$^{144}$Nd, and $^{144}$Sm($\alpha$,$\gamma$)$^{148}$Gd. Although the potential does not describe scattering data (at higher energy) it was found to work surprisingly well for low-energy cross sections for target nuclei across a large mass range $70\leq \tilde{A} \leq 151$.\cite{koe04,galaviz,gy06,tombranch,rappalpha,yalcin,gyeu151} Sharing the same imaginary part with the potential of Ref.~\refcite{mcf66}, it predicts systematically lower cross sections at low energy due to a shallower real part. Although not fully satisfactory yet, this shows that the main change required by a global $\alpha$-potential is to reduce the predicted low-energy cross sections by factors of about $2-3$.

The only and remarkable exception to these factors known so far is in the comparison to experimental data for $^{144}$Sm($\alpha$,$\gamma$)$^{148}$Gd.\cite{somorjai} This reaction is important in the astrophysical context not just generally in p-process calculations\cite{arngorp} but specifically in deriving astronomical timescales from the Nd/Sm abundance ratios measured in meteoritic inclusions or, vice versa, to determine production ratios of these elements in supernovae.\cite{woosmlett,rausmlett} The astrophysical S-factors were measured from $10.2-13.0$ MeV, the astrophysically relevant energy window extends from 9 MeV downwards. Above 11.5 MeV the S-factors obtained with the potential of Ref.~\refcite{mcf66} are too high by a factor of three but the data is well described with the potential of Ref.~\refcite{frohdip} and a potential derived from scattering data at 20 MeV.\cite{mohrrau} However, below 11.5 MeV the energy-dependence changes and requires further modifications of the potentials. At the lowest measured energy of 10.2 MeV, the measured S-factor differs by a factor of 10 from the one predicted with the potential of Ref.~\refcite{mcf66}. In Ref.~\refcite{somorjai} a potential with energy-dependent imaginary part was fitted to describe the data. It predicts S-factors at and below 9 MeV which are orders of magnitude lower than those predicted with the global potentials. Although the original problem was a too low calculated Nd/Sm ratio as compared to what is found in meteorites, such low S-factors yield much too high ratios and additional astrophysical dilution effects have to be invoked in an ad hoc manner.\cite{somorjai} The S-factors are shown in Fig.~\ref{fig:sm144a}, where also new results obtained with the recent potential of Ref.~\refcite{avri10} are included. It should be noted that this reaction may be a special case because it is strongly endothermic. Nevertheless, it was shown that the stellar enhancement factor is lower in the capture direction (see also Sec.~\ref{sec:stellarexp}).\cite{seftom} Since the extrapolation to astrophysical energies strongly hinges on the data points at the lowest energies, an independent remeasurement of this reaction at comparable or lower energies is highly desireable.

\begin{figure}
\centerline{\includegraphics[angle=-90,width=0.64\columnwidth,clip]{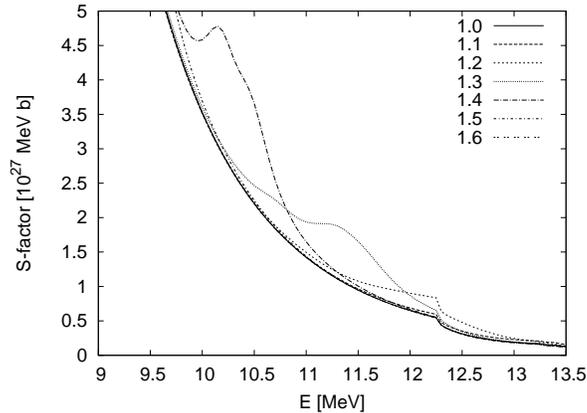}}
\vspace*{8pt}
\caption{\label{fig:sm144rcoul1} Dependence of predicted astrophysical S-factors of $^{144}$Sm($\alpha$,$\gamma$)$^{148}$Gd on the Coulomb radius parameter used with the potential of Ref.~\protect\refcite{somorjai}.}
\end{figure}

\begin{figure}
\centerline{\includegraphics[angle=-90,width=0.64\columnwidth,clip=true]{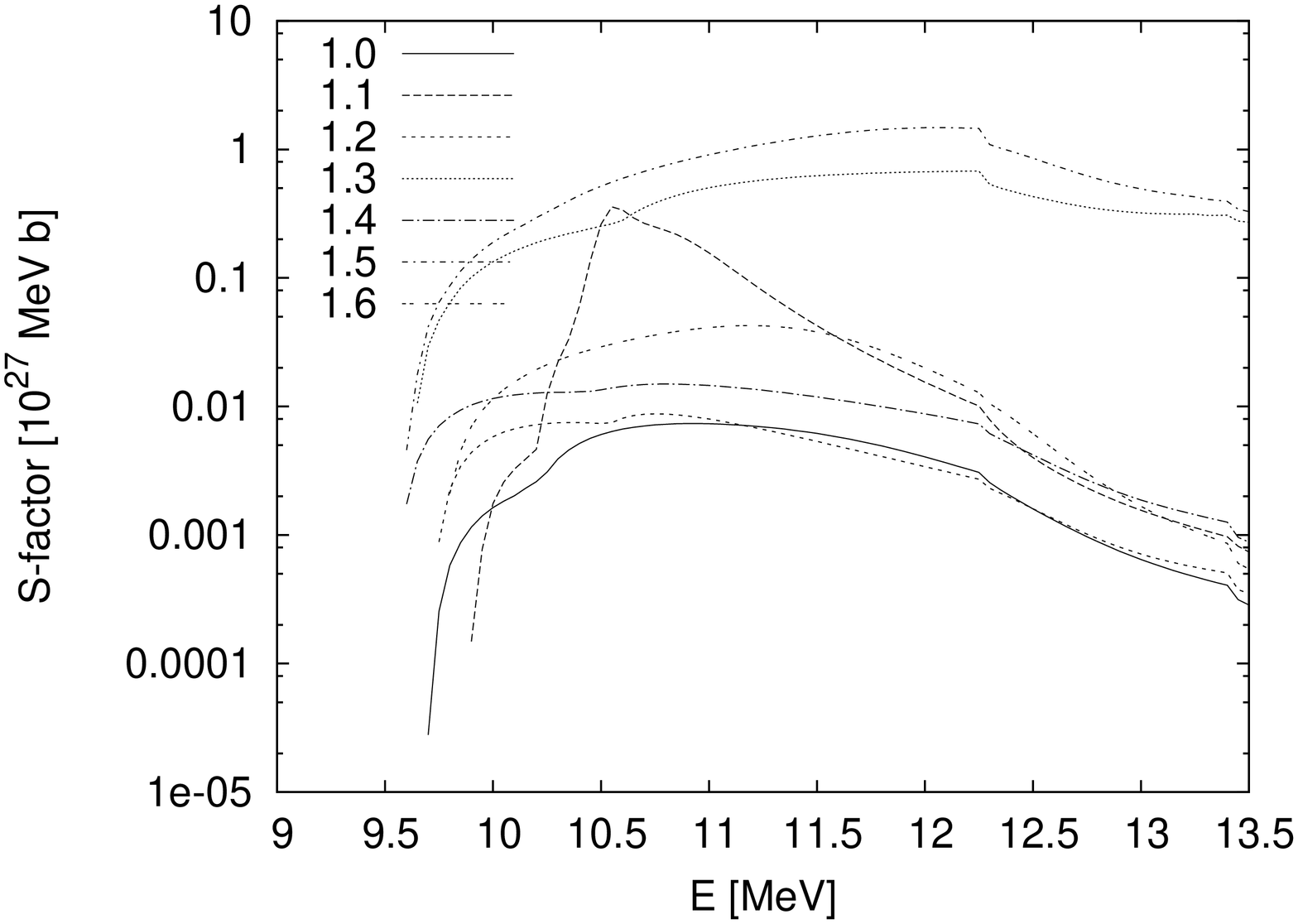}}
\vspace*{8pt}
\caption{\label{fig:sm144rcoul2} Same as Fig.~\protect\ref{fig:sm144rcoul1} but with the potential of Ref.~\protect\refcite{avri10}. Note the logarithmic scale of the plot.}
\end{figure}

Further issues in the determination of S-factors below the Coulomb barrier are illustrated in Figs.~\ref{fig:sm144a}$-$ \ref{fig:sm144rcoul2}. Figure \ref{fig:sm144a} shows the results obtained when using the routine of Ref.~\refcite{smithroutine} (as also used in Refs.~\refcite{ctt,rath,frohdip,somorjai,arnouldthiel,most}) for the solution of the Schr\"odinger equation and the determination of the $\alpha$-particle transmission coefficients. It does not work properly when applied at energies at or below the Coulomb barrier as a comparison to results obtained with a modern routine shows (Fig.~\ref{fig:sm144b}). The discrepancy between predictions and measurement is even enhanced when using an appropriate method. Since the S-factors are very sensitive also to the wavefunction far outside the nuclear center, they show a strong dependence to the shape and width of the effective barrier (determined by the sum of nuclear potential and Coulomb potential). Results obtained with the optical potentials of Refs.~\refcite{somorjai} and \refcite{avri10} with fine-tuned energy-dependences of the real and/or imaginary part also have a strong sensitivity to the Coulomb radius parameter, as shown in Figs.~\ref{fig:sm144rcoul1} and \ref{fig:sm144rcoul2}. This is often overlooked because the sensitivity is much lower at higher energies. Also, when using the potentials of Refs.~\refcite{mcf66,frohdip} there is no sensitivity to the Coulomb radius in the investigated energy range. These additional complications show that any extrapolation to subCoulomb energies has to be performed very carefully and that it is difficult to construct a global potential.

Further experimental data (on scattering and reactions) are especially in demand for improving the optical potentials (see also the discussion of additional reaction mechanisms contributing to the absorptive part of the potential in Sec.~\ref{sec:directintro}). Currently, progress is hampered by the lack of systematic reaction (and scattering) data at astrophysically relevant energies, even at stability.

\paragraph{Electromagnetic transitions:}

\begin{figure}
\centerline{\includegraphics[angle=-90,width=0.5\columnwidth]{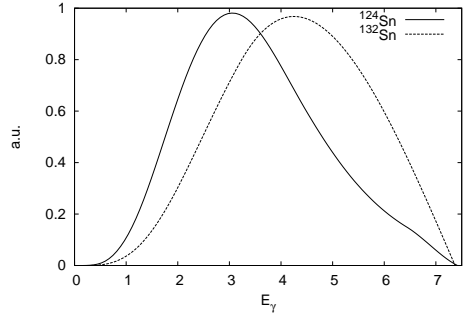}}
\vspace*{8pt}
\caption{Shape of the integrand in (\protect\ref{eq:tottrans}) for photon transmission coefficients in two nuclei. The two functions have been renormalized to yield comparable maxima (Ref.~\protect\refcite{tomgam}).\label{fig:gammapeak}}
\end{figure}

\begin{figure}
\centerline{\includegraphics[angle=-90,width=0.64\columnwidth]{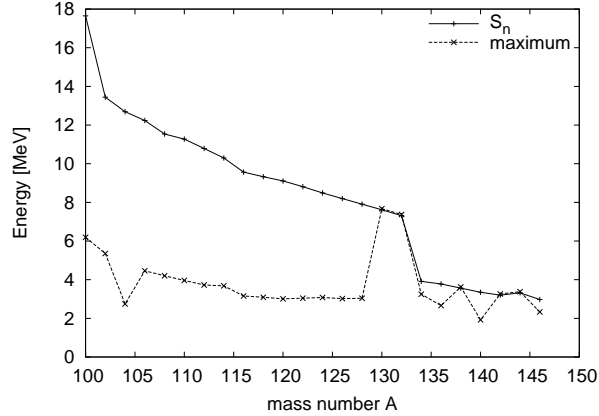}}
\vspace*{8pt}
\caption{Maximally contributing $\gamma$-energies $E_\gamma$ compared to the projectile separation energies $E_\mathrm{sep}$
when capturing $E_\mathrm{proj}=60$ keV neutrons on Sn isotopes. The mass number $A$ is the one of the final (compound) nucleus. See Ref.~\protect\refcite{tomgam} for details. \label{fig:gammaenergies1}}
\end{figure}

\begin{figure}
\centerline{\includegraphics[angle=-90,width=0.64\columnwidth]{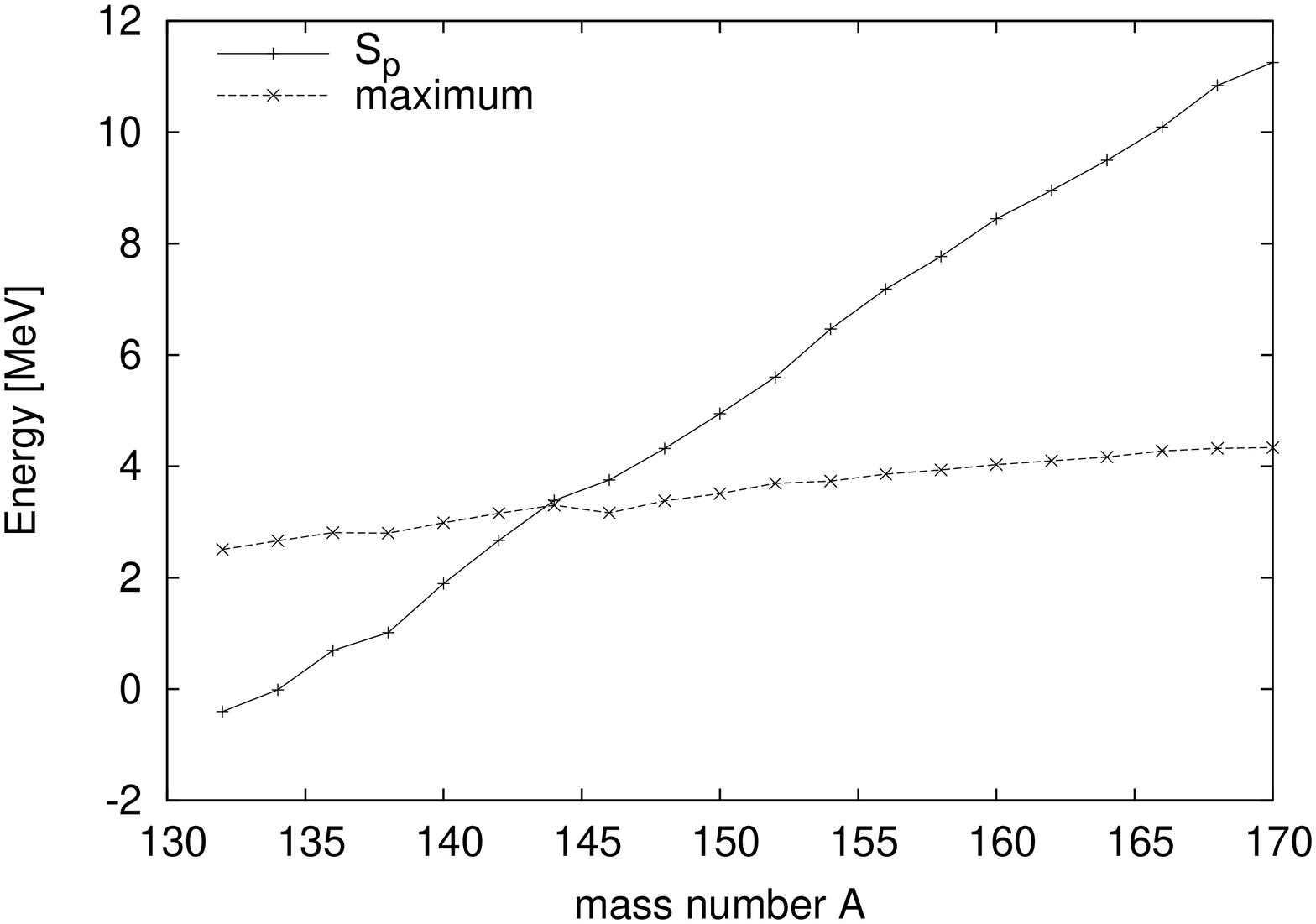}}
\vspace*{8pt}
\caption{Same as Fig.~\protect\ref{fig:gammaenergies1} but for $E_\mathrm{proj}=5$ MeV protons. See Ref.~\protect\refcite{tomgam} for details. \label{fig:gammaenergies2}}
\end{figure}

\begin{figure}
\centerline{\includegraphics[angle=-90,width=0.64\columnwidth]{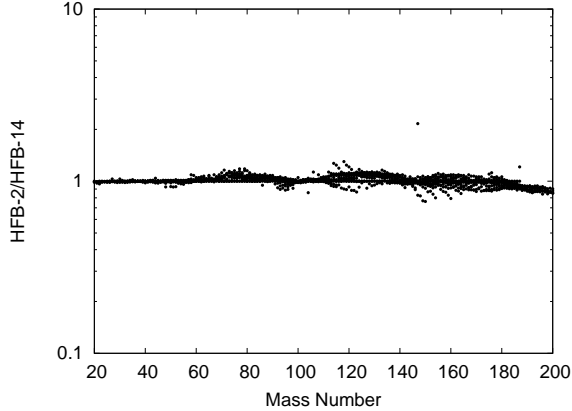}}
\vspace*{8pt}
\caption{\label{fig:pgdens1} Ratios of (p,$\gamma$) rates at $T=3$ GK on stable and proton-rich nuclei, obtained by using the HFB-02 and HFB-14 nuclear density distributions in the calculation of the optical potentials.\protect\cite{hfb-02,hfb-14}}
\end{figure}

\begin{figure}
\centerline{\includegraphics[angle=-90,width=0.64\columnwidth]{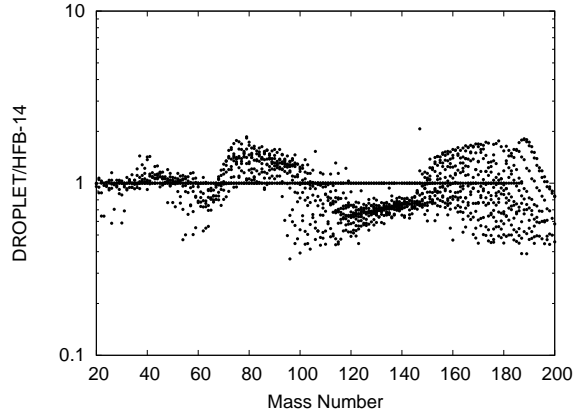}}
\vspace*{8pt}
\caption{\label{fig:pgdens2} Ratios of (p,$\gamma$) rates at $T=3$ GK on stable and proton-rich nuclei, obtained by using the HFB-14 and droplet model nuclear density distributions in the calculation of the optical potentials.\protect\cite{hfb-14,droplet}}
\end{figure}

\begin{figure}
\centerline{\includegraphics[angle=-90,width=0.64\columnwidth]{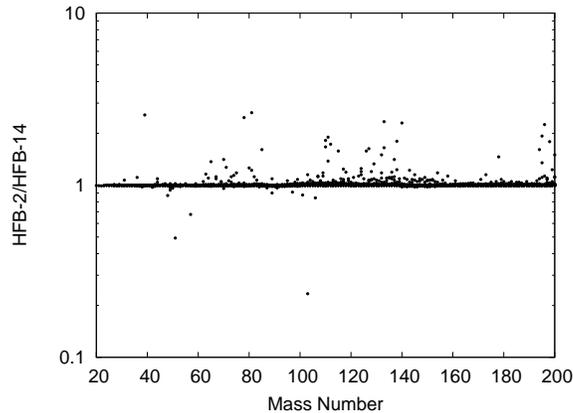}}
\vspace*{8pt}
\caption{\label{fig:ngdens1} Ratios of (n,$\gamma$) rates at $T=1$ GK on stable and neutron-rich nuclei, obtained by using the HFB-02 and HFB-14 nuclear density distributions in the calculation of the optical potentials.\protect\cite{hfb-02,hfb-14}}
\end{figure}

\begin{figure}
\centerline{\includegraphics[angle=-90,width=0.64\columnwidth]{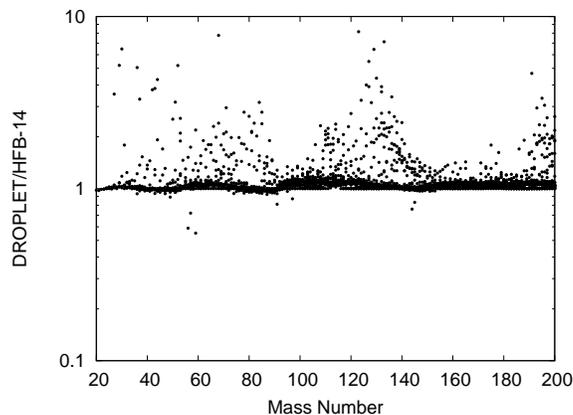}}
\vspace*{8pt}
\caption{\label{fig:ngdens2} Ratios of (n,$\gamma$) rates at $T=1$ GK on stable and neutron-rich nuclei, obtained by using the HFB-14 and droplet model nuclear density distributions in the calculation of the optical potentials.\protect\cite{hfb-14,droplet}}
\end{figure}

First, a few words on the energies of the emitted $\gamma$-rays and their significance for changes in $\gamma$-ray strength
functions (drawn from experiment or from theory) are in order. As shown in Fig.~\ref{fig:CNscheme}, the energies of emitted $\gamma$-rays
are in the range $0\leq E_{\gamma}\leq E_{\mathrm{form}}$. Therefore,
the behavior of $\gamma-$strength functions at low energy have to
be known. Since the strength of the $\gamma$-transition scales with
some power of $E_{\gamma}$, $\gamma$-transitions with higher energies
are favored. On the other hand, the number of available endpoints
of the transitions increases with increasing excitation energy of
the nucleus because the NLD increases rapidly. This competition between transition strength and NLD gives rise to a peak in the
$\gamma$-emission energies as shown in Fig.~\ref{fig:gammapeak}. This peak is fragmented when certain transitions to
discrete excited states are dominating. This is mainly the case far off stability for captures with low $Q$-values, forming a
compound nucleus with low NLD.\cite{tomgam} Figures \ref{fig:gammaenergies1}, \ref{fig:gammaenergies2} show examples for the $\gamma$-energies which maximally contribute to the reaction rate integral.
Interestingly, it can be seen that for astrophysically relevant projectile energies,
the $\gamma$-energies with the strongest impact are between $2-4$
MeV unless the level density is so low that only few transitions are
allowed (usually for highly unstable nuclei).\cite{tomgam} Then the relevant $E_\gamma$ is $E_\gamma=E_\mathrm{proj}+E_\mathrm{sep}-E^\mathrm{x}_\nu$ (in the figures this is the ground state with $E^\mathrm{x}_\nu=0$) instead of the almost constant value below $E_\mathrm{form}$.

The $\gamma$-emission peak defines the range of $\gamma$-energies at which changes in the strength function have largest impact
as well as the excitation energies at which the NLD is most important. This also holds for the reverse reaction (photodisintegration) under stellar
conditions because the additional, linear weight $1-E^\mathrm{x}/(E_\mathrm{form}-E_\gamma)$ from (\ref{eq:effweights})
has a much weaker $E_\gamma$ dependence then both the NLD and the $\gamma$-strength. In consequence, changes in the strength function around this energy
have the largest impact. Testing strength function models outside the energy range will not be relevant to astrophysics.
Unfortunately, such low energies cannot be probed by photodisintegration experiments because they are below the particle separation energies, at least close to stability. Such experiments would allow to study strength functions in the most direct way (but they cannot test the astrophysically relevant rates, either, see Sec.~\ref{sec:stellarexp}). Other types of experiments are complicated by the fact that the observables are generated by a convolution of different nuclear properties (such as the dependence on the NLD or different spin selectivities of transitions) which have to be known and disentangled.

At least the dominant $\gamma$-transitions (E1 and M1) have to be included
in the calculation of the total photon width for astrophysics. Some codes offer the possibility of including higher order transitions.
There are two issues involved: 1) Obtaining, understanding, and modeling photon strength functions (PSF) at stability; 2) predicting strength functions far from stability by using parameterized or microscopic models, predicting the nuclear properties (e.g., deformation) entering the descriptions. Despite of decades of experience in studying nuclear reaction data, the understanding of the
electromagnetic transitions between nuclear states is limited and predictions are subject to considerable uncertainties, even at stability.\cite{krbecrev,ru08mo}

The paradigm in the field of studying electromagnetic transitions is the validity of the reciprocity theorem (\ref{eq:reci_single}) also when applied to photon emission and absorption, and the independence of the PSFs from the nuclear structure of the initial and final states (except for spin and parity, selecting the allowed multipolarity of the radiation).\cite{krbecrev} This is called the Brink hypothesis.\cite{brink} This is also the basis for the construction of the effective cross section (\ref{eq:effcs}), the expression for the reverse rate (\ref{eq:revphoto}), and the introduction of equilibrium abundances (Sec.~\ref{sec:equilibria}). The Brink hypothesis has been studied extensively in experiments and its violation would have grave consequences not only in nuclear reaction theory but also for astrophysical reaction rates and network calculations.

Among the collective modes of nuclei the electric dipole
(E1) excitation has the special property that most of its
strength is concentrated in the isovector giant dipole resonance
(GDR). Macroscopically, this strong resonance is described
as a vibration of the charged (proton) matter in the nucleus
against the neutral matter (neutrons). The transmission coefficient in a nucleus with charge number $\tilde{Z}$, neutron number $\tilde{N}$, and mass number $\tilde{A}=\tilde{Z}+\tilde{N}$
can be parameterized as\cite{krbecrev}
\begin{equation}
\label{GDR}
T_{E1}(E_{\gamma })=\frac{8}{3}\frac{\tilde{N}\tilde{Z}}{\tilde{A}}\frac{e^{2}}{\hbar c}\frac{1+\chi }{m_\mathrm{p}c^{2}}\sum _{i=1}^{2}\frac{i}{3}\frac{\Gamma _{\mathrm{GDR},i}E^{4}_{\gamma }}{{(E_{\gamma }^{2}-E^{2}_{\mathrm{GDR},i})^{2}+\Gamma ^{2}_{\mathrm{GDR},i}E^{2}_{\gamma }}}\quad .
\end{equation}
Here, $m_\mathrm{p}$ is the proton mass, $\chi (=0.2)$ accounts for the
neutron-proton exchange contribution,\cite{LS89}
and the summation over $i$ includes two terms which correspond to
the split of the GDR in statically deformed nuclei, with oscillations along
($i=1$) and perpendicular ($i=2$) to the axis of rotational symmetry. In this deformed case, the two resonance energies
are related to the mean value calculated by the relations\cite{D58}
\begin{align}
E_{\mathrm{GDR},1}+2E_{\mathrm{GDR},2}&=3E_\mathrm{GDR} \quad, \nonumber \\
E_{\mathrm{GDR},2}/E_{\mathrm{GDR},1}&=0.911\eta +0.089 \quad.
\end{align}
The deformation parameter $\eta$ is the ratio of the
diameter along the nuclear symmetry axis to the diameter perpendicular
to it, and is obtained from the experimentally known deformation or from mass
model predictions.
Many microscopic and macroscopic models have been devoted to the calculation
of the GDR energies $E_\mathrm{GDR}$ and widths $\Gamma _\mathrm{GDR}$.

Of special interest here is the low-energy tail of the GDR.
It is a long-standing
question of nuclear physics to specify how much of the
E1 strength is still present at energies far below the GDR
maximum, which also encompasses the astrophysically relevant energy region. Theoretically it has been shown that it is
justified to describe the GDR by a Lorentzian also below
the particle emission thresholds.\cite{krbecrev,ru08mo,dov72}
Various experimental attempts to determine the low-energy
extension of the GDR for heavier nuclei have led to conflicting
results. Neutron-capture experiments often have indicated an
overshoot of the Lorentzian over the observed E1 strength at
the low-energy tail of the GDR.\cite{krbecrev,kop90,uhl,becedp} On the basis of these data
theoretical explanations have been proposed to explain the
differences.\cite{krbecrev,McC81,kad82,becedp} Photon-scattering experiments, however,
are in many cases in good agreement with the Lorentzian
extrapolation.\cite{axel,Las79,ru08mo,grosse,nair} Unfortunately, they do not access the astrophysically relevant energies.
Also other experiments are used to extract PSFs, e.g., $^3$He-induced reactions.\cite{becedp}
Recent investigations have shown that the E1 strength can be described by Lorentzians for a large range of nuclei.\cite{grosse,nair,junghans08}
It was also shown that indirect determinations of PDFs are prone to large uncertainties due to the experimental difficulties and some claims
of enhancement\cite{osloprl,oslo} at low $E_\gamma$ have been premature.\cite{krbecrev,becedp,krbecmoly,dancemoly}

However, in some nuclei extra strength at low energy
with respect to the smooth Lorentzian was found consistently and
denoted as “pygmy dipole resonance” (PDR).\cite{bar73} The PDR has experimentally been studied so far
in spherical nuclides around $\tilde{Z},\tilde{N} = 20, 28$, $\tilde{Z} = 50,\tilde{N} = 82$, and in the doubly magic $^{208}$Pb.\cite{knzil}
Theoretical approaches describe the PDR
as caused by an oscillation of excessive neutrons against the
symmetric proton-neutron system (see, e.g., Refs.~\refcite{rye,litvi07,tson,gorpyg}).
Other oscillation modes were also proposed (e.g., scissor modes) which may also add strength beyond the Lorentzian tail of the GDR. In any case,
extra strength in the low-energy tail only has an astrophysical impact if it is within the relevant energy range defined above. Although a PDR may lead to an
increase by several orders of magnitude in the astrophysical capture rate,\cite{gorpyg,gorkhanpygmy} this depends sensitively on its location and width. Different models give varying predictions of these crucial properties. Depending on the microscopic model used, the pygmy resonance is sometimes predicted at too high or too low an energy as to have any astrophysical consequence.\cite{vrete,ring,vret2,litvi} Further investigation of this issue is required. It is important to note that the uncertainties in predicting the PDR enter additionally to the general uncertainties still present in the prediction of GDR energies and widths. Together with the predictions of optical potentials for the particle transmission coefficients, these are among the largest uncertainties in the determination of astrophysical rates.

PSFs of higher multipole order are even less studied due to their small contributions to cross sections. There are several descriptions available for M1 transitions, starting from PSFs from the simple single particle approach to more sophisticated (but less thoroughly tested) models.\cite{BW52,hwfz,krbecrev,becedp} Unless energy-independent PSFs are employed, a relevant energy window similar to the one for E1 transitions will arise.

\paragraph{Isospin:}

Isospin conservation restricts transitions to certain final states with the same isospin as the initial and compound states, i.e.\ $\Delta I=0$. Isospin conservation is not absolute and cross section measurements of isospin-forbidden reactions give an estimate of the size of the isospin breaking (or isospin mixing). Internal isospin mixing due to the Coulomb interaction and external mixing via the other reaction channels have to be distinguished.
The HFM equation as shown in (\ref{eq:hf}) with the transmission coefficients (\ref{eq:tottrans}) does not account for isospin conservation unless it is included in the transmission coefficients. In other words, complete
isospin mixing is assumed. The calculation of the transmission coefficients can be generalized to explicitly treat the
contributions of the dense background states with isospin $I^<=I^{\rm
g.s.}$ and the isobaric analog states with
$I^>=I^<+1$.\cite{sar82,gri71,har77,har86} In reality, compound nucleus
states do not have unique isospin and for that reason an isospin mixing
parameter $\mu\downarrow$ was introduced which is the
fraction of the width of $I^>$ states leading to $I^<$ transitions.\cite{gri71}
For complete isospin mixing $\mu^\downarrow=1$, for pure $I^<$ states
$\mu^\downarrow=0$. In the case of overlapping resonances for each
involved isospin, $\mu^\downarrow$ is directly related to the level
densities $\rho^<=\rho$ and $\rho^>$, respectively. Isolated resonances can
also be included via their internal spreading width
$\Gamma^{\downarrow}$ and a bridging formula was derived to cover both
regimes.\cite{lan78}

In order to determine the mixing parameter
$\mu^\downarrow=\mu^\downarrow(E)$, experimental information for
excitation energies of $I^>$ levels can be used where available.\cite{rei90}
Experimental values for spreading widths are also
tabulated.\cite{har86,rei90} Inspection of the tables shows that internal mixing dominates and that the associated spreading width
is nearly independent of mass number and excitation energy, facilitating the extrapolation to unstable nuclei.\cite{gh92,har86}
Similarly to the standard treatment for
the $I^<$ states (the regular transmission coefficients as shown above), a NLD can be invoked
above the last experimentally known
$I^>$ level. Since the $I^>$ states in a
nucleus ($\tilde{Z}$,$\tilde{N}$) are part of
multiplet, they can be approximated by the levels (and NLD) of the nucleus ($\tilde{Z}$$-$1,$\tilde{N}$+1), only shifted by a certain
energy $E_{\rm d}$. This displacement energy can be
calculated and it is dominated by the Coulomb displacement
energy $E_{\rm d}=E_{\rm d}^{\rm Coul}+\epsilon$.\cite{aue72}
Thus, the uncertainties involved are the same as in the prediction of the NLD and discrete excited states.

The inclusion of the explicit treatment of isospin has two major effects
on statistical cross section calculations in astrophysics which will be discussed below:\cite{nonsmoker}
the suppression of $\gamma$-widths for reactions involving
self-conjugate nuclei and the suppression of the neutron emission
in proton-induced reactions. Non-statistical effects, i.e.\ the
appearance of isobaric analog resonances, can be included in the treatment
of the mixing parameter $\mu^\downarrow$ but will not be
further discussed here.\cite{lan78}


The isospin selection rule for E1 transitions is $\Delta I=0,1$ with
transitions $0\rightarrow0$ being forbidden.\cite{jon87} An approximate
suppression rule for $\Delta I=0$ transitions in self-conjugate nuclei
can also be derived for M1 transitions.\cite{jon87}

In the case of ($\alpha$,$\gamma$) reactions on targets with $\tilde{N}=\tilde{Z}$, the
cross sections will be heavily suppressed because $I=1$ states cannot be
populated due to isospin conservation. A suppression will also be found
for capture reactions leading into self-conjugate nuclei, although
somewhat less pronounced because $I=1$ states can be populated according
to the isospin coupling coefficients. This cross section suppression can be implemented as a suppression of the photon transmission
coefficient.
Some older reaction rate calculations treated this
suppression of the $\gamma$-widths completely
phenomenologically by dividing by rather arbitrary factors of 5 and 2, for
($\alpha$,$\gamma$) reactions and nucleon capture reactions,
respectively.\cite{whfz,ctt,sch98} This can be improved by explicitly accounting for
population and decay of $T^<$ and $T^>$ states, and considering isospin mixing
by the parameter $\mu^\downarrow$.\cite{RGW99} An astrophysically important reaction with pronounced isospin suppression effect
is the reaction $^{40}$Ca($\alpha$,$\gamma$)$^{44}$Ti which is responsible for the production of the long-lived $^{44}$Ti in supernovae.\cite{RGW99,hoff10}
Decay $\gamma$-emission of $^{44}$Ti is observed in supernova remnants and can be used to test supernova models.\cite{radiobook}

Furthermore, assuming incomplete isospin mixing, the strength of the neutron channel
will be suppressed in comparison to the proton channel in proton-induced reactions.\cite{gri71,sar82,har86} This leads to a smaller
cross section for (p,n) reactions and an increase in the cross section
of (p,$\gamma$) reactions above the neutron threshold, as compared to
calculations neglecting isospin (i.e.\ implicitly assuming complete
isospin mixing with $\mu^\downarrow=1$).
The isospin mixing parameter was varied in the theoretical investigation
of a $^{51}$V(p,$\gamma$)$^{52}$Cr experiment.\cite{zys80} It was
found that complete isospin mixing closely reproduced the
measured cross sections when width fluctuation corrections were
considered. Width fluctuation corrections affect the (p,$\gamma$) cross
sections above as well as below the neutron threshold, whereas
incomplete isospin mixing only reduces the cross sections above the
threshold. Thus, the two corrections can be discriminated. Mainly from
this result it was concluded that -- contrary to width fluctuation
corrections -- isospin can be neglected.
However, a closer investigation of the $I^>$ levels in $^{52}$Cr
(using results from Refs.~\refcite{lan78,rtk97}) shows that
isospin mixing should be rather complete already at the neutron threshold
(since the first $I^>$ state is almost 1 MeV below the
threshold).\cite{nonsmoker} This is also true for lighter targets. For reactions on
heavier nuclei ($\tilde{Z}>30$), however, the neutron and proton threshold, respectively,
will still be in a region of
incomplete isospin mixing and therefore isospin effects should be
detectable there. On the other hand, this effect is not as important
in the calculation of astrophysical reaction rates as the
suppression of the $\gamma$-width
because of the averaging over an energy range in the
calculation of the rate, washing out the cusp effect.

\paragraph{Nuclear matter density distribution:}

The density distribution $\rho^\mathrm{matt}$ of neutrons and protons inside a nucleus
is needed to calculate the optical potentials for some choices of
potentials (see above). There are charge density distributions available from electron scattering
experiments\cite{devries} on stable nuclei but the majority of density distributions for application across the nuclear chart comes from theoretical predictions. Rates are mostly sensitive to optical potentials, and thus densities, at large radii because of the low astrophysical energies.

Density distributions are required in the determination of optical potentials and the potentials used in direct capture calculations. Some optical potentials rely on nuclear density distributions from a certain model which were used when fitting the remaining open parameters to experimental data.\cite{bauge98,bauge}

Although modern microscopic models have considerably improved in predicting nuclear masses and radii, the differences between different models still are large. Because of the sensitivity of the transmission coefficients to the optical potentials, one would expect that small differences in the nuclear density distributions can give rise to large differences in the rates, especially at low plasma temperatures. As an example, Figs.~\ref{fig:pgdens1}, \ref{fig:pgdens2} show comparisons of (p,$\gamma$) rates obtained when using densities from Hartree-Fock-Bogolyubov (HFB) models and from the droplet model in the JLM potentials while leaving all other input unchanged.\cite{hfb-02,hfb-14,droplet} Densities from HFB-2 were included in the Recommended Input Parameter Library for Hauser-Feshbach calculations RIPL-2 (Ref.~\refcite{ripl-2}) and the HFB-14 densities are in its successor RIPL-3 (Ref.~\refcite{ripl-3}). The left panel shows the comparison of rates obtained within the ``family'' of HFB models. The right panel shows a comparison of rates obtained with droplet model and HFB-14 densities. While the ratios stay well within a factor of two even with the seasoned droplet densities, it is interesting to note that they reach Unity when approaching the proton dripline.

Figures \ref{fig:ngdens1}, \ref{fig:ngdens2} are the same as above but for (n,$\gamma$) rates at $T=1$ GK with stable and neutron-rich target nuclei. Here, the largest ratios appear for very neutron-rich nuclei but again recede to Unity when approaching the neutron dripline. Overall, the maximal ratios are higher than for proton capture but stay within a factor of ten. This is partly due to the fact that nuclei further from stability are involved where the disagreement between different models becomes larger. The main reason, however, is the lower relevant temperature because the differences in the rates due to the density distributions decrease with higher $T$, as the average transmission coefficients become less sensitive to the nuclear surface region.

A further example of the impact of nuclear density distributions on calculated reaction $S$-factors is shown in Figs.~\ref{fig:dens1}, \ref{fig:dens2} in the section on optical potentials above.

\paragraph{Deformation:}

Deformations are implicitly present when taking excited states from experiments or theory, or nuclear density distributions from theoretical models. There the problem lies in the fact that most results of microscopic models available for large-scale calculations assume sphericity.

The HFM using transmission coefficients as described in (\ref{eq:tottrans} and (\ref{eq:parttrans}) cannot describe reactions on deformed nuclei because it assumes $\ell$ to be a good quantum number. The coupled-channel model has to be invoked for a rigorous treatment.\cite{gh92,tam65} It is computationally very expensive and thus not suited for large-scale calculations. Furthermore, the effective cross sections (\ref{eq:effcs}) require to include even more transitions than in standard, laboratory nuclear reactions. Fortunately, it has been shown that experimental data can be well described in a spherical HFM using an effective optical potential which is obtained by averaging over nuclear orientation.\cite{gh92} This leads to a spherical potential with larger diffuseness. To compute this modified potential, the explicit inclusion of the nuclear deformation is required and is usually taken from microscopic or macroscopic-microscopic models.

A deformation parameter may also enter the description of the NLD employed in the HFM calculation. But certain NLD descriptions include the deformation implicitly, like the one of Ref.~\refcite{rtk97} where the deformation is contained in the microscopic correction.

The splitting of the GDR in deformed nuclei can also be accounted for phenomenologically by a dependence on a deformation parameter. This has the largest impact on the rates among the possibilities of the inclusion of deformation discussed here.

Finally, fission transmission coefficients (see below) are also very sensitive to deformation parameters. Usually, the deformation is already included in an effective fission barrier, leading to double-humped fission barriers.\cite{gh92}

\paragraph{Width fluctuation corrections:}

The width fluctuation coefficients (WFC) defined in (\ref{eq:wfc}) impact the reaction cross section only closely around channel openings, with a few keV to tens of keV. Contrary to isospin competition cusps (see above), the modify the cross section above and below the channel threshold.\cite{sar82} Generally, they enhance the elastic channel and reduce the other open channels accordingly to obey flux conservation. Above the neutron threshold the behavior may be not so obvious because of the dominance of the neutron channel with respect to other channels. The transfer of strength from the dominant neutron channel to the elastic channel results in a marked reduction in competition with other exit channels and actually increases the strengths in those other channels.

There are different ways to implement the calculation of the WFC, depending on the assumptions taken.\cite{gh92} An explicit form for the $\tilde{W}$, requiring the knowledge of the level width distribution in the compound nucleus, is obtained by assuming that all transmission coefficients are small.\cite{mol76,mol80,grupp77} This approach has been widely used, even when the transmission coefficients were not small.\cite{gh92} An alternative approach is to recognize that the main effect of the correlations is in the elastic channel, with smaller effects on the other channels. This leads to a modification of the transmission coefficients and an additional factor applied in the elastic channel only. This is the more general HRTW method which has been used in most astrophysical applications.\cite{hof75,tep74,hof80,sar82} Another implementation of the HRTW method gives a general formula without restricting assumptions for the additional factor but is complicated to apply.\cite{verba}

Overall, in astrophysical rates the impact of the WFC is small because of the low energies encountered. Even when the relevant energy window covers the neutron threshold, the difference when choosing one or the other description is barely noticeable because of the energy averaging taking place in the rate integration. The WFC are important, however, when trying to compare theoretical and experimental cross sections close to channel openings.

\paragraph{Fission:}

Neutron-induced (and $\beta^-$-delayed) fission of extremely neutron-rich nuclides is important in determining the endpoint of the r-process and the amount of intermediate nuclei produced by the fission process.\cite{ctt,freiburg,arngorr,rauprimordial,igor1,gabrielfiss,igor2,gorfiss} It can be included in the HFM by using an additional exit channel in (\ref{eq:tottrans}) describing the fission process. The fission transmission coefficient is calculated from the penetration probability through a fission barrier. Since most of the astrophysical fission occurs at energies below or close above the barrier, the resulting rates are very sensitive to the height and width of the fission barrier. Barrier predictions from various models show large differences and thus there are considerable uncertainties (reaching several orders of magnitude) in the resulting fission rates. These uncertainties have been explored in Ref.~\refcite{igor2} and recommendations for comparative rate sets were given. The prediction of fission barriers remains a challenge to current microscopic models.

Another issue concerns the fission fragment distribution. Earlier studies of the r-process have used simple fission barriers and assumed symmetric fission.\cite{ctt,freiburg,rauprimordial} Recent years have seen the advent of improved predictions using more sophisticated statistical models and their results are being included in rate calculations.\cite{igor2,kelic05,pan_fragm08,schmidt08,ABLA} As for the barriers, considerable uncertainties may exist for extremely neutron-rich nuclei, though.

On one hand, fission determines how far the r-process can synthesize elements and whether it could reach the region of long-lived superheavy nuclei. On the other hand, the fission rate together with the fragment distribution impacts the abundances of intermediate and heavy r-process nuclei. Intermediate r-process nuclei (including the rare earth peak) have a contribution from fission fragments.\cite{freiburg,gabrielfiss} If the fission timescale is short compared to the r-process timescale, fission cycling can occur whereby the fission fragments capture neutrons and follow an r-process evolution until they fission again.\cite{freiburg,rauprimordial} This exponentially enhances the final r-abundances. The number of possible fission cycles depends on the fission rates and thus is very sensitive to the fission barriers. The final abundance level is less sensitive to the fragment distribution but the distribution will determine the details of which nuclei receive contributions from fission.

\subsubsection{A remark on HFM codes}
\label{sec:codes}

In principle, one would think that any code implementing the HFM should give the same result. This is obviously true when using the same implementation of the HFM \textit{and} the same descriptions of the required input. However, both may differ among different codes. When quoting results it is therefore essential to not only always specify the exact version of the used program but also what selections regarding the properties described above have been made.

There is a variety of codes which have been and are used in data evaluation. These usually focus on higher energies than astrophysically relevant. They also may include further reaction mechanisms which may not be relevant for astrophysical application (see Sec.~\ref{sec:statmodmod}). Finally, they use experimental knowledge (either directly or by renormalizing theoretical results) or local parameterizations of nuclear properties. This way, high accuracy may be achieved locally for one or a few nuclei but no global prediction, essential for astrophysics, can be made.

Astrophysical codes are especially written for global predictions of reaction rates and thus focus on low-energy cross sections. This includes using different internal numerics (see also Figs.~\ref{fig:sm144a}$-$\ref{fig:sm144rcoul2} for issues regarding the calculation of subCoulomb S-factors) but also different choices of the used input. Global treatments by global parameterizations or (semi-)microscopic models are preferred. Experimental information may be used where available to test those global approaches and, of course, to locally improve astrophysical reaction rates. However, a fair comparison of the global predictions of different astrophysical and other codes is only possible when a similar philosophy is used in determining their input (and by using the same experimental input where unavoidable). Most importantly, however, astrophysical codes directly account for the additional transitions required for the calculation of the true astrophysical reaction rate with thermally excited nuclei in the stellar plasma. They properly include the effective cross section defined in (\ref{eq:effcs}) and thus implicitly use the correct weighting factors of transitions from excited states as derived in (\ref{eq:effweights}).

Early large-scale reaction rate and cross section predictions made use of a code developed at Caltech.\cite{hwfz,whfz} The original code\cite{holmesthesis,oap-422} was developed further\cite{kghfp} but no further tables of reaction rates were published, ready for use in astrophysical reaction networks (see also Ref.~\refcite{sar82} for a comparison of these early codes). The work of Refs.~\refcite{hwfz,whfz} was not only important for astrophysical modeling but also in nuclear physics. Prior to these calculations, all hitherto experimentally studied reactions had featureless excitation functions and tests of the statistical theory were hindered by this. The large-scale calculations allowed to identify the cases suited to study competition between different channels.\cite{sar82}

Another influential development is the one of the SMOKER code.\cite{arnouldthiel,fktthesis} It went beyond the previous codes by including a more sophisticated calculation of the transmission coefficients in all channels by explicitly solving the Schr\"odinger equation with optical potentials and including several new global parameterizations of the further inputs. A set of neutron-capture rates for r-process nucleosynthesis was published in Ref.~\refcite{ctt} and complete sets of neutron-induced as well as charged-particle induced reactions calculated with this code became included in the first version of REACLIB, a library of theoretical and experimental cross sections which can directly be used in astrophysical reaction networks.\cite{cyburt,kadonis,reaclib,jina} The REACLIB format, using fits of reaction rates to a function of 7 temperature-dependent terms (see, e.g., Ref.~\refcite{rath}), has become a standard widely used in the astrophysics community.\cite{rath,rhhw02,wooxray}

A series of codes -- some closely related, other only loosely based on SMOKER -- has appeared since then. The code NON-SMOKER extended the functionality by not only updating the input data but also including an improved, global NLD description and isospin effects.\cite{nonsmoker,rtk97,schatz} Although it already included many possibilities for descriptions of nuclear properties to use, often the term ``NON-SMOKER calculations'' is used synonymously for the extended tables in Refs.~\refcite{rath,nonsmokerphoto,rath2} calculated with a chosen input set. In a parallel development, the code MOST also updated the input physics and provided a different selection of treatments of nuclear properties.\cite{most} Tables of reaction rates (but no fits in REACLIB format) were provided online for several versions of the code.\cite{mostonline} The newly written, but still closely related, NON-SMOKER$^\mathrm{WEB}$ code included several changes.\cite{nonsmokerweb} Apart from several updates of included nuclear data, also the internal numerical calculations were modified, the isospin suppression was improved, additional choices for microscopic and macroscopic predictions of nuclear properties were offered. The innovative web interface allows access to the code from anywhere through a simple web browser. Additional switches can be set and different nuclear properties provided in an optional input file which is uploaded to the server running the code. The resulting cross sections and reaction rates are immediately displayed. Certain nuclear properties, e.g.\ optical potentials, cannot only be uploaded as data but also as formulae because the code includes a simple equation parser. NON-SMOKER$^\mathrm{WEB}$ has been used for the astrophysical analysis of a large number of experimental results, especially for p-process nucleosynthesis (see, e.g., Refs.~\refcite{galaviz,gy06,tombranch,rappalpha,yalcin,gyeu151} and references therein). Its development has been frozen at version v5.8.1w but it is still available and used for calculations.

Most recently, the code SMARAGD (Statistical Model for Astrophysical Reactions And Global Direct reactions) continues and extends the development initiated with NON-SMOKER$^\mathrm{WEB}$ of a user-friendly, easily extensible code tailored for astrophysical reaction rates.\cite{smaragd,osaka,cyburt} The code is written completely in FORTRAN90/95 (with exception of the routines handling the web interface and the function parser, which are written in C) and has a modular structure, making changes easy. The latest nuclear data can also swiftly be included through files or web downloads. Internally, the numerics and solvers for the Schr\"odinger equation have been improved to be more accurate at low, subCoulomb energies and to consistently also calculate direct processes as required for astrophysics (see Sec.~\ref{sec:directintro}). Recent developments in parameterized or microscopic predictions of global nuclear properties (masses, NLDs, density distributions, optical potentials, photon strength functions) have been included. It also uses the modified HFM discussed in Sec.~\ref{sec:statmodmod}. Reaction rates are provided in tabular form as well as in the REACLIB format through an automated fit routine. Future versions will include multiple particle emission, follow $\gamma$-cascades explicitly, and allow the calculation of fission rates. Code versions below v1.0s are not public, those below v2.0s do not include direct reactions, yet. Direct reactions are included in a number of ways, as discussed in Sec.~\ref{sec:direct} below. At later stages of the code development it is planned that users may upload modules providing nuclear data, numerical methods to compute required properties, or even altering the functionality of the program. However, before these improvements in versatility are made accessible, a new large-scale calculation will provide a new set of published reaction rates between the driplines, intended to improve on and supersede the NON-SMOKER rates\cite{rath,nonsmokerphoto,rath2} which currently are used by the majority of astrophysical modelers worldwide.\cite{rauprep}

\subsection{Direct reactions}
\label{sec:direct}

\subsubsection{General remarks}
\label{sec:directintro}

In its most general definition, the term ``direct reaction'' includes all processes directly connecting the initial and final states of a nuclear reaction without formation of an intermediate compound system. This includes elastic scattering as described in the optical model, and inelastic scattering which predominantly excites collective states.\cite{sat83,gh92} The latter includes Coulomb excitation which has been found to be important in heavy ion collisions due to the high Coulomb barriers involved.\cite{glen83,gh92,alder} In astrophysically relevant reactions, especially with $\alpha$-particles, energies may also be close to or below the Coulomb barrier and Coulomb excitation may also become important, depending on the structure of the target nucleus.\cite{mohrcoulex}

Here, we focus on direct reactions when some (if it is a stripping reaction) or all (if it is a capture or charge-exchange reaction) nucleons of the projectile are incorporated in the target nucleus. In a pick-up reaction, one or more nucleons from the target nucleus are added to the projectile to form the ejectile, again in a direct manner. Pick-up and stripping reactions are subsumed under the term ``transfer reactions''.
In contrast to the HFM model, direct reactions excite only few degrees of freedom because most of the nucleons included in the system of target nucleus plus projectile remain spectators. A nucleon of the projectile reaches its final state without sharing any energy with any of the other nucleons present and the excess energy is emitted as a discrete photon carrying the energy difference between initial and final state. Direct reactions can be identified experimentally because of their angular dependence of the differential cross sections, being peaked in forward direction. Direct processes are also faster by at least 5 orders of magnitude than compound reactions, with reaction timescales of the order of $10^{-22}$ s. This is comparable to the time the projectile requires to cover a distance of the size of a nucleus. Therefore, direct reactions are important at high projectile energies when compound formation is disfavored.

\begin{figure}
\centerline{\includegraphics[angle=-90,width=0.63\columnwidth,clip]{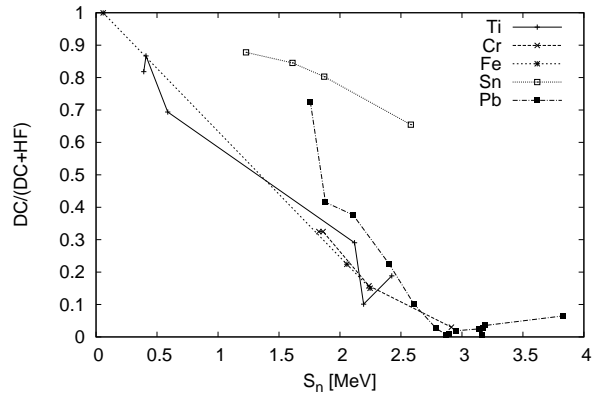}}
\vspace*{8pt}
\caption{\label{fig:dchf}Relation between direct neutron capture and compound capture in the HFM as function of neutron separation energy.\protect\cite{raujpg} The result for Ti, Cr, Fe are taken from Ref.~\protect\refcite{rauenam95}, the ones for Sn from Ref.~\protect\refcite{balogh94}, and the ones for Pb from Ref.~\protect\refcite{microdirect}.}
\end{figure}

Although the notion of direct processes was inspired originally by angular distributions of low-energy reactions, it was assumed for a long time that higher energies are the domain where they are dominating.\cite{glen83,gh92} In resonant reactions at lower energy, it is sometimes necessary to include a non-resonant background (which may show interference with resonances) but experimentally it is often difficult to distinguish between a direct component and contributions from tails of resonances. However, in systems with low NLD, and thus widely spaced resonances, direct reactions become important even at astrophysically low interaction energies because compound formation is suppressed.\cite{ohu89,rai90,her91,ohu91,ohu96,rau92,winkler92,krauss93,krausmann} This even applies to intermediate and heavy nuclei far off stability, e.g., for neutron capture in nuclei with low neutron separation energy.\cite{MaMe83,microdirect} The direct capture cross section can become considerably larger than the compound cross section. Figure \ref{fig:dchf} compares (n,$\gamma$) cross sections at 30 keV  from direct capture and from HFM for a number of isotopes. With decreasing neutron separation energy, the direct component plays an increasingly important role because the compound nucleus is formed at lower excitation energy and thus also at lower NLD. Elements with inherently low NLD, such as Sn, show a larger direct contribution for all isotopes. Also for nuclei at shell closures the NLD is low and the importance of direct reactions enhanced relative to the HFM.\cite{microdirect} Similar considerations may apply to proton-induced reactions on proton-rich nuclei.

Direct processes are not only important to be included in reaction rate predictions. All possible processes (elastic and other direct ones, compound-elastic and compound nucleus reactions) have to be taken into account in the analysis of scattering and reaction data when extracting optical potentials. The elastic scattering cross section $\sigma_\mathrm{s}$ (which is a direct process), the one for reactions $\sigma_\mathrm{r}$, and the total $\sigma_\mathrm{T}$ are related by
\begin{align}
\sigma_\mathrm{s} &=\sigma_\mathrm{s}^\mathrm{opt}+\sigma_\mathrm{s}^\mathrm{comp} \\
\sigma_\mathrm{r} &=\sigma_\mathrm{x}-\sigma_\mathrm{s}^\mathrm{comp} \\
\sigma_\mathrm{T} &=\sigma_\mathrm{s}^\mathrm{opt}+\sigma_\mathrm{x} \quad,
\end{align}
where we distinguished between elastic scattering at the optical potential $\sigma_\mathrm{s}^\mathrm{opt}$ and compound-elastic scattering $\sigma_\mathrm{s}^\mathrm{comp}$. The cross section $\sigma_\mathrm{x}$ includes all inelastic processes, i.e.\ reactions. Often, one reaction mechanism is dominating by far and then $\sigma_\mathrm{x}$ can be identified with the cross section for that mechanism, e.g.\ the compound cross section $\sigma_\mathrm{s}^\mathrm{comp}$ for resonant processes (including the HFM) as discussed in Secs.~\ref{sec:reso} and \ref{sec:statmod}. But this is not always the case as -- depending on the nucleus and the projectile energy -- there may be additional mechanisms contributing to $\sigma_\mathrm{x}$ in some cases.
Therefore, using an optical potential derived from scattering implies that the absorption term is due to some reaction(s) but does not define the reaction mechanism(s). Using such a potential in a pure HFM implicitly assumes that the missing flux from the elastic channel is due to the compound mechanism only. This may not be appropriate when direct processes are non-negligible (this comprises direct reactions as discussed below but also Coulomb excitation at low energy) and will require a modification of the optical potential depending on which mechanism is to be studied. This is also implicitly contained in the idea of the modified HFM briefly discussed in Sec.~\ref{sec:statmodmod}. Earlier calculations of direct neutron capture have made use of a simple hard sphere capture model (see Sec.~\ref{sec:dc}) in a combination of direct and HFM capture, not just to simplify calculations but also because it also allows the assumption that the contribution to direct capture potential absorption due to the tail of distant resonances is already included in the statistical model averaging.\cite{MaMe83}

Astrophysical rates can be calculated from cross sections by applying (\ref{eq:distrirate}), (\ref{eq:rate}), and (\ref{eq:effrate}), regardless of the reaction mechanism. Since each discrete transition appearing in a direct process obeys the reciprocity relation (\ref{eq:reci_single}), a similar effective cross section (\ref{eq:effcs}) can be derived as for the compound case, resulting in the same weighting factors (\ref{eq:effweights}) of transitions from the excited states. Also the same reciprocity relations (\ref{eq:revrate}) and (\ref{eq:revphoto}) apply, provided that thermal population of the states in all participating nuclei is valid.

In the following only a few methods are outlined which have been used to calculate direct reactions for astrophysics. Only the basic equations for the reaction from one initial state to one final state are given but the actual rate equations can straighforwardly be obtained by using the methods described in Secs.~\ref{sec:stellar} and \ref{sec:stellarexp}. It should be noted that here not only light targets are implied but also intermediate and heavy ones (see Fig.~\ref{fig:dchf}), for which microscopic models are not feasible (see Sec.~\ref{sec:mechintro}). Furthermore, low projectile energies are implied as these are required for calculating astrophysical reaction rates. Nevertheless, direct reactions at higher energies can be used to extract certain properties, such as spin assignments and spectroscopic factors (see, e.g., Ref.~\refcite{kate}), of stable and unstable nuclei which are required for the calculation of the cross sections and the rates.

\subsubsection{DWBA}
\label{sec:dwba}

Direct transfer reactions can be treated by solving the time-independent Schr\"odinger equation with optical potentials in the entrance and exit channels. A simple implementation of this is the Distorted Wave Born Approximation (DWBA). The differential cross section for the one-nucleon or cluster
transfer $a + A \rightarrow b+ B$ with $a- x = b$, $A+ x = B$ for $\tilde{A}_a\leq 4$ and $\tilde{A}_x=1$ or 3
is given in zero-range DWBA by\cite{sat83,ohu96}
\begin{equation}
\frac{d\sigma^{\mu \nu}}{d\Omega}=\frac{m_{Aa}m_{Bb}}{\left(2\pi \hbar^2\right)^2} \frac{k_{Bb}g_B}{k_{Aa}g_A}\sum_{\ell sj} C^2 \mathcal{S}_{lj} N_0 \frac{\sigma_{\ell s j}(\theta)}{2s+1}
\end{equation}
with the zero-range normalization constant $N_0$.
The reduced cross section without spin-orbit coupling is given by
\begin{equation}
\sigma_{\ell s j}(\theta)=\sum_m \left| t^m_{\ell s j}\right|^2 \quad,
\end{equation}
with the reduced transition amplitude
\begin{equation}
t^m_{\ell s j}=\frac{1}{2\ell +1}\int \chi_{Bb}^{(-)*} \left(k_{Bb},\frac{\tilde{A}_A}{\tilde{A}_B} r \right) u_{\ell j}(r)
\left[ i^\ell Y_\ell^m(\vec{r})\right]^* \chi_{Aa}^{(+)}\left(k_{Aa},r \right)\, dr \quad.
\end{equation}
As before, the quantities $m_{Aa}$, $m_{Bb}$ and $k_{Aa}$, $k_{Bb}$ are the reduced masses and wave numbers in
the entrance and exit channel, respectively. The orbital
angular momentum quantum number $\ell$, the spin quantum number $s$, and the total
angular momentum quantum number $j$ refer to the nucleon or cluster $x$ bound in
the residual nucleus $B$. The spectroscopic factor and the isospin Clebsch-Gordan
coefficient for the partition $B = A + x$ are given by $C$ and $\mathcal{S}_{\ell j}$, respectively. The
optical wavefunctions in the entrance and exit channels are given by $\chi^{(+)}$ and the
time-reversed solution $\chi^{(-)}$. The bound state wave function is denoted by $ u_{\ell j}$ and the $Y_\ell^m$ are the usual spherical harmonics.
Expressions similar to the above are obtained when the finite range of the interaction potential is taken into account.\cite{sat83}

Important for the successful application is to keep the number of open parameters as small as possible. For this reason, folding potentials (see the paragraphs on optical potentials in Sec.~\ref{sec:relevance}) were used in many astrophysical applications of the model, with $\lambda$ either determined from scattering data or from global dependences.\cite{ohu91,ohu96} This leaves the spectroscopic factor $\mathcal{S}$ which is usually determined by simply comparing the calculated magnitude of the differential cross section to measurements. This is mainly done with (d,p) or (d,n) reactions at energies above the astrophysically relevant ones.\cite{kate} The partial width $\Gamma^\mu$ appearing in (\ref{eq:breit}) can be related to spectroscopic factors for a particle $z$ in a state $\mu$ by\cite{ilibook,ohu96,ili97}
\begin{equation}
\label{eq:spwidth}
\Gamma^{\mu}=C^2\mathcal{S}_\mu^z\Gamma_\mu^z \quad.
\end{equation}
The single particle width $\Gamma_\mu^z$ can be derived from scattering phase shifts and this offers a different experimental access to spectroscopic factors (see Sec.~\ref{sec:avdc} for further methods to determine spectroscopic factors).\cite{ohu96,mohrscat}

Further required input includes, of course, masses or separation energies and nuclear spectroscopic information which have a similar importance as for the HFM, discussed in Sec.~\ref{sec:relevance}.

In the absence of experimental data, the spectroscopic factor can be calculated microscopically from the overlap between initial and final state wave functions, e.g., in the shell model.\cite{sat83,glen83,rauprimordial,kwas94,herndl95,herbro97} However, there is some ambiguity because this overlap is not well defined in different microscopic approaches.\cite{Bertu2010,gh92} It has been shown that the spectroscopic factors for depositing or picking up a single nucleon are related to the occupation factors of the participating quasi-particle states (see also Sec.~\ref{sec:avdc}).\cite{yosh61,cohen}

As in the case of the HFM, many codes have been used for DWBA calculations over the years, especially for the analysis of data at intermediate and high energies. The code TETRA has been written especially for application at astrophysically relevant low energies.\cite{tetra} It has been applied successfully to astrophysically relevant reactions with light and intermediate target nuclei (see, e.g., Refs.~\refcite{ohu89,rai90,ohu91,ohu96,rau92,winkler92,raube9,rauli7dn,raurai,ili96,vancrae98}, and references therein).

Is it necessary to go beyond the DWBA? There are three fundamental assumptions contained in the DWBA treatment:\cite{glen83}
\begin{enumerate}
\item The reaction proceeds directly from initial to final state and all particles except the transferred one(s) remain unaffected spectators.
\item The wave function for the relative motion between the reactands is assumed to be correctly described by the optical potential.
\item The reaction is assumed to be sufficiently weak to be treated in lowest order.
\end{enumerate}
To relax the first two assumptions, the coupled-channel Born approximation (CCBA) was introduced.\cite{sat83,glen83} Under rare circumstances the transfer amplitudes may be large and the third assumption has to be relaxed. This leads to a full coupled-channels treatment for the reaction.\cite{sat83,glen83,gh92}

One has to be aware of the fact that the relevant energies remain low for astrophysical reaction rates, also because transitions from excited states contribute considerably. This is contrary to what one is used to in the investigation of reactions proceeding at several tens of MeV. Due to the low energies involved, the reaction channel is weak (compared to, e.g., elastic scattering) and the third assumption is valid. The usual concern with the second assumption is that the optical potential also has to describe well the wave function even deep in the nuclear interior. The deep region, however, is crucial for reactions at higher energy whereas at the low astrophysical energies most contributions to the overlap integrals stem from regions close to the surface of the nucleus or even from outside of the nuclear radius. As long as these regions are described well by the optical potentials, the DWBA should work. The first assumption implies that either no indirect processes exist or that they can be treated separately (incoherently) as was suggested by the above, separate discussion of compound reactions and other mechanisms. In the HFM it is assumed that interference terms cancel and thus also interference with direct reactions should cancel on average. Interference with isolated resonances can be treated explicitly by adding an interference term, e.g., between the S-factor of the direct reaction $S_\mathrm{direct}$ and the one of a Breit-Wigner resonance $S_\mathrm{BW}$ (see Sec.~\ref{sec:reso})
\begin{equation}
S=S_\mathrm{direct}+S_\mathrm{BW}-2(\sqrt{S_\mathrm{direct}S_\mathrm{BW}})\cos \delta_\mathrm{inter} \quad,
\end{equation}
where
\begin{equation}
\delta_\mathrm{inter}=\arctan \left( \frac{2(E-E_0^\mathrm{res})}{\hat{\Gamma}^\mathrm{tot}} \right)
\end{equation}
is the energy-dependent, relative phase shift.\cite{raurai}

\subsubsection{Direct capture}
\label{sec:dc}

A potential model can also be used to calculate direct capture (DC). Although microscopic models are an alternative for light systems (see Sec.~\ref{sec:mechintro}), a DC potential model has the advantage that it can be applied also to heavier nuclei. The DC cross section for a particular transition is determined
by the overlap of the scattering wave function in the
entrance channel, the bound-state wave function in the exit
channel, and the electromagnetic multipole transition operator.

The DC cross section is then given by\cite{kim87,mohrcap}
\begin{align}
\sigma_{{\rm DC}}^{\mu \nu} & =
\int d\Omega \:\frac{d \sigma_{{\rm DC}}^{\mu \nu}}{d \Omega} \nonumber \\
& =
\int d\Omega \: 2 \left( \frac{e^2 m_{Aa} c^2}{\hbar c} \right)
\left( \frac{ k_\gamma}{k_{Aa}} \right)^3 \frac{1}{g_Ag_a}
\sum_{M_A M_a M_B \sigma}
\left| t_{M_A M_a M_B, \hat{\sigma}} \right| ^2 \quad.
\end{align}

The polarisation
$\hat{\sigma}$ of the electromagnetic radiation can be $\pm 1$. The wave
number in the entrance channel and for the emitted radiation
is given by $k_{Aa}$ and $k_{\gamma}$, respectively.

The multipole expansion of the transition matrices $T_{M_A M_a M_B, \sigma}$
including electric dipole (E1) and quadrupole (E2)
transitions as well as magnetic dipole (M1) transitions is given by
\begin{equation}
t_{M_A M_a M_B, \sigma}=
t_{M_A M_a M_B, \sigma}^{\rm E1}
\, d_{\delta \sigma}^1(\theta) + t_{M_A M_a M_B, \sigma}^{\rm E2}
\, d_{\delta \sigma}^2(\theta) + t_{M_A M_a M_B, \sigma}^{\rm M1}
\, d_{\delta \sigma}^1(\theta) \quad .
\end{equation}
The rotation matrices depend on the angle between
$\vec{k}_{Aa}$ and $\vec{k}_\gamma$ which is denoted by $\theta$,
where $\delta = M_A + M_a - M_B$.

Defining
\begin{align}
C({\rm E}1) & =
i m_{Aa} \left( \frac{\tilde{Z}_a}{\tilde{A}_a} - \frac{\tilde{Z}_A}{\tilde{A}_A} \right) \quad ,\nonumber\\
C({\rm E}2) & =
\frac{k_\gamma}{\sqrt{12}} m_{Aa}^2
\left( \frac{\tilde{Z}_a}{\tilde{A}_a^2} + \frac{\tilde{Z}_A}{\tilde{A}_A^2} \right) \quad ,
\end{align}
we can write for the transition matrices for the electric
dipole (E${\cal L}$ = E1) or quadrupole (E${\cal L}$ = E2) transition
\begin{align}
t_{M_A M_a M_B, \sigma}^{{\rm E}{\cal L}}
 &=
        \sum_{l_a j_a}
        i^{l_a} (l_a\, 0\, S_a\, M_a \mid j_a\, M_a)
        (j_b\, M_B\! -\! M_A\, I_A\, M_A \mid I_B\, M_B) \nonumber\\
&\times
         ({\cal L}\, \delta\, j_b\, M_B\! -\! M_A \mid j_a\, M_a)
        \, C({\rm E}{\cal L})
        \, \hat{l}_a\, \hat{l}_b\, \hat{j}_b \nonumber\\
&\times
        \, (l_b\, 0\, {\cal L}\, 0 \mid l_a\, 0)
        \: {\cal W}({\cal L}\, l_b\, j_a\, S_a;\, l_a\, j_b)
        \: I_{l_b j_b I_B; l_a j_a}^{{\rm E}{\cal L}}~~.\label{eq:kim3}
\end{align}
In the above expressions the
quantum numbers for the channel spin in the entrance channel
and for the transferred angular momentum are denoted by $j_{a}$ and
$l_{a}$, respectively. The quantities $I_{A}$, $I_{B}$ and $S_{a}$
($M_{A}$, $M_{B}$ and $M_{a}$)
are the spins (magnetic quantum numbers) of the target nucleus $A$,
residual nucleus $B$ and projectile $a$, respectively.

For magnetic dipole transitions (M${\cal L}$ = M1) we obtain
\begin{align}
t_{M_A M_a M_B, \sigma}^{{\rm M}{\cal L}}
&=
        \sum_{l_a j_a}
        i^{l_a}\, \sigma \, \Biggl\{ (l_a\, 0\, S_a \, M_a \mid j_a\, M_a)
        (j_b\, M_B\! -\! M_A\, I_A\, M_A \mid I_B\, M_B) \nonumber\\
&\times
        (1\, \delta \, j_b\, M_B\! -\! M_A \mid j_a\, M_a) \nonumber\\
&\times
        \Biggl[ \mu \Biggl( \frac{Z_A}{m_A^2} + \frac{Z_a}{m_a^2}
        \Biggr) \, \hat{l}_b\,
        \hat{j}_b\, \sqrt{\, l_a(l_a+1)}
         \: {\cal G}(1\, l_a\, j_a\, S_a;\, l_a\, j_b) \nonumber \\
&+
        2 \mu_a (-1)^{j_b-j_a}\, \hat{S}_a\, \hat{j}_b\,
        \sqrt{\, S_a(S_a+1)}\:
        {\cal G}(1\, S_a\, j_a\, l_a;\, S_a\, j_b) \Biggr] \nonumber\\
&-
        (l_a\, 0\, S_a\, M_a \mid j_a\, M_a)
        (j_a\, M_a\, I_A\, M_B\! -\! M_a \mid I_B\, M_B) \nonumber\\
&\times
        (I_A\, M_B\! -\! M_a\, 1\, \delta\, \mid I_A\, M_A) \nonumber \\
&\times
        \mu_A \, \delta_{j_a j_b}
        \sqrt{\, (I_A+1)/I_A} \Biggr\}
        \Biggl\{ \frac{\hbar c}{2 m_{\rm p} c^2}
         \Biggr\}\, \delta_{l_a l_b}
        \: \hat{l}_a\: I_{l_b j_b I_B; l_a j_a}^{\rm M1}
        \quad,\label{eq:kim4}
\end{align}
where ${\cal G}$ is the Racah coefficient, the $\mu_i$ are the magnetic moments
and
$m_{\rm p}$ is the mass of the proton.

The overlap integrals in (\ref{eq:kim3}) and (\ref{eq:kim4}) are given as
\begin{equation}
I_{l_b j_b I_B; l_a j_a}^{{\rm E}{\cal L}} = \int dr\: u_{NLJ}(r) \:
{\cal O}^{{\rm E}{\cal L}}(r) \: \chi_{l_a j_a} (r) \label{eq:overlapE}
\end{equation}
for the electric
dipole (E${\cal L}$ = E1) or quadrupole (E${\cal L}$ = E2)
transition, and by
\begin{equation}
I_{l_b j_b I_B; l_a j_a}^{\rm M1} = \int dr\: u_{NLJ}(r) \:
{\cal O}^{\rm M1}(r) \: \chi_{l_a j_a} (r)
\end{equation}
for the magnetic dipole transition (M${\cal L}$ = M1).

The radial part of the bound state wave function
in the exit channel and the scattering wave function
in the entrance channel is given by $u_{NLJ}(r)$
and $\chi_{l_a j_a} (r)$, respectively.
The radial parts of the electromagnetic multipole operators
are\cite{bail67}
\begin{align}
{\cal O}^{\rm M1}(r)&=
\frac{1}{2 \hat{\rho}}\left[
\sin \hat{\rho} + \hat{\rho} \cos \hat{\rho}\right] \quad ,\nonumber\\
{\cal O}^{\rm E1}(r)&=
\frac{3}{\hat{\rho}^3}\left[ (\hat{\rho}^2 - 2) \sin \hat{\rho} +
2 \hat{\rho} \cos \hat{\rho} \right] r \quad ,\nonumber\\
{\cal O}^{\rm E2}(r)&=
\frac{15}{\hat{\rho}^5}\left[ (5 \hat{\rho}^2 - 12)
\sin \hat{\rho} + (12 - \hat{\rho}^2) \hat{\rho} \cos \hat{\rho} \right] r^2 \quad .\label{11}
\end{align}
In the long wave-length approximation -- applicable
as long as $\hat{\rho}=k_\gamma r \ll 1$ -- these
quantities reduce to
\begin{align}
{\cal O}^{\rm M1}(r)&\simeq  1 \quad ,\nonumber\\
{\cal O}^{\rm E1}(r)&\simeq  r \quad ,\nonumber\\
{\cal O}^{\rm E2}(r)&\simeq  r^2 \quad .\label{14}
\end{align}

Usually, only the dominant E1 transitions have to be taken into account. Possible exceptions are captures far from stability with very low reaction $Q$-values because for these cases no final states may be energetically accessible through E1 transitions. However, because the astrophysical reaction rate involves summing over transitions originating from excited states, a larger spin range may be available and E1 (from excited target states) may again dominate.
For E1 transitions, the above expressions reduce to\cite{Chr61}
\begin{align}
\label{DCE1}
\sigma_{\rm E1}^{\mu \nu} & =  \frac{16\pi}{9}
\left(\frac{E_\gamma m_{Aa}}{k_{Aa}\hbar c }\right)^3
\left(\frac{e}{\hbar}\right)^2
\frac{3}{g_a g_A}
\left(\frac{\tilde{Z}_{\rm a}}{\tilde{A}_{\rm a}} -\frac{\tilde{Z}_{\rm A}}{\tilde{A}_{\rm A}}\right)^2
C^{2}\mathcal{S}_{\ell_ \beta J_\beta} \nonumber \\
& \times \sum_{\ell_ \alpha J_\alpha}
\left(2J_\beta+1\right)\left(2J_\alpha+1\right)
\max\left(\ell_\alpha,\ell_\beta\right) \nonumber \\
&\times \left\{
\begin{array}{ccc}1 & \ell_{\beta} & \ell_{\alpha}\\
I & J_{\alpha} & J_{\beta}
\end{array}
\right\}^2
a_{I}^2 \left\vert \int u_{\beta}^{*}(r)\chi_{\alpha}(r) r\,dr \right\vert^2 \quad.
\end{align}
The coefficients $a_{I}^2$ are calculated in LS coupling to
\begin{equation}
a_I^2  =  g_A (2I+1)
\left(2L_{\rm B}+1\right)\left(2S_{\rm B}+1\right)
 \left\{\begin{array}{ccc}I & L_{\rm A} & S_{\rm B} \cr
L_{\rm B} & I_{\rm B} & \ell_{\beta}\end{array}\right\}_6^2
\left\{\begin{array}{ccc}I & L_{A} & S_{B}\\
I_{\rm A} & I_{\rm a} & I_{\rm A}\end{array}\right\}_6^2 \quad.
\end{equation}
In the above expressions, the energy of the emitted photon
is $E_{\gamma}$. The orbital
and total angular momentum quantum numbers of the nuclei
in the entrance and exit channels are $\ell_{\alpha}$,
$J_{\alpha}$, $\ell_{\beta}$ and $J_{\beta}$, respectively.
The spin quantum number, the orbital
and total angular momentum quantum numbers are characterized by
$S$, $L$ and $I$, respectively, with indices $a$, $A$ and $B$
corresponding to the projectile, target and
residual nucleus, respectively. The notation
$\{\cdots\}_6$ stands for the $6j$ symbol. The radial wave functions in the entrance and
exit channels are given by $\chi_{\alpha}$ and $u_{\beta}$,
respectively. The spectroscopic factor and the isospin
Clebsch-Gordan coefficient for the partition $B=A+a$ are given
by $C$ and $\mathcal{S}_{\ell_{\beta} J_{\beta}}$, respectively.

The DC potential model has been successfully applied to many reactions with light target nuclei at astrophysical energies, e.g., for $^7$Be(p,$\gamma$)$^8$B and $^7$Li(n,$\gamma$)$^8$Li.\cite{haribo93} For further calculations see, e.g., Refs.~\refcite{Bertu2010,rauprimordial,ohu91,ohu96,krauss93,vancrae98,mohrcap,ohucluster,mohr3,mohrprop,mohrndst,herndlomeg,herndl99} and references therein. Many calculations were performed with the DC potential model code TEDCA, which is tailored to treat low-energy reactions of astrophysical interest.\cite{tedca} Again, folding potentials (see Sec.~\ref{sec:relevance}) were the key to reduce the number of parameters. Spectroscopic factors were taken from experiment or shell model calculations.
It was also used to extend calculations to intermediate and heavy target nuclei.\cite{raufrogu,ohu96,krausmann,rauenam95,balogh94,herndl95,ili96,beerbalogh,beercoceva,ejnis,forst,guber02,rauni62,beer03,guber03}

Other astrophysical calculations have not made use of the full DC equations but used simplifying assumptions.
Resonant and DC rates on the proton-rich side based on hard sphere scattering wavefunctions in the entrance channel -- with resonance properties, final states, and spectroscopic factors taken from shell model calculations -- were provided for target nuclei with mass number $44\leq \tilde{A} \leq 63$.\cite{fisker} In these calculations, however, the resonant part exceeds the direct part of the cross section by several orders of magnitude.
Astrophysical neutron capture on stable and neutron-rich nuclei was calculated in the hard-sphere model for E1 capture by Refs.~\refcite{MaMe83,raujpg,raurep96,raurejected,holzer97}. In this model, the E1 neutron capture cross section can be written as\cite{MaMe83,lanelynn}
\begin{equation}
\sigma_\mathrm{E1,hard}^{\mu \nu}= \frac{8\pi}{3}\left( \frac{\tilde{Z}}{\tilde{A}}\right)^2 \frac{r_\mathrm{hard}e^2}{c^3 \sqrt{2m_{A+\mathrm{n}}^3 E}} \xi \frac{g_{J^\nu}}{2g_{J_A^\mu}(2l^\nu +1)} \left(\frac{\tilde{Y}+3}{\tilde{Y}+1}\right)^2 (Q_{J^\nu}+E) C^2\mathcal{S}_{J^\nu} \quad, \label{eq:lanelynn}
\end{equation}
where $r_\mathrm{hard}$ is the hard sphere radius and the multiplicity $\xi$ is the number of incident channel spins which can lead to the same final state with spin $J^\nu$. It is $\xi=1$ for $J_A^\mu=0$, or for $J_A^\mu \neq 0$ and $J^\nu=J_A^\mu \pm 3/2$. The value $\xi=2$ applies for $J_A^\mu \neq 0$ and $J^\nu=J_A^\mu \pm 1/2$. The quantity $l^\nu$ is the orbital angular momentum of the final bound state $\nu$ and $\mathcal{S}_{J^\nu}$ is the spectroscopic factor of this state. The dimensionless parameter $\tilde{Y}$ is given by
\begin{equation}
\tilde{Y}=\frac{r_\mathrm{hard}\sqrt{2m_{A+\mathrm{n}}(Q_{J^\nu}+E)}}{\hbar} \quad.
\end{equation}
The advantage of this approach is that no explicit wave functions for scattering and bound states are required. On the other hand, the correct overlap of the wave functions may yield more accurate cross sections, especially for low projectile energies when considerable contributions to (\ref{eq:overlapE}) are coming from far outside the nuclear radius. Ref.~\refcite{MaMe83} showed the importance of direct neutron capture in the r-process based on the hard-sphere model.

For the validity of the low-order potential model approach for DC while neglecting higher-order processes, similar arguments can be made as were presented for the DWBA towards the end of Sec.~\ref{sec:dwba}.

\subsubsection{Sensitivity of DWBA and DC in astrophysical calculations}
\label{sec:directsensi}

For astrophysics, a large number of rates for highly unstable nuclei have to predicted. Similar to the uncertainties of the HFM discussed in Sec.~\ref{sec:relevance} the DWBA and DC predictions are sensitive to certain inputs, such as masses ($Q$-values), excited and bound state properties, spectroscopic factors, optical and bound state potentials. When folding potentials are used, a (weak) dependence on nuclear matter density distributions appears additionally. The discussion of these quantities in Sec.~\ref{sec:relevance} also applies here. However, some of the quantities appearing in the HFM are sums over individual transitions and averaged quantities whereas individual transitions are determining the direct reaction cross sections. Thus, direct reactions are more sensitive to nuclear properties impacting those individual transitions, including spins, parity, energy of bound and excited states, $Q$-values, and spectroscopic factors. Due to the angular momentum barrier and the low projectile energies, low partial waves contribute most to the E1 cross section, i.e.\ s-waves when initial and final states have differing parity, p-waves when they have the same parity. Assuming equal spins and parities, low-lying states are contributing more than higher ones. The strong sensitivity to individual transitions, however, may be reduced in the astrophysical rate, when using the weighted sums (\ref{eq:rate}), (\ref{eq:effrate}), (\ref{eq:effweights}) over transitions from excited target states to available final states.

\begin{figure}
\centerline{\includegraphics[angle=-90,width=0.63\columnwidth,clip]{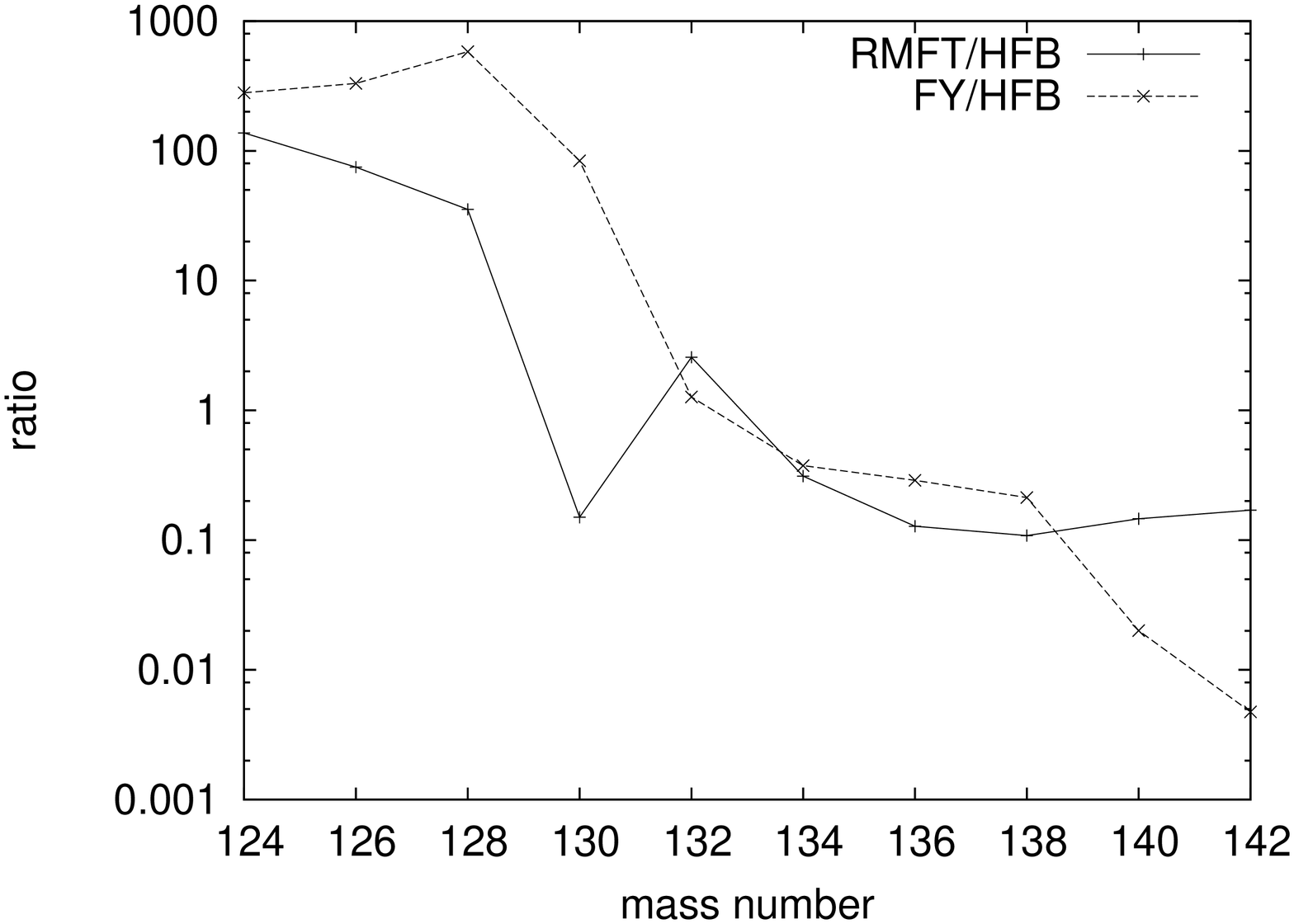}}
\vspace*{8pt}
\caption{\label{fig:dccompsn}Comparison of direct neutron capture cross sections of Sn isotopes calculated with input taken from a Hartree-Fock-Bogolyubov model (HFB, Refs.~\protect\refcite{doba1,doba2}), a Relativistic Mean Field model (RMFT, Refs.~\protect\refcite{sharma1,sharma2}), and a semi-microscopic model using folded Yukawa potentials (FY, Refs.~\protect\refcite{moll1,moll2,moll3}). For details, see Ref.~\protect\refcite{microdirect}.}
\end{figure}

\begin{figure}
\centerline{\includegraphics[angle=-90,width=0.63\columnwidth,clip]{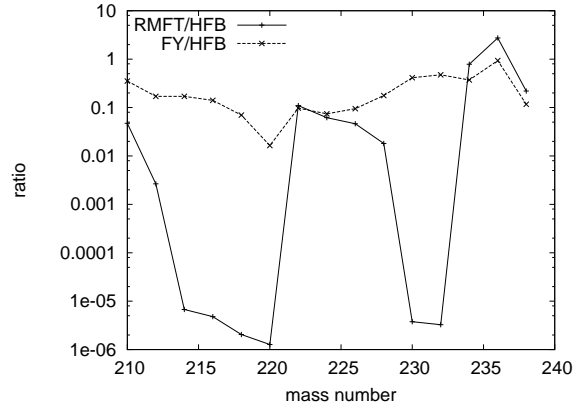}}
\vspace*{8pt}
\caption{\label{fig:dccomppb}Same as Fig.~\protect\ref{fig:dccompsn} but for Pb isotopes.}
\end{figure}

For neutron-rich Sn and Pb isotopes, predictions for 30 keV neutron capture employing input from different microscopic or semi-microscopic approaches were compared in detail.\cite{microdirect} The DC calculation was tested in comparison to experimental data for $^{208}$Pb(n,$\gamma$)$^{209}$Pb. In this case, a discrepancy between the data obtained in an activation measurement and the data from a high-resolution resonance counting experiment was resolved by showing that the difference is due to the DC contribution to the cross section which is only included in the cross section from the activation measurement. For very neutron-rich isotopes of Sn and Pb it was found
that the resulting cross sections differ by orders of magnitude with the different inputs (see Figs.~\ref{fig:dccompsn}, \ref{fig:dccomppb}). This is mainly due to the sensitivity of the cross section to the predicted location of the low-spin bound states with respect to the neutron separation energy.\cite{microdirect} The DC calculation can be tested for $^{132}$Sn(n,$\gamma$) because there is experimental information on the bound states in $^{133}$Sn.\cite{kate,hoff96} This nucleus is predicted to be close to or directly in the r-process path.\cite{microdirect} An independent calculation confirmed the original work.\cite{chiba08} Fortunately for astrophysics, very similar cross sections (within a factor of 3) are computed for this reaction with input from the different microscopic approaches. Unfortunately for nuclear physics, the reaction is not a good case to select a preferred microscopic model for the same reason. Spectroscopic data for neutron-richer isotopes would be necessary. Furthermore, spectroscopic factors were set to Unity in these calculations. This is a good assumption for the states in $^{133}$Sn (and was recently confirmed by Ref.~\refcite{kate}) but is not valid in mid-shell. However, the uncertainty introduced by the predicted bound state energies exceeds by far the one introduced by the spectroscopic factors when only particle states are considered and hole states are neglected.\cite{microdirect} The latter have very small spectroscopic factors and are negligible because they involve a reordering process in the final nucleus (see also Ref.~\refcite{rauenam95} for another example of a reordering process in $^{45}$S and its dependence on deformation).

\subsubsection{Averaged direct capture}
\label{sec:avdc}

Regular DC cross sections are obtained by summing over all allowed transitions to energetically accessible final states,
\begin{equation}
\sigma_\mathrm{DC}^\mu=\sum_\nu \sigma_\mathrm{DC}^{\mu \nu} \quad,
\end{equation}
where each summand contains the appropriate spectroscopic factor. Following the derivations in Sec.~\ref{sec:stellar}, the astrophysical rate will contain an effective cross section and a partition function. Similarly to the HFM, it was suggested (Refs.~\refcite{raurep96,raurejected,holzer97,gordc97}) that the sum over final states may be (partially) replaced by an integration over the NLD in the final nucleus $\rho_f$,
\begin{align}
\bar{\sigma}_\mathrm{DC}^\mu(E)&=\left[ \sum_\nu^{\nu^f} \sigma_\mathrm{DC}^{\mu \nu}(E) \right] + \nonumber \\
&+\int_{E^\mathrm{x}_{\nu^f}}^{S_\mathrm{n}} \sum_{J^f\pi^f} \rho_f(E^f,J^f,\pi^f) \sigma_\mathrm{DC}^{\mu\rightarrow f}(E,E^f,J^f,\pi^f,\bar{\mathcal{S}}_{E^fJ^f\pi^f}) \, dE^f \quad. \label{eq:avdcmu}
\end{align}
Similarly, the summation over initial states $\mu$ in (\ref{eq:effcs}) can be (partially or fully) replaced by an integration over the NLD in the target nucleus $\rho_i$,
\begin{align}
\bar{\sigma}_\mathrm{DC}^\mathrm{eff}&=\left[ \sum_\mu^{\mu^f} \bar{\sigma}_\mathrm{DC}^{\mu} \right] + \nonumber \\
&+ \int_{E^\mathrm{x}_{\mu^f}}^{E_\mathrm{proj}(T)} \sum_{J^i\pi^i} \rho_i(E^i,J^i,\pi^i) \bar{\sigma}_\mathrm{DC}^{i\rightarrow (\nu,f)}(E,E^i,J^i,\pi^i,\bar{\mathcal{S}}_{E^iJ^i\pi^iE^fJ^f\pi^f}) \, dE^i \quad. \label{eq:avdceff}
\end{align}
Through the NLD, the cross sections $\bar{\sigma}_\mathrm{DC}^\mu$ and $\bar{\sigma}_\mathrm{DC}^\mathrm{eff}$ not only include transitions to discrete final states but also ``average'' transitions to states described by the NLD. Therefore I call this \textit{averaged direct capture} (ADC). As in the HFM, the ADC cross section may only include the ground state transitions and NLDs above the ground state when all excited state properties are unknown. The advantage of this is that the sensitivity to the location of discrete states relative to the projectile separation energy (as seen in Ref.~\refcite{microdirect}) is washed out and the change in cross section from one isotope to the next is smoother. Of course, this may not properly describe nuclear structure details in the cross sections when only few states are available but it can give a more appropriate estimate of the magnitude of the rate for astrophysics (which involves an averaging over a relevant energy range and includes more transitions than the laboratory rate) than relying on just a single (semi-)microscopic approach.

Obviously, the ADC cross section will be sensitive to the NLD. Contrary to the HFM, where the NLD mostly impacts the $\gamma$-widths and has its strongest effect at several MeV excitation energy, captures to low-lying states (assuming spins and parities are favorable) are dominating and thus the NLD at low excitation energy will be relevant. This is why it is important to use a proper spin- \textit{and} parity-dependent NLD description.
It was found recently, however, that thermal excitation of target nuclei reduces the
sensitivity to the parity dependence in the NLD for astrophysical rates in the HFM.\cite{darko07} We can
expect a similar effect for direct capture, although the number of possible transitions is
more limited.\cite{darko07}

Appropriate spectroscopic factors are a further important ingredient in ADC calculations. In the integral of (\ref{eq:avdcmu}) these are \textit{averaged} spectroscopic factors $\bar{\mathcal{S}}_{E^fJ^f\pi^f}$, describing the average overlap between the initial state $\mu$ and the final bound states with given spin and parity at an excitation energy $E^f$. The doubly averaged spectroscopic factors $\bar{\mathcal{S}}_{E^iJ^i\pi^iE^fJ^f\pi^f}$ appearing in the integrand of (\ref{eq:avdceff}) are even more complicated, as they involve the average overlap between all initial states with given spin and parity at excitation energy $E^i$ and all final states. Currently, the only spectroscopic factors to be found in literature (from experiment or theory) are for transitions connecting the ground state $\mu=0$ of the target nucleus with the final states. Spectroscopic factors for transitions from excited states have yet to be calculated. They are needed not just for the averaged DC model employing NLDs but also for the regular DC model when applied to compute astrophysical reaction rates (see also the discussion of the stellar enhancement and the effective weights in Sec.~\ref{sec:stellarexp}).

Spectroscopic factors for one-nucleon capture (or transfer) on a target in the ground state can easily be computed from the occupation numbers $v^2$ as calculated, e.g., from BCS or Lipkin-Nogami pairing.\cite{glen83,yosh61,darko07,rauenam95,moll2} Then the spectroscopic factor for putting a nucleon in state $j$ with spin $J_j$ is just
\begin{equation}
\label{eq:occeven}
\mathcal{S}_j=1-v_j^2
\end{equation}
for a target nucleus with an even number of nucleons of the same type as the projectile. The occupation probability is the one of the target nucleus. In a chain of linked reactions, the total processing efficiency is given by the slowest reaction(s). Therefore, when considering sequences of capture reactions, e.g., in the s-, r-, rp-processes, the rates on such target nuclei with even nucleon number will be the slowest (and their reverse photodisintegrations the fastest) and thus they will have the largest astrophysical impact.

When the target nucleus has an odd number of nucleons of the projectile type, the spectroscopic factor is
\begin{equation}
\label{eq:occodd}
\mathcal{S}_j=(2J_j+1)v_j^2 \quad.
\end{equation}
The occupation probability $v_j^2$ is always taken in the nucleus with even number of nucleons. In (\ref{eq:occodd}) this is the final nucleus of the reaction.
The expressions for the extraction of a nucleon from a given state follow from the fact that the time-reversed reactions have to show the same spectroscopic factor. The occupation probabilities can be calculated from microscopic theory, e.g., using BCS or Lipkin-Nogami pairing on a single-particle basis.\cite{glen83,darko07,rauenam95,moll2}

In the absence of calculated spectroscopic factors several different approaches have been used in the past. Often, spectroscopic factors were set constant to 1.0 (e.g., Refs.~\refcite{microdirect,ohu91,krausmann}) or to
 0.1 (e.g., Refs.~\refcite{gordc97,cf88}). These values can already be seen as averaged spectroscopic factors $\bar{\mathcal{S}}_{E^fJ^f\pi^f}$ for low-lying particle-states.

Useful for the application to the ADC approach is the construction of an excitation energy-dependent, averaged spectroscopic factor to be employed along with the NLD. As an example for this, in studying neutron capture on the astrophysically important nucleus $^{44}$Ti the DC component was estimated implementing a distributed spectroscopic strength.\cite{ejnis} In this case, transitions to $1/2^-$ and $3/2^-$ states in $^{45}$Ti are dominating. A distribution $\phi_{\bar{\mathcal{S}}}$ of the $1/2^-$ and $3/2^-$ strengths was assumed, reaching from the (experimentally known) location of the lowest $1/2^-$ and $3/2^-$ state, respectively, to the neutron separation energy. With this ``smearing out'' of the states and of the strength and due to the $E_\gamma^3$ dependence of the E1 transition probability, then the calculation is reduced to computing a transition to an effective bound state with full spectroscopic strength at an energy of
\begin{equation}
\bar{E}_\mathrm{bound}=\sqrt[3]{\widetilde{E^3_\gamma}}\quad,
\end{equation}
where the average transition energy is given by
\begin{equation}
\widetilde{E^3_\gamma}=\frac{\int_0^{E_{\gamma,\mathrm{max}}}E_\gamma^3\phi_{\bar{\mathcal{S}}}(E_\gamma)\,dE_\gamma}{\int_0^{E_{\gamma,\mathrm{max}}}\phi_{\bar{\mathcal{S}}}(E_\gamma)\,dE_\gamma}
\quad. \label{eq:spectdist}
\end{equation}
Such an approach accounts for the uncertainties in spectroscopic strength and location of excited states and can be viewed as a zeroth approximation to the ADC.

\begin{figure}
\centerline{\includegraphics[width=0.63\columnwidth,clip]{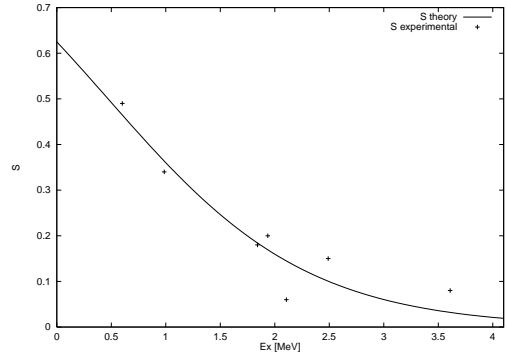}}
\vspace*{8pt}
\caption{\label{fig:avdcspect}Comparison between experimental and averaged (Eq.~\protect\ref{eq:holzerspectfact})
spectroscopic factors
as function of excitation energy for $^{136}$Xe+n.\protect\cite{holzer97} }
\end{figure}

Another suggestion for the functional form of the energy-dependence of the average spectroscopic factors was made in Refs.~\refcite{raurep96,raurejected,holzer97}.
The spectroscopic factors describe the overlap between the antisymmetrized wave
functions of target+nucleon and the final state.
The number of final state configurations increases with increasing excitation energy
$E^\mathrm{x}$ and the overlap of initial and final state wavefunctions decreases. Thus, also the
spectroscopic factor decreases. In a simple approximation, the energy dependence of the
spectroscopic factor for single nucleon transfer can be parameterized by a Fermi function with
\begin{equation}
\label{eq:holzerspectfact}
\mathcal{S}=f_\mathrm{Fermi}(E^\mathrm{x})=\frac{1}{1+e^\frac{E^\mathrm{x}-E^*}{\Delta^*}}
\end{equation}
and the parameters $E^*,\Delta^*$. This is motivated by the excitation-energy dependence of the occupation probabilities. Figure \ref{fig:avdcspect} shows how well averaged spectroscopic factors with the functional dependence (\ref{eq:holzerspectfact}) compare to experimental ones for $^{136}$Xe+n.\cite{holzer97}

There is also a connection between single-particle spectroscopic factors $\mathcal{S}_\mathrm{sp}$ and the partial resonance widths $\Gamma=\Gamma^\mu,\hat{\Gamma},\dots$ appearing in the BWF (see Sec.~\ref{sec:reso})
\begin{equation}
\label{eq:reduwidth}
\Gamma=2P_\ell C^2 \mathcal{S}_\mathrm{sp} \theta_0^2 \frac{\hbar^2}{mr_\mathrm{nuc}^2} \quad,
\end{equation}
where $C$ is the isospin Clebsch-Gordan coefficient, $m$ the reduced mass of the system nucleus+particle, and $\theta_0^2$ is the dimensionless single-particle reduced width.\cite{cauldrons,ilibook,ili97,lanethomas,schiff63} The penetrability $P_\ell$ for the relative angular momentum $\ell$ can be expressed in terms of the regular and irregular Coloumb wavefunctions $F_\ell$ and $G_\ell$
\begin{equation}
P_\ell=\left(\frac{kr}{F_\ell^2(r)+G_\ell^2(r)}\right)_{r=r_\mathrm{nuc}} \quad.
\end{equation}
In literature, a value of $\theta_0^2=0.6$ is often assumed for an average single-particle reduced width.
Comparing (\ref{eq:spwidth}) and (\ref{eq:reduwidth}), it can be seen that an average $\theta_0^2$ can be calculated from solutions of the radial Schr\"odinger equation (giving $\Gamma$) with an optical potential.\cite{ili97,schiff63} Since $\langle \Gamma \rangle=T_\mathrm{trans}/(2\pi \rho)$, there is a direct connection to the average transmission coefficients $T_\mathrm{trans}=T^\mu,\hat{T},\dots$ appearing in the HFM (see Sec.~\ref{sec:statmod}) which can be used to consistently derive the average combined value $\langle \mathcal{S}\rangle=\langle \mathcal{S}_\mathrm{sp} \theta_0^2 \rangle$. The averaged spectroscopic factors for transitions from excited target states in (\ref{eq:avdceff}) can be estimated in this way.

\begin{figure}
\centerline{\includegraphics[width=0.63\columnwidth,clip]{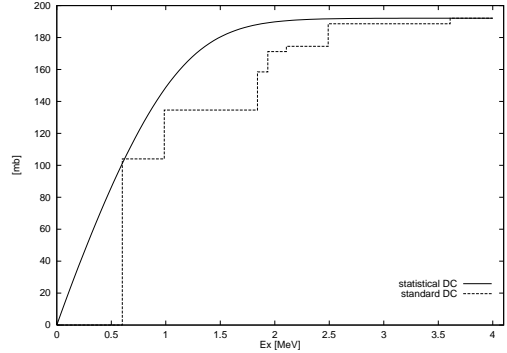}}
\caption{\label{fig:holzeravdc}Sums of the contributions to the total direct capture
cross sections with the averaged DC model and the standard DC
for $^{136}$Xe(n,$\gamma$) as function of excitation energy of the final nucleus.\protect\cite{holzer97} }
\end{figure}

The integrals appearing in (\ref{eq:avdcmu}) and (\ref{eq:avdceff}) always contain products $\rho \langle \mathcal{S} \rangle$ of NLD and averaged spectroscopic factors. With the relation between strength function $s_f=T_\mathrm{trans}/(2\pi)$ and reduced width (see, e.g., Ref.~\refcite{mugha}) we can derive
\begin{equation}
\rho \langle \mathcal{S} \rangle = \frac{mr_\mathrm{nuc}}{\hbar^2} s_f = \frac{2\pi m r_\mathrm{nuc}}{h} T_\mathrm{trans}
\label{eq:avspect}
\end{equation}
and thus (partially) eliminate the NLD within the integrals as it is implicitly contained in the $s_f$ or $T_\mathrm{trans}$.
The strength function $s_f$ is only defined for resonance states above the projectile separation energy. For estimating transitions to bound states it was suggested to construct an ``internal strength function'' $s_\mathrm{int}$, extending the regular strength function below the separation energy.\cite{raurep96,raurejected,holzer97} This is motivated by the relation (\ref{eq:avspect}) and the fact that the spectroscopic factors describe the structure of bound states as well. The ADC was investigated in Ref.~\refcite{holzer97} for the reactions $^{136}$Xe(n,$\gamma$)$^{137}$Xe and $^{144}$Sm(n,$\gamma$)$^{145}$Sm, using (\ref{eq:lanelynn}) with an internal strength function. The ADC cross section (for the target in the ground state)
was compared to a calculation summing over transitions to bound states in the standard potential model (as described in Sec.~\ref{sec:dc}) and with experimental data. The energy-dependence of $s_\mathrm{int}$ was chosen very similar to the one found for the average spectroscopic factors $s_\mathrm{int}=C^* f_\mathrm{Fermi}$, with independent parameters $C^*$, $E^*$, $\Delta^*$. The parameters were determined by requiring $s_\mathrm{int}=s_f$ at the neutron separation energy. Simultaneously, it was required that the ADC cross section integrated up to the excitation energy of the last included state yields the same cross section as obtained with the standard potential model. Although this is not suited for a prediction, it can be used to assess the validity of the assumptions. Including only s-wave neutrons, the experimental cross sections were reproduced within 25\% for $^{136}$Xe(n,$\gamma$) and 2\% for $^{144}$Sm(n,$\gamma$). Figure \ref{fig:holzeravdc} displays a comparison between the results from the
averaged direct neutron capture and standard direct neutron capture for $^{136}$Xe.

For predictions across the nuclear chart, the spectroscopic factors and/or internal strength functions can be obtained from optical model single-particle states and/or occupation numbers of quasi-particle states as shown in (\ref{eq:occeven}) and (\ref{eq:occodd}). For simpler application and to increase computational speeds, these can be parameterized according to (\ref{eq:holzerspectfact}). Spectroscopic factors for transitions from excited target states or the doubly averaged factors required in (\ref{eq:avdceff}) remain an open problem.

\begin{figure}
\centerline{\includegraphics[angle=-90,width=0.63\columnwidth,clip]{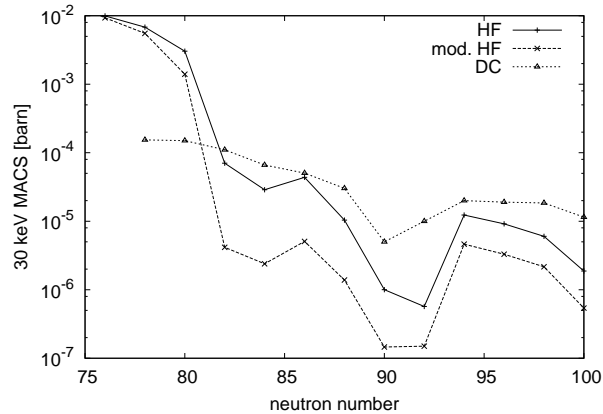}}
\vspace*{8pt}
\caption{\label{fig:avdc}Comparison between averaged direct capture with energy-dependent spectroscopic factor (Sec.~\protect\ref{sec:avdc}), modified Hauser-Feshbach capture (Sec.~\protect\ref{sec:statmodmod}), and standard Hauser-Feshbach capture (Sec.~\protect\ref{sec:statmod}) on even Sn isotopes (preliminary results).\protect\cite{raujpg}}
\end{figure}

The code SMARAGD (see Sec.~\ref{sec:codes}) will also include a global DC treatment using an ADC model and energy-dependent spectroscopic factors, making use of a combination of the above approaches intended to also yield consistency with the HFM.\cite{bon07,raujpg,osaka} Such an ADC approach aims at providing robust predictions despite of considerable differences between microscopic predictions.\cite{microdirect} Preliminary results for this ADC are shown in Fig.~\ref{fig:avdc}, along with results from the HFM (Sec.~\ref{sec:statmod}) and a modified HFM (Sec.~\ref{sec:statmodmod}).
The final rate (or cross section) is the sum of the modified HFM value and the ADC one. Interestingly, for the isotopes shown here (except for $N=92$) this sum is approximated by the unmodified HFM result within a factor of 10. This is in accordance with Ref.~\refcite{gordc} (see figure 3 therein), where the DC contribution also almost replaces the standard HFM values. This shows that it seems justified to use unmodified HFM rates as crude estimate of the total rates for exotic nuclei.

\section{Conclusion}
\label{sec:conclusion}

There is no fast highway to improvements in predictions of reaction rates, not even a wide road. Historically, the fields of standard nuclear physics and nuclear data for applications have taken another direction. Therefore -- instead of following a beaten track -- rather a new, narrow path has to be driven step by step through a jungle of complications and details. This is only possible in a concerted effort of theory and experiment, the latter involving both large-scale rare-isotope production sites and smaller facilities. There is not one ``most important'' nucleus or ``most important'' reaction in Nuclear Astrophysics. Which nuclei, nuclear properties, and reactions are at the center of attention depends on the astrophysical process studied. Therefore, systematical studies are needed as well as information on specific nuclei and reactions. On the theory side large-scale studies of general trends and dependences are required as well as detailed predictions of individual nuclear properties.

As shown above, the calculation of astrophysical rates, even when experimental information is present, involves a number of specialities not encountered in usual nuclear physics investigations. Thus, Nuclear Astrophysics is heavily relying on advances in nuclear theory and experiment but also requires its own special developments in theory and experiment which justify its existence as a separate field. Nuclear physicists working at the boundary to astrophysics have to be aware of these special requirements and it is one of the aims of this work to having outlined a number of them. A further noteable fact is that it is necessary to not only point out special effects or important possible improvements but to actually apply them across the nuclear chart and produce large-scale sets of reaction rates which can be readily used by astrophysicists. This implies that they are made accessible in a form suitable to be implemented in astrophysical models.

It should not be forgotten that the other essential aspect of Nuclear Astrophysics is the astrophysical modeling using reaction networks. Only in conjunction with this part of the field progress can be made. The models set the stage and define the ranges of conditions within which nuclear processes occur. Nevertheless, it can be treacherous to rely too strongly on a certain model. Reliable nuclear models and astrophysical reaction rates should cover a large range of possibilities and provide a sound base for pinpointing the sites of nucleosynthesis processes or even for discovering new types of nucleosynthesis in different astrophysical models. Moreover, the modifications of the cross sections and rates in a stellar plasma are an interesting topic in itself and warrant an independent study even without connection to a specific astrophysical site.

The prediction of astrophysical reaction rates takes nuclear physics in a new direction and tests nuclear theory at the limits. Due to the finite number of nuclei, however, this is a finite task. This is also why parameterizations or phenomenological models may still have their justification, if designed in an appropriate manner. We do not have to extrapolate to infinity but within rather limited ranges of nucleon numbers. It is reassuring that the overall abundance distributions obtained by combining several postulated nucleosynthesis processes are already closely resembling what we find in nature, even when explosive events and highly unstable nuclei are involved. This tells us that the most important properties seem to be described acceptably well. There are remarkable exceptions, however, both in explaining abundance distributions (e.g., of the p-nuclei, the light s-process elements, isotopic anomalies in meteorites, the heaviest nuclei at the endpoint of the r-process, and many more) and in assigning astrophysical sites (e.g., to the r-process, probably also to parts of the p-process). A detailed understanding and reliable reaction rates are also essential for using nuclear cosmochronometry to determine astronomical timescales. To go beyond previous estimates and reach a new level of detail, however, requires a large, dedicated effort of both experiment and theory.

\section*{Acknowledgments}

This work was supported by the Swiss National Science Foundation, grant 200020-105328, and by the European Commission within the ENSAR/THEXO project.


\end{document}